\documentclass{aa}

\usepackage{newtxtext,newtxmath}
\usepackage[T1]{fontenc}
\usepackage{ae,aecompl}

\usepackage{natbib}
\bibpunct{(}{)}{;}{a}{}{,} 

\usepackage{graphicx}   
\usepackage{amsmath}    
\usepackage{amssymb}    
\usepackage{xcolor}
\usepackage[normalem]{ulem}
\usepackage{subcaption}
\usepackage{booktabs}
\usepackage[colorlinks, citecolor=blue, linkcolor=blue]{hyperref}
\usepackage{orcidlink}

\providecommand{\gaia}{\textsl{Gaia }}
\providecommand{\gaianospace}{\textit{Gaia}}

\providecommand{\degrnospace}{$^{\circ}$}

\defcitealias{belokurov16}{BK16}
\defcitealias{navarrete19}{N19}
\defcitealias{Lilleengen2023}{L23}

\begin{document}

\title{Beyond the Clouds: S3 as the most \\ distant extended Milky Way stream, not of LMC origin}

\author{Ó. Jiménez-Arranz\inst{1}
   \and S. Lilleengen\inst{2}
   \and M. S. Petersen\inst{3}
}

\institute{{Lund Observatory, Division of Astrophysics, Department of Physics, Lund University, Box 43, SE-22100, Lund, Sweden\\  \email{oscar.jimenez\_arranz@fysik.lu.se}}
\and
{Institute for Computational Cosmology, Department of Physics, Durham University, South Road, Durham DH1 3LE, UK}
\and
{Institute for Astronomy, University of Edinburgh, Royal Observatory Edinburgh, Blackford Hill, Edinburgh EH93HJ, UK}
}

\date{Received <date> / Accepted <date>}

\abstract 
{While the LMC’s influence on Milky Way (MW) stellar streams has been extensively studied, streams associated with the Clouds have received far less attention. Beyond the Magellanic Stream, only four stream candidates (S1--S4) have been reported.}
{We focus on the S3 stream, a long ($\sim$30\degrnospace) and narrow ($\sim$1.2\degrnospace) structure at 60--80 kpc, nearly aligned with the LMC. Our goals are: 1) to validate the stream through a kinematic analysis of S3 candidates with \gaia DR3 data; 2) to enlarge the sample of potential members with machine-learning methods; and 3) to model the stream in order to test its association with either the MW or the LMC.} 
{We selected new S3 candidates with a supervised neural network classifier trained on \gaia DR3 astrometry and photometry, and further reduced contamination through a polygon cut in the proper-motion space. To investigate the origin of S3, we evolve stream models within time-dependent, deforming MW and LMC haloes, thereby accounting for possible effects of the MW--LMC interaction.} 
{We identify 1,542 high-confidence new S3 stream candidates and find that the stream’s apparent width has grown from $\sim$1.2$^\circ$ to $\sim$3--4$^\circ$ compared to previous studies. We also present a list of 440 potential S3 red clump stars, which are valuable targets for spectroscopic follow-up thanks to their well-defined luminosities and ability to yield precise distances. Both modelling and a comparison of S3 stars’ closest approach distance and velocity with the LMC’s escape velocity indicate that S3 is unlikely to originate from the LMC, instead representing a distant ($\sim$75 kpc) MW stream.}
{S3 is the most distant ($\sim$75 kpc) extended ($\sim$30$^\circ$ long, $\sim$3–4$^\circ$ thick) MW stream known, offering a unique probe of the outer halo and the LMC’s recent influence. Its angular width corresponds to a physical thickness of $\sim$4–5 kpc, making S3 among the thickest streams discovered.}



\maketitle

\section{Introduction}
\label{sec:introduction}

Stellar streams are elongated structures of stars that originate from the tidal disruption of globular clusters or dwarf galaxies as they interact with the gravitational potential of the host galaxy. These coherent structures are among the most powerful tracers of the host galaxy's assembly history and its dark matter distribution \citep[e.g.,][]{springel_white99,dubinski99,johnston99,johnston05,binney08,koposov10,law-majewski10,price_whelan14,erkal16,bovy16,shipp21,pearson22,koposov23,ibata24}. In the Local Universe \citep[e.g.,][]{martínez-delgado10,bilek20,martínez-delgado23,miro-carretero24,martínez-delgado25} and beyond \citep[e.g.,][]{kado-fong18}, dozens of tidal streams have been found thanks to facilities designed to detect low surface-brightness features.

Closer, in our own Galaxy, the Milky Way (MW), we now count nearly 150 candidate stellar streams \citep{mateu23}, with an explosion of stream discoveries over the past decade from surveys such as the Dark Energy Survey \citep[\textsl{DES},][]{dark_energy_survey05}, the Sloan Digital Sky Survey \citep[\textsl{SDSS},][]{kollmeier17,kollmeier25}, and, in particular, the \gaia mission \citep{gaiadr2summary,gaiaedr3summary,gaiadr3summary}. Thanks to \gaianospace, over the past ten years, we have not only increased the number of known MW streams, but also, for the first time, uncovered their kinematics. This constrains their exact orbits, as well as the origins and the complex structure of stellar streams, pointing to dynamical histories shaped by significant perturbations. In the future, we anticipate that telescopes like the Vera Rubin Observatory \citep{ivezic19} and the Nancy Grace Roman Space Telescope \citep{spergel15} will find dozens more streams in both the MW and other external galaxies \citep[e.g.,][]{pearson22,bonaca_pricewhelan25}.

When detecting stellar streams, the primary difficulty comes from their intrinsic low stellar densities and brightness, which make them challenging to observe and study. Historically, the central approach to locating and characterizing stellar streams focused on enhancing contrast against the foreground MW stars -- either by targeting rare tracers more commonly associated with streams than with the field, or by filtering datasets to boost the relative presence of stream stars. This approach, which focused on photometric data from the Galactic halo -- where the background stellar density is naturally lower than in the disc or central regions -- enabled the discovery of the first stellar streams and substructures around the Galaxy \citep[e.g.,][]{rockosi02,newberg02,grillmair_dionatos06,belokurov06}.

However, even though a stream can be detected using matched filtering on photometry alone, its density is usually not well constrained because the stream stars are frequently low signal-to-noise features over the background stellar density. Shortly after \gaia DR2, \citet{price_whelan_bonaca18} demonstrated the power of combining kinematic and photometric data to identify stream members and estimate stream's density in the MW  -- applying this approach specifically to the GD-1 stream. They showed how the exceptional astrometric precision provided by \gaia enabled the development of a wide range of methods for probing the density structure of known streams \citep[e.g.,][]{koposov19,ferguson22,tavangar22,patrick22,starkman23,tavangar25}, and also paved the way for an entirely new class of techniques dedicated to discovering previously unknown streams in the MW’s halo \citep[e.g.,][]{malhan-ibata18,borsato20,necib20,gatto20,shih22,pettee24}. All of these methods take advantage of the kinematic data provided by \gaia to identify stellar over-densities across various projections or transformations of phase-space.

Moreover, the advent of \gaia kinematics has not only enabled the discovery of numerous new stellar streams, but has also provided a powerful means to reassess and validate previously identified ones with unprecedented precision. Among the most compelling and overlooked environments for such studies are the LMC and SMC (hereafter also referred to as the Clouds), located at a distance of approximately 50–60 kpc \citep{graczyk14, pietrzynski19}. As the most massive satellite galaxies of the MW, the Clouds offer a valuable opportunity to explore a range of dynamical phenomena. These include tidal interactions \citep[e.g.,][]{besla12, vasiliev23b, jimenez-arranz24b, jimenez-arranz25b}, dynamical perturbations \citep[e.g.,][]{Vasiliev2018, jimenez-arranz23a, kacharov24, jimenez-arranz24a, jimenez-arranz25a, rathore25, scholch25}, and stream formation \citep[e.g.,][]{nidever08, lucchini20, lucchini21,chandra23,zaritsky25}, among others. Crucially, the Clouds occupy a regime that is external to the MW, yet close enough to allow detailed studies of resolved stellar populations and coherent dynamical structures.

While a considerable amount of effort has been devoted to understanding the impact of the LMC on the dynamics and morphology of MW stellar streams \citep[e.g.,][]{erkal19,koposov19,shipp21,vasiliev21,koposov23,Lilleengen2023,brooks24}, comparatively little work has focused on the streams that are themselves associated with the LMC and SMC. To the best of our knowledge, only four potential stellar streams associated to the Clouds have been reported in the literature -- besides the well-known Magellanic Stream \citep[e.g.,][]{bajaja85,putman00,putman03b,nidever08,donghia-fox16,lucchini21,Petersen.lmc.2022,chandra23,zaritsky25}. In a seminal study, \citet[][hereafter BK16]{belokurov16} identified several narrow stellar streams and diffuse debris clouds, cataloguing the (four) most prominent -- labelled from S1 to S4 according to the distance modulus bin they occupy -- by applying photometric filtering to blue horizontal branch (BHB) stars detected in \textsl{DES} \citep[][]{diehl15,koposov15}. Following up on \citetalias{belokurov16}, \citet[][hereafter N19]{navarrete19} conducted a spectroscopic follow-up program of the four stellar stream candidates, where only two of the four (S1 and S2) were confirmed to have an LMC/SMC origin. Discovered in the pre-\gaia era (2016, prior to \gaia DR2 in 2018), these streams have since been largely overlooked by the community, with no follow-up studies incorporating astrometric data to date.

In this work, we build upon these previous analyses, focusing on the S3 stream, a long ($\sim$30\degrnospace) and narrow ($\sim$1.2\degrnospace) stream at distances ranging from 60 to 80 kpc that points nearly exactly in the direction of the LMC. We pursue three primary objectives: 1) to extend the kinematic analysis of the S3 stellar candidates identified by \citetalias{navarrete19} by incorporating astrometric data from \gaia Data Release 3 (DR3), with the goal of reassessing and validating the stream’s existence; 2) to expand the sample of potential S3 members using machine learning techniques; and 3) to generate stream models to determine S3's association to either the MW or the LMC and to gain a better understanding of the data and future observation needs.

This paper is organised as follows: In Sect. \ref{sec:data}, we describe the datasets used to (kinematically) confirm the existence of the S3 stellar stream. In Sect. \ref{sec:classifier}, we present the methodology developed to identify new candidate members using machine learning techniques and to characterize the expanded sample of S3 candidates. In Sect. \ref{sec:simulations}, we present stream models to better understand discrepancies in the data and to confirm that S3 is a MW stream. In Sect. \ref{sec:discussion}, we contextualise and discuss our results. Finally, in Sect. \ref{sec:conclusions}, we summarise the main conclusions of this work.

\section{Data}
\label{sec:data}

In this work, we make use of two distinct datasets. First, in Sect. \ref{subsec:s3_candidates}, we present the S3 candidates from \citetalias{navarrete19}. Second, in Sect. \ref{subsec:gaiadr3}, we describe the \gaia DR3 bulk catalogue, which we use both to kinematically confirm the existence of the stream and to identify new S3 candidates (see Sect. \ref{subsec:neural_network}). 

Hereafter, every on-sky density figure is displayed in the orthographic projection $(x,y,v_x,v_y)$ -- namely, a method of representing 3D objects where the object is viewed along parallel lines that are perpendicular to the plane of the drawing -- of the usual celestial coordinates $(\alpha,\delta)$ and proper motions $(\mu_{\alpha*},\mu_\delta)$, centred in the LMC photometric center, defined as $(\alpha_c,\delta_c)$ = (81.28\degrnospace, --69.78\degrnospace) by \citet{vandermarel01}. Please refer to Eqs. 1 and 2 of \citet{luri20} and Fig. 11 of \citet{jimenez-arranz23a} for additional details on the coordinate transformation.

\subsection{BHB and BS S3 candidates from \citetalias{navarrete19}}
\label{subsec:s3_candidates}

The detection of an extended and lumpy stellar debris distribution around the Clouds was reported by \citetalias{belokurov16} using BHB stars found in \textsl{DES} Year 1 data. The authors of that work reported the discovery of several narrow streams and diffuse debris clouds, and they catalogued the (four) most important of the stellar substructures that were discovered -- labelled from S1 to S4 according to the distance modulus bin they occupy. Among them, the BHBs traced the long ($\sim$30\degrnospace) and narrow ($\sim$1.2\degrnospace) S3 stream, at distances ranging from 60 to 80 kpc, which runs along the great circle with the pole at $(\alpha,\delta)$ = (250.15\degrnospace, 152.35\degrnospace) and points nearly exactly in the direction of the LMC. In that work, the authors postulated that the S3 stream could conceivably be a by-product of the LMC–SMC interaction because of its alignment with the proper motion of the Clouds and its overlap with its gaseous Stream.

As a logical extension of the project, and using the medium-resolution spectrograph FORS2 installed at the Very Large Telescope (VLT), \citetalias{navarrete19} conducted a spectroscopic follow-up program of the four stellar stream candidates found on the outskirts of the LMC by \citetalias{belokurov16}. In that work, a quarter of the stellar stream candidates (25 out of 104) were found to be contaminants, primarily white dwarfs and quasi-stellar objects (QSO). However, for the other 79 stellar stream candidates, the authors used the Balmer lines to create a classification system that distinguished the BHB stars from blue stragglers (BSs). According to their classification, 24 stars are of BHB type, 45 are BSs, and 10 have uncertain classification. 

\begin{figure*}
    \centering
    \includegraphics[width=0.9\textwidth]{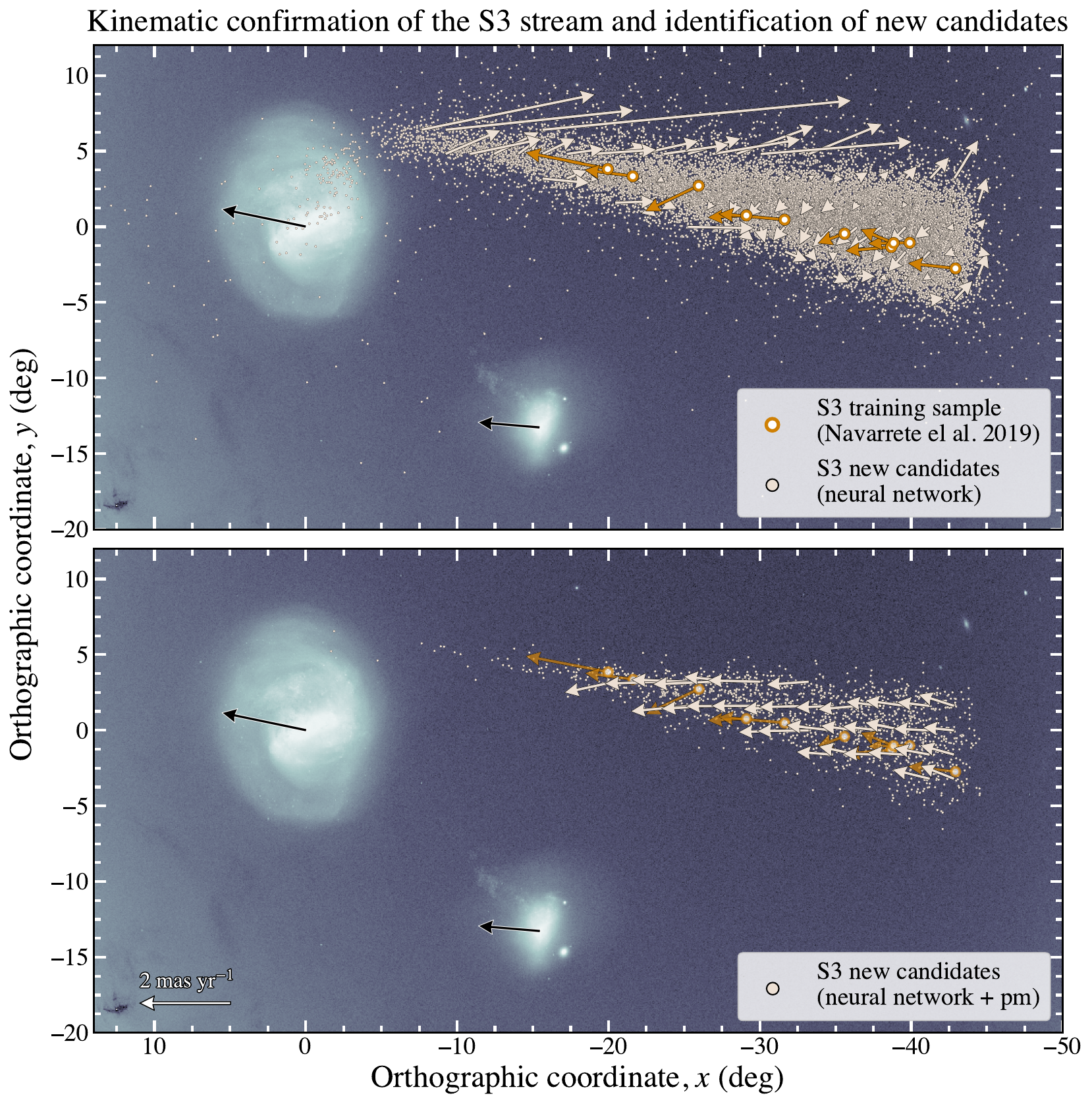}
    \caption{Comparison of the on-sky distribution of the \citetalias{navarrete19}'s BHBs/BSs S3 training sample (orange circles) to the newly identified S3 candidates (beige circles), shown in the top panel using the neural network classifier alone, and in the bottom panel using both the neural network classifier and a polygon selection in proper motion space (see Sect. \ref{subsec:neural_network}). The arrows’ orientation and length indicate the direction and magnitude of the stars’ motion across the sky, with a 2~mas~yr$^{-1}$ white arrow shown as a reference in the bottom panel. For the newly identified S3 candidates, we compute the median direction and magnitude of their motion across the sky within $1.6 \times 1.6$ deg$^2$ bins, displaying the results only for bins containing more than 20 stars. The black arrows indicate the systemic motion of the LMC and SMC. The background image corresponds to a two-dimensional histogram of the \gaia DR3 sample utilized in this study (see Sect. \ref{subsec:gaiadr3}), consisting of 28 million stars that include both stars from the Clouds and foreground halo stars of the MW. Both panels are displayed using the orthographic projection $(x,y,v_x,v_y)$ of the standard celestial coordinates $(\alpha,\delta)$ and proper motions $(\mu_{\alpha*},\mu_\delta)$, centred in the LMC photometric center, defined as $(\alpha_c,\delta_c)$ = (81.28\degrnospace, --69.78\degrnospace) by \citet{vandermarel01}.}
    \label{fig:S3_skymap}
\end{figure*}

In this study, we begin with a sample of 11 BHB and BS stars identified as members of the S3 stream associated with the Clouds (labelled as ``MCs-M1'' or ``MCs-M2'' in Table 1 of \citetalias{navarrete19}). This classification, originally proposed by \citetalias{navarrete19}, is based exclusively on distance and should therefore be interpreted with caution. After crossmatching with \gaia DR3 data (see Sect.~\ref{subsec:gaiadr3}), we remove one source (S3~05) exhibiting near-zero proper motion $(\mu_{\alpha*}, \mu_\delta) \sim (0,0)$ mas yr$^{-1}$, which is indicative of still potential QSO contamination. This sample of 10 S3 stellar candidates, consisting of BHB and BS stars identified by \citetalias{navarrete19}, is first used to reassess and kinematically confirm the presence of the stream using \gaia DR3 data (see Sect. \ref{subsec:gaiadr3}), and subsequently serves as the training set for identifying additional S3 candidates (see Sect. \ref{subsec:neural_network}). Figure \ref{fig:S3_skymap} top panel displays the on-sky distribution of the clean S3 candidates (orange circles) from \citetalias{navarrete19} along with an arrow representing their corresponding proper motion after crossmatching with \gaia data (see Sect. \ref{subsec:gaiadr3}). This figure highlights the directionality and grouping of its member stars and demonstrates how the stream structure is coherent in density. 

It is important to acknowledge some limitations of the \citetalias{navarrete19} dataset. First, the classification between BHB and BS stars is uncertain, which directly impacts the inferred distances (see Sect. \ref{sec:discussion} for more details). Additionally, the reported line-of-sight velocities exhibit significant variation \citepalias[see Fig. 10 of][]{navarrete19} that cannot be fully accounted for at this stage. While these issues introduce some ambiguity, we defer a detailed treatment to the modelling section (see Sect. \ref{sec:simulations}), where we show that, across a reasonable range of line-of-sight velocity assumptions, the stars still trace a coherent stream.

\subsection{\gaia DR3 data: kinematic confirmation of the S3 stream}
\label{subsec:gaiadr3}

Building on the spectroscopic efforts of \citetalias{navarrete19}, which provided missing line-of-sight velocity measurements for stars scattered across the outskirts of the LMC identified by \citetalias{belokurov16}, our study has two main objectives. First, we use \gaia DR3 data to incorporate proper motion information for the known S3 stellar candidates in order to reassess and validate the existence of the stream (see this section). Second, we aim to identify new S3 candidates within the \gaia dataset (see Sect. \ref{subsec:neural_network}).

The \gaia mission is a primarily astrometric (with also photometric and spectroscopic instruments) survey whose main goal is to create the most precise and detailed 3D map of our Galaxy. Insofar, it has catalogued and determined astrometric and photometric data for almost two billion stars \citep{gaiadr2mission,gaiadr2summary,gaiaedr3summary,gaiadr3summary}, representing around $1\%$ of all stars of the MW. Among the vast number of sources observed by \gaianospace, approximately 15 million stars are associated with the Clouds \citep{jimenez-arranz23a,jimenez-arranz23b}. This dataset has proven effective for investigating the internal kinematics of the LMC \citep[e.g.,][]{jimenez-arranz23a,navarrete23,jimenez-arranz24a,kacharov24,dhanush2024,jimenez-arranz25a,rathore25}. Within the field of view considered in this study (see Fig. \ref{fig:S3_skymap}), there are approximately 28 million \gaia DR3 stars, comprising both stars from the Clouds and MW foreground halo stars. This \gaia DR3 sample forms the background for the two panels shown in Fig. \ref{fig:S3_skymap}.

In this section, we crossmatch the sample of 10 BHB and BS stars identified by \citetalias{navarrete19} as potential S3 stream members (see Sect. \ref{subsec:s3_candidates}) with the \gaia DR3 catalogue. We use this sample aiming to reassess and kinematically confirm the presence of the stream structure. Figure \ref{fig:S3_skymap} (top panel) displays the on-sky distribution of the clean S3 candidates (orange circles), with arrows indicating their respective proper motions. The orientation and length of the arrows represent the direction and magnitude of the stars’ motion across the sky. This visualization underscores both the spatial alignment and the coherent motion of the candidate members, illustrating that the stream is not only continuous in position but also coherent in proper motion space. The consistency in their motion provides strong kinematic evidence that S3 is a genuine stellar stream.

\section{Identification of new S3 candidate members with \gaia data}
\label{sec:classifier}

In this section, we present the methodology developed to identify new S3 candidate members within the \gaia DR3 dataset (see Sect. \ref{subsec:gaiadr3}) and to characterize the expanded candidate sample. First, in Sect. \ref{subsec:neural_network}, we introduce the neural network classifier used for the initial selection of new S3 candidate stars. Then, in the same section, we examine the proper motion space of these candidates and identify an additional cut in the proper motion space that helps remove potential contaminants, yielding the final new list of 1,542 S3 candidates based on \gaia data. Finally, in Sect. \ref{subsec:characterisation}, we characterise the new resulting sample of highly reliable S3 stream candidates.

\subsection{Neural network classifier and polygon selection in the proper motion space}
\label{subsec:neural_network}

Since our goal is to develop a classifier capable of identifying stars belonging to the S3 stream within the \gaia DR3 sample (see Sect. \ref{subsec:gaiadr3}), we employ a machine learning approach -- specifically, supervised learning. This requires a well-constructed, labelled training sample so that the classifier can learn to distinguish the characteristics of stars associated with the S3 stream from field stars. As introduced earlier in the manuscript, the training sample combines the 10 S3 stellar candidates -- composed of BHB and BS stars identified by \citetalias{navarrete19} (see Sect. \ref{subsec:s3_candidates}) -- with stars from the \gaia DR3 catalogue (see Sect. \ref{subsec:gaiadr3}). Given the strong imbalance between the two datasets (10 stars vs. 28 million stars, respectively), we replicate the S3 sample 20 times to create a training set of 200 S3 stars, increasing the representation of this class in the training sample. From the \gaia DR3 catalogue, we randomly select a subsample of 10,000 stars to represent the field population. While it is possible that a small number of genuine S3 members may be included in this sample, their presence is expected to be negligible and unlikely to significantly affect the performance of the classifier.
The results in this work have been confirmed to be robust against reasonable variations in these sample sizes.

The neural network architecture employed in this work closely follows the design used by \citet{jimenez-arranz23a,jimenez-arranz23b}. It consists of 11 input neurons, corresponding to 11 parameters either derived from or directly measured by \gaia (detailed below), and three hidden layers containing six, three, and two nodes, respectively. The network outputs a single value, $P$, representing the probability that a given star belongs to the S3 stream. A probability close to 1 indicates a high likelihood of S3 membership, whereas a value near 0 suggests the star is more likely associated with the MW halo or the Clouds. The activation function used in all hidden layers is the rectified linear unit (ReLU), and the model is optimized using the ``adam'' stochastic gradient descent algorithm \citep{kingma17}, with a constant learning rate. Training is performed by minimizing the log-loss function, and to mitigate overfitting, we apply L2 regularization with a strength of 1e-5. 

As input variables, we use the orthographic positions ($x$, $y$), parallax and its uncertainty ($\varpi$, $\sigma_{\varpi}$), orthographic proper motions and their uncertainties\footnote{To compute the uncertainties in the orthographic proper motions $(\sigma_{v_x}, \sigma_{v_y})$, we apply Gaussian error propagation, taking into account both the individual uncertainties in $(\mu_{\alpha*},\mu_\delta)$ and their correlation. It calculates the partial derivatives of $v_x$ and $v_y$ with respect to $\mu_{\alpha*}$ and $\mu_\delta$, and uses them -- along with the covariance -- to propagate the errors into the orthographic frame.} ($v_x$, $v_y$, $\sigma_{v_{x}}$, $\sigma_{v_y}$), along with \gaia photometry ($G$, $G_{\text{BP}}$, $G_{\text{RP}}$). After testing various coordinate systems for the neural network input -- such as equatorial coordinates $(\alpha,\delta,\mu_{\alpha*},\mu_\delta)$ and galactocentric coordinates $(l,b,\mu_l,\mu_b)$ -- we chose to adopt the orthographic projection $(x,y,v_x,v_y)$. This choice avoids issues associated with coordinate singularities at the poles, which affected both the equatorial and galactocentric systems, and resulted in better performance for the classifier. The \textsc{sklearn} Python package \citep{sklearn} was used to create the S3 classifier.

To convert the classifier's output probabilities into a binary classification, we must define a probability threshold. Then, a star is considered a candidate member of the S3 stream if its probability exceeds this threshold, i.e., $P>P_{\text{cut}}$. Choosing a lower $P_{\text{cut}}$ increases completeness by ensuring that few, if any, true S3 members are missed, but this comes at the expense of higher contamination from non-members (field stars). In contrast, a higher threshold improves the purity of the selected sample by reducing contaminants, though it risks excluding genuine S3 stars and thus lowers completeness. In this work, we adopt $P_{\text{cut}}=0.8$, as our priority is to obtain a cleaner, less contaminated sample of S3 candidates, even at the cost of missing some true members\footnote{However, the catalogue of S3 stars released with this work includes the 2,177 stars with $P>0.5$ that also satisfy the polygon selection in proper motion space (see the end of Sect. \ref{subsec:neural_network} for further details). This allows other researchers interested in further studying the S3 stream to choose their own balance between completeness and purity based on their specific scientific goals.}. However, the main results of this work remain unchanged across the probability threshold range of $P_{\text{cut}}=0.5-0.8$. We refer the reader to Appendix \ref{sec:appendix_p05} for the on-sky distribution of the S3 clean samples when using $P_{\text{cut}}=0.5$.

To train and evaluate the performance of the classifier, we split the sample of 10,200 stars (including both S3 and field stars) into two subsets: 60\% for training the algorithm and 40\% for testing it. The classifier’s performance is assessed by computing the receiver operating characteristic (ROC) curve, the precision-recall curve, and their respective areas under the curve (AUC). All these metrics indicate an almost perfect classifier -- see Appendix \ref{sec:appendix_nn} for full details. Nonetheless, these results should be interpreted with caution, as they reflect performance on the test portion of our simulated sample, not on the full \gaia DR3 dataset. In addition, also in Appendix \ref{sec:appendix_nn}, we present an analysis of the SHAP (SHapley Additive exPlanations) values to gain insight into the internal decision-making process of the classifier and to better understand the contribution of each feature to the model’s output.

\begin{figure}
    \centering
    \includegraphics[width=\columnwidth]{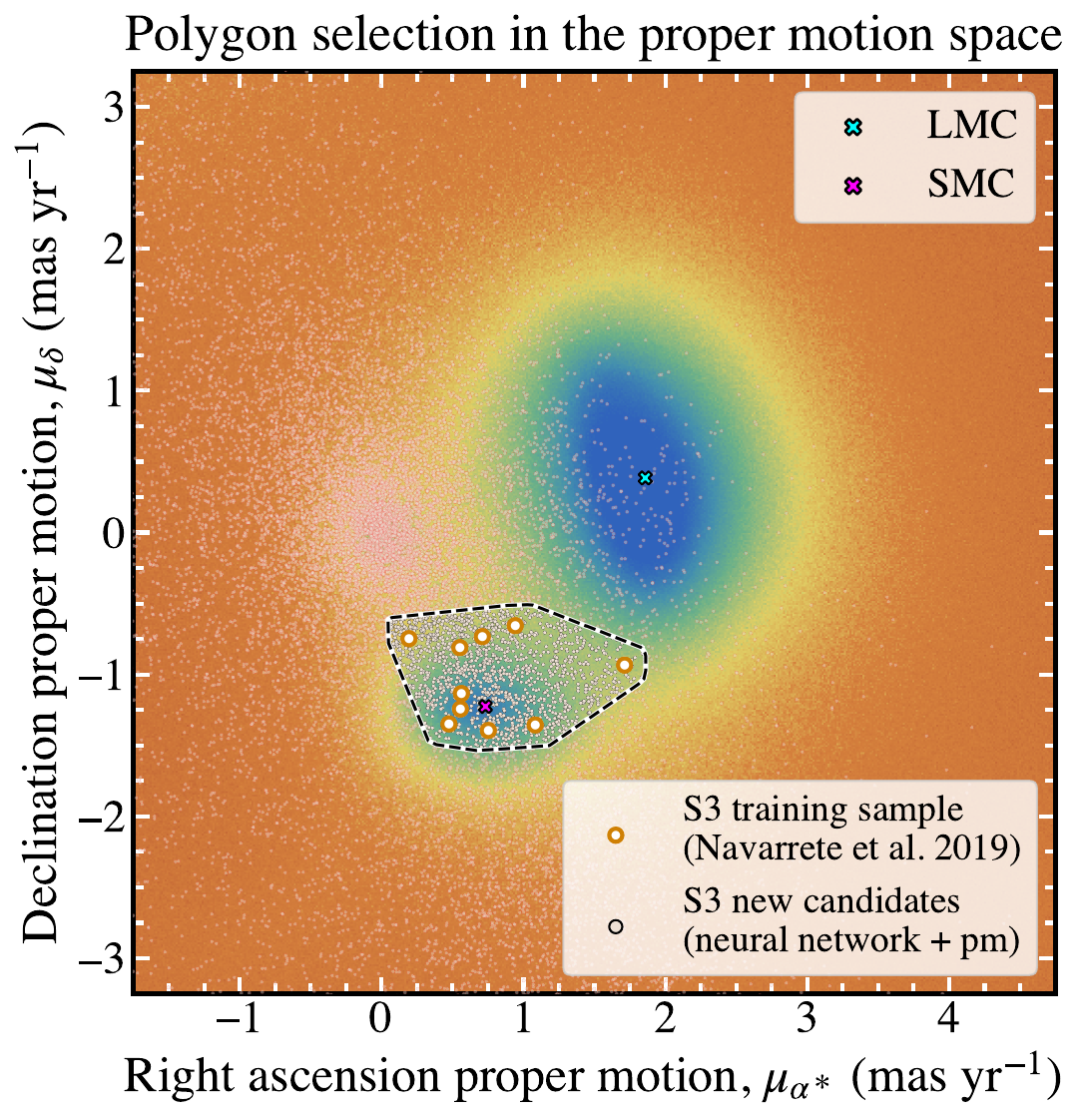}
    \caption{Comparison of the proper motion distribution between the BHBs/BSs S3 training sample from \citetalias{navarrete19} (orange circles) and the newly identified S3 candidates (beige and red transparent circles), overlaid on the \gaia DR3 sample (background histogram, see Sect. \ref{subsec:gaiadr3}), which consists of 28 million stars, including both Clouds members and MW halo stars. The black and white dashed line indicates the polygon selection applied in proper motion space (see Sect. \ref{subsec:neural_network}). The newly identified S3 candidates are shown in beige if they lie inside the polygon selection, and in red transparent if they fall outside it. The cyan and magenta crosses indicate the systemic motions of the LMC and SMC, respectively. In the background, regions of higher (lower) density are shown in bluer (redder) color.}
    \label{fig:polygon_pm}
\end{figure}

When applied to the \gaia DR3 sample (28 million sources, see Sect. \ref{subsec:gaiadr3}), the neural network classifier identifies 25,536 potential S3 stream candidates (beige circles in the top panel of Fig. \ref{fig:S3_skymap}). The spatial distribution of these candidates shows a reasonable alignment with the BHBs/BSs training sample from \citetalias{navarrete19} (orange circles, see Sect. \ref{subsec:s3_candidates}). However, their proper motion vectors are not well aligned with the stream track or with those of the training sample. Figure \ref{fig:polygon_pm} compares the proper motion distribution of the BHBs/BSs S3 training sample from \citetalias{navarrete19} (orange circles) with that of the newly identified S3 candidates selected by the neural network (beige and red transparent circles). In the newly identified S3 candidates sample, we observe two distinct overdensities. The first, centred around $(\mu_{\alpha*},\mu_\delta) \sim (1,-1)$ mas yr$^{-1}$, is consistent with the expected kinematics of the S3 stream. While there is some overlap with the kinematics of the SMC, the implications of this are addressed in the Discussion (see Sect. \ref{sec:discussion}). The second (and more prominent) overdensity, located near $(\mu_{\alpha*},\mu_\delta) \sim (0,0)$ mas yr$^{-1}$, is not associated with the stream and likely represents contamination from unrelated field stars -- or QSO, which typically exhibit very small proper motions.

This contaminating component is filtered out to produce a cleaner and more reliable sample of S3 candidates by applying a polygon selection in the proper motion space (black and white dashed line), computed from a convex hull encompassing the $1-\sigma$ uncertainties of all S3 members' of the training sample to define the region occupied by more reliable stream members. After applying the polygon selection in proper motion space, we retain 1,542 highly reliable ($P>0.8$) S3 stream candidates (beige circles). This is the refined sample that we characterise in Sect. \ref{subsec:characterisation} and use for modelling in Sect. \ref{sec:simulations}. 

Figure \ref{fig:S3_skymap} bottom panel shows the on-sky distribution of the newly identified S3 candidates (beige circles) using both the neural network classifier and a polygon selection in proper motion space, where the arrows’ orientation and length indicate the direction and magnitude of the stars’ motion across the sky. In comparison to the neural network classifier only (top panel), we can see that the stream-like spatial distribution is preserved while the proper motions are more aligned to the stream track -- except the part closer to the LMC, at the left of the stream, where the proper motion of the new S3 candidates do not align with the training sample.

We observe a significant increase in the apparent width of the S3 stream compared to previous studies. In \citetalias{belokurov16}, S3 stream stars are traced out to $\sim1.2^\circ$, while in this work, we identify members extending up to $\sim3^\circ$, and in some regions as wide as $\sim4^\circ$. In \citetalias{navarrete19}, the S3 stream appears even narrower, but this is likely a consequence of selection effects inherent to the spectroscopic sample. A key factor contributing to the broader extent in our analysis may be the difference in selection methodology. \citetalias{belokurov16} relied on photometric criteria targeting specific stellar types such as BHB and BS stars, which naturally limited the sample. In contrast, our neural network approach is more inclusive, allowing a wider range of stellar populations to be identified (see Sect. \ref{subsec:characterisation}), potentially revealing a more complete and extended picture of the stream. If this broader structure is confirmed, S3 would rank among the thickest stellar streams discovered to date, with an apparent width of up to $\sim3-4^\circ$. At the median distance of the S3 training sample ($\sim$73.5 kpc), this corresponds to a physical thickness of $\sim4-5$ kpc.

\subsection{Characterisation of the new S3 candidate sample}
\label{subsec:characterisation}

\begin{figure*}
    \centering
    \includegraphics[width=0.9\textwidth]{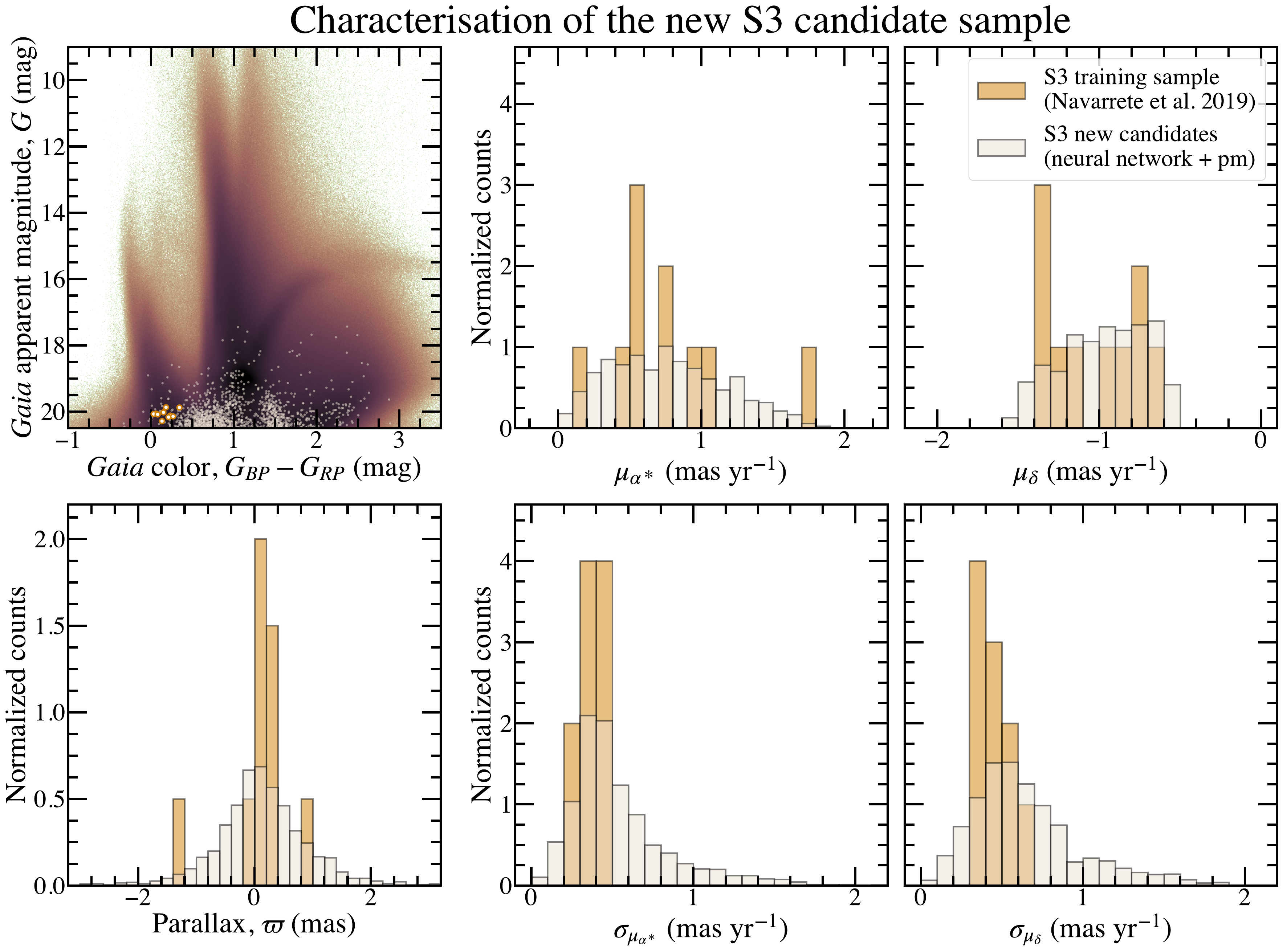}
    \caption{Characterisation of the refined sample of 1,542 new S3 stellar candidates (see Sect. \ref{subsec:neural_network}). Top left panel: CMD of the S3 training sample by \citetalias{navarrete19} (orange circles) and the new S3 candidates (beige circles). The background image corresponds to the CMD of the \gaia DR3 sample utilized in this study (see Sect. \ref{subsec:gaiadr3}), consisting of 28 million stars that include both stars from the Clouds and foreground halo stars of the MW. Top center and right panels: proper motion normalized distributions in right ascension $\mu_{\alpha*}$ and declination $\mu_{\delta}$, respectively. Bottom from left to right: parallax $\varpi$ and proper motion error normalized distributions in right ascension $\sigma_{\mu_{\alpha*}}$ and declination $\sigma_{\mu_{\delta}}$. In the histograms, the S3 training sample from \citetalias{navarrete19} is shown in orange, while the new S3 candidates are displayed in beige.}
    \label{fig:S3_charact}
\end{figure*}

In this section, we analyse the refined sample of 1,542 new S3 stellar candidates, identified through the combined application of the neural network classifier and a polygonal selection in proper motion space. Figure \ref{fig:S3_charact} top panel compares the color-magnitude diagram (CMD) of the training sample by \citetalias{navarrete19} (orange circles) with the new 1,542 S3 candidates (beige circles). We observe that the training sample is concentrated around $G_{\rm BP} - G_{\rm RP} \sim 0-0.5$ and $G \sim 20$, as expected given its composition of BHB and BS stars. However, the first step of our selection process -- the neural network classifier -- successfully generalizes the search for new S3 stars, identifying candidates belonging to a broader range of stellar populations. When compared to the LMC evolutionary phases proposed in \cite{luri20}, shifted to a distance of $\sim$73.5 kpc -- the median distance of the \citetalias{navarrete19} training sample -- we find that the sample of 1,542 new S3 stellar candidates is (as a first indication) predominantly composed of red clump (RC, $29\%$) and RR Lyrae ($25\%$) stars -- further details are provided in Appendix \ref{sec:appendix_stellar_types}. We consider the overdensity at $G_{\rm BP} - G_{\rm RP} \sim 1.5$ and $G \sim 19-20.5$ to be of particular interest, as it may correspond to the RC population of the S3 stream, an especially valuable target for spectroscopic follow-up due to the RC stars’ well-defined luminosities and their potential to provide precise distance measurements\footnote{These 440 RC candidate stars are identified in the catalogue of S3 candidates released alongside this work.} (see further discussion in Sect. \ref{subsec:follow_up_obs}). Given that the CMD polygon cut proposed by \citet{luri20} indicated the possible presence of RR Lyrae stars within our S3 clean sample, we attempted to crossmatch this sample with the \gaia DR3 RR Lyrae catalogue \citep[\texttt{gaiadr3.vari\_rrlyrae};][]{clementini23}, aiming to identify any overlap between the datasets. However, we found only 3 (7) RR Lyrae stars at distances greater than 50 kpc within the neural network S3 sample after (before) applying the proper motion cut. The distances are computed using the absolute magnitude derived in \citet{Iorio.RRlyrae.2021}, with approximate uncertainties of 10\%. The individual distances of those 3 RR Lyrae stars are 60, 55, and 73 kpc, placing some of them on the nearer side of the S3 stream. We refer the reader to Appendix \ref{sec:appendix_rrl} for details.

The top center and right panels of Fig.~\ref{fig:S3_charact} show the proper motion normalized distributions in right ascension $\mu_{\alpha*}$ and declination $\mu_{\delta}$, respectively. We can observe that the proper motion distribution of the new S3 candidates appears smoother and more Gaussian-like, closely resembling that of the training sample. The bottom left panel shows the parallax $\varpi$ distribution where, again, the new S3 candidates show a Gaussian-like distribution. Finally, the bottom center and right panels proper motion error normalized distributions in right ascension $\sigma_{\mu_{\alpha*}}$ and declination $\sigma_{\mu_{\delta}}$, respectively. Here, we observe that the training sample is clustered around $\sigma_{\mu_{\alpha*}}$ and $\sigma_{\mu_{\delta}} \sim 0.5$ mas yr$^{-1}$, whereas the new S3 sample exhibits a tail extending up to approximately $\sim 2$ mas yr$^{-1}$.

\section{Association of the distant S3 stream to the MW through dynamical modelling}
\label{sec:simulations}
Although the data are too sparse to confidently fit the stream, dynamical stream models can be used to gain a better understanding of the stream.
To explore the association of the stream with either the MW or the LMC, and possible effects by either galaxy on the stream, we create models following a similar approach to \citet[hereafter L23]{Lilleengen2023}.
We evolve the stream models in time-dependent and deforming MW and LMC potentials to take into account possible effects of the MW--LMC interaction \citep[see e.g.,][]{erkal19,garavito-camargo19,petersen.lmc.2020,garavito-camargo21,petersen.lmc.2021,Petersen.lmc.2022, koposov23, Arora2024, brooks24,brooks25a,brooks25b, Weerasooriya2025,Yaaqib.lmc.2025}. 

\subsection{Modelling approach}\label{subsec:modelling}
The MW--LMC simulation is evolved using the \textsc{exp} method \citep{Petersen2022,Petesen.joss.2025}, where potential and density are modelled as a sum of orthogonal basis functions with an associated weight quantifying the contribution of the function to the total system.
The coefficients vary over time to describe the time-dependent system, while the functions remain constant. 
This provides tabulated, i.e. fast and lightweight, access to force-replay for integrating orbits. 
The simulation consists of three components: the MW halo, the MW stellar component (disc and bulge), and the LMC halo. 
Further details can be found in \citetalias{Lilleengen2023}, and a Python package for accessing the simulation is available at \url{https://github.com/sophialilleengen/mwlmc}.

\begin{figure*}
    \centering
    \includegraphics[width=\textwidth]{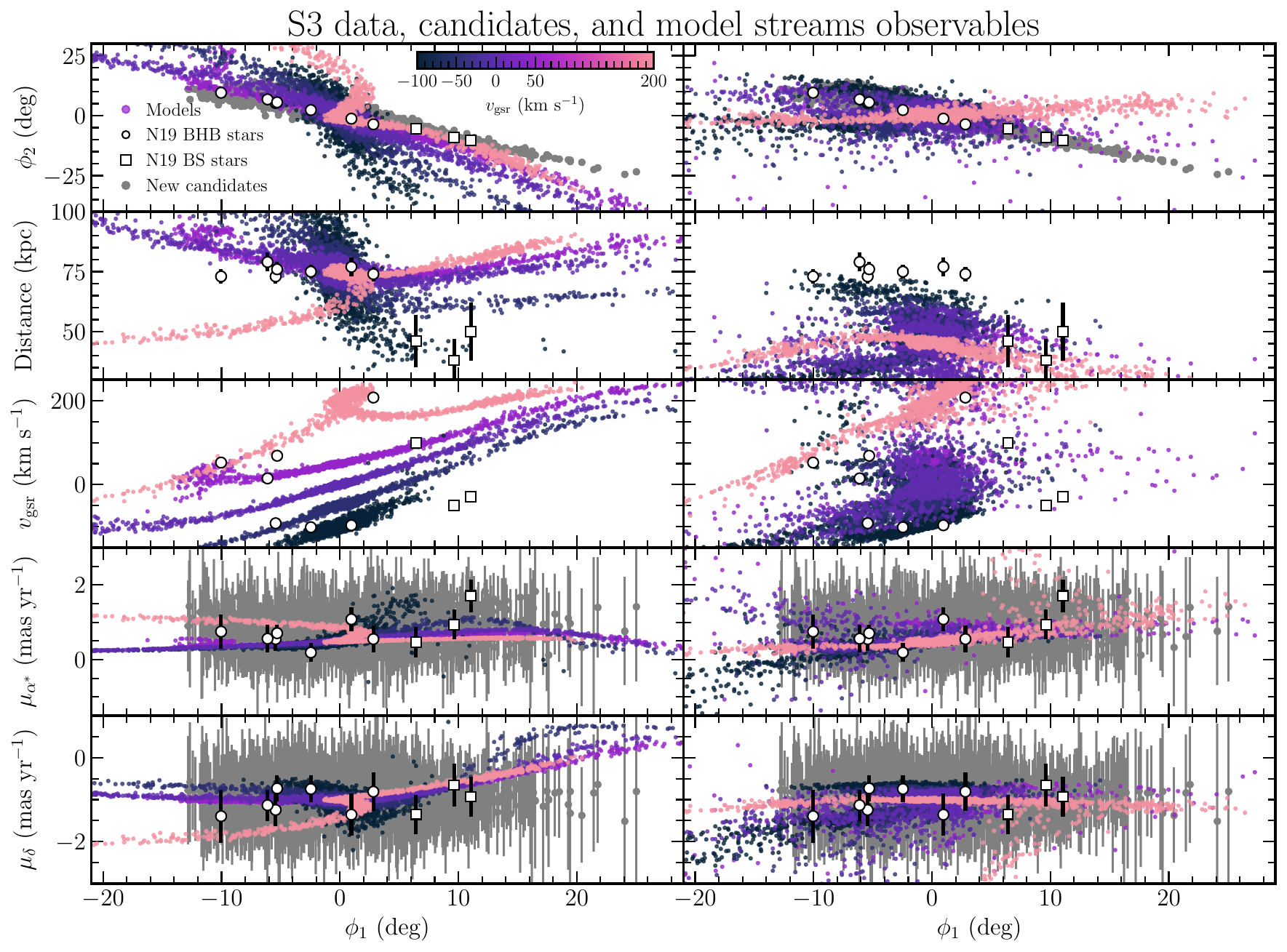}
    \caption{Observables of S3 BHB (BS) data from \citetalias{navarrete19}, represented by white points (squares) symbols with black error bars; neural network candidates, shown as grey points and error bars; and the stream models, shown as coloured points.
    The rows show the stream track, its heliocentric distance, radial velocity in the Galactic standard of rest, and proper motions, not reflex-corrected, respectively. 
    The left column shows model streams with a progenitor distance of 75\,kpc, and the right column with a distance of 45\,kpc.
    The colours refer to the progenitor's Galactic standard of rest radial velocities and are given in the colorbar in the top left panel.
    The streams at the larger distance (75\,kpc) match well with the data, particularly with a radial velocity of 50\,km\,s$^{-1}$.
    The closer progenitors (45\,kpc) fail to produce streams that cover the whole range of all observables in $\phi_1$.    
    }
    \label{fig:S3_stream_observables}
\end{figure*}

We create a set of stream models using the modified Lagrange Cloud Stripping technique \citep{Gibbons2014,erkal19}. 
The progenitor is modelled as a Plummer sphere with a range of masses between $10^5 - 10^7\,\mathrm{M}_\odot$ and scale radii between $0.001 - 0.1$\,kpc.
Present-day phase-space coordinates for the progenitor are estimated from the \citetalias{navarrete19} candidates and the potential members identified in Sect.~\ref{sec:classifier}.
We set the present-day phase-space position of the progenitor at fixed $\alpha=18$\,deg, $\delta=-50$\,deg, $\mu_{\alpha^*}=0.5$\,mas\,yr$^{-1}$, $\mu_\delta=-1.0$\,mas\,yr$^{-1}$.
The stream's distance and radial velocities are more uncertain as discussed in Sect.~\ref{subsec:s3_candidates} -- Figure~\ref{fig:S3_stream_observables} shows the \citetalias{navarrete19} candidates as white points, with distances and radial velocities in the second and third row, respectively. 
We test two distances for the stream, 75\,kpc as indicated by the BHB stars, and 45\,kpc, the approximate mean distance of the BS stars. 
The \citetalias{navarrete19} line-of-sight velocities $v_\mathrm{los}$ are converted into Galactic standard of rest radial velocities $v_\mathrm{gsr}$ using \textsc{Astropy} \citep{astropy:2013, astropy:2018, astropy:2022} conversions. 
Since there is no clear trend in $v_\mathrm{gsr}$ (see the third row in Fig.~\ref{fig:S3_stream_observables}), we try five values for the progenitor that are in the regime of the data: $v_\mathrm{gsr} \in \{200, 50, 0, -50, -100\}$\,km\,s$^{-1}$.
A coordinate system aligned with the stream provides the stream track coordinates $(\phi_1, \phi_2)$.
It follows a great circle with a pole at $(\alpha_\mathrm{S3}, \delta_\mathrm{S3}) = (18^\circ, 40^\circ)$, and has its origin at $(\alpha_0, \delta_0) = (18^\circ, -50^\circ)$ which we chose as the progenitor's position.

The progenitor is rewound in the time-evolving MW--LMC potential for 4\,Gyr.
Then, the system is evolved forward, with tracer particles being released from the progenitor’s Lagrange points, generating a stream.
These Lagrange points are generally calculated with respect to the MW, but with the possibility of S3 being an LMC stream, we also calculate them with respect to the LMC in another run.
However, given the results presented in the remainder of this section, we do not further explore streams evolved around the LMC.
A more in-depth description of the modelling approach is provided in \citetalias{Lilleengen2023}.

\subsection{Modelling results}
\subsubsection{Distance comparison and radial velocities}
We present models with a progenitor mass of $10^7\,\mathrm{M}_\odot$ and a scale radius of 0.1\,kpc, as these best match the observed ranges. 
Figure~\ref{fig:S3_stream_observables} shows the stream observables for all generated models stripping around the MW as coloured points, BHB stars and BS stars from \citetalias{navarrete19} as white circles and squares, respectively, and new candidates as grey points.
The newly identified candidates do not have any radial velocity measurements and only unresolved parallaxes, i.e. no reliable distance measurements.
The colours of the models are set by the progenitors' Galactic standard of rest radial velocities, ranging from 200\,km\,s$^{-1}$ in light pink to $-100$\,km\,s$^{-1}$ in dark blue. 
The left column shows streams at larger distances with the progenitors initialised at 75\,kpc, and the right column shows streams at smaller distances with the progenitors at 45\,kpc. 
This addresses the ambiguity in the data between the BHB and BS streams. 

\begin{figure*}
    \centering
    \includegraphics[width=0.7\textwidth]{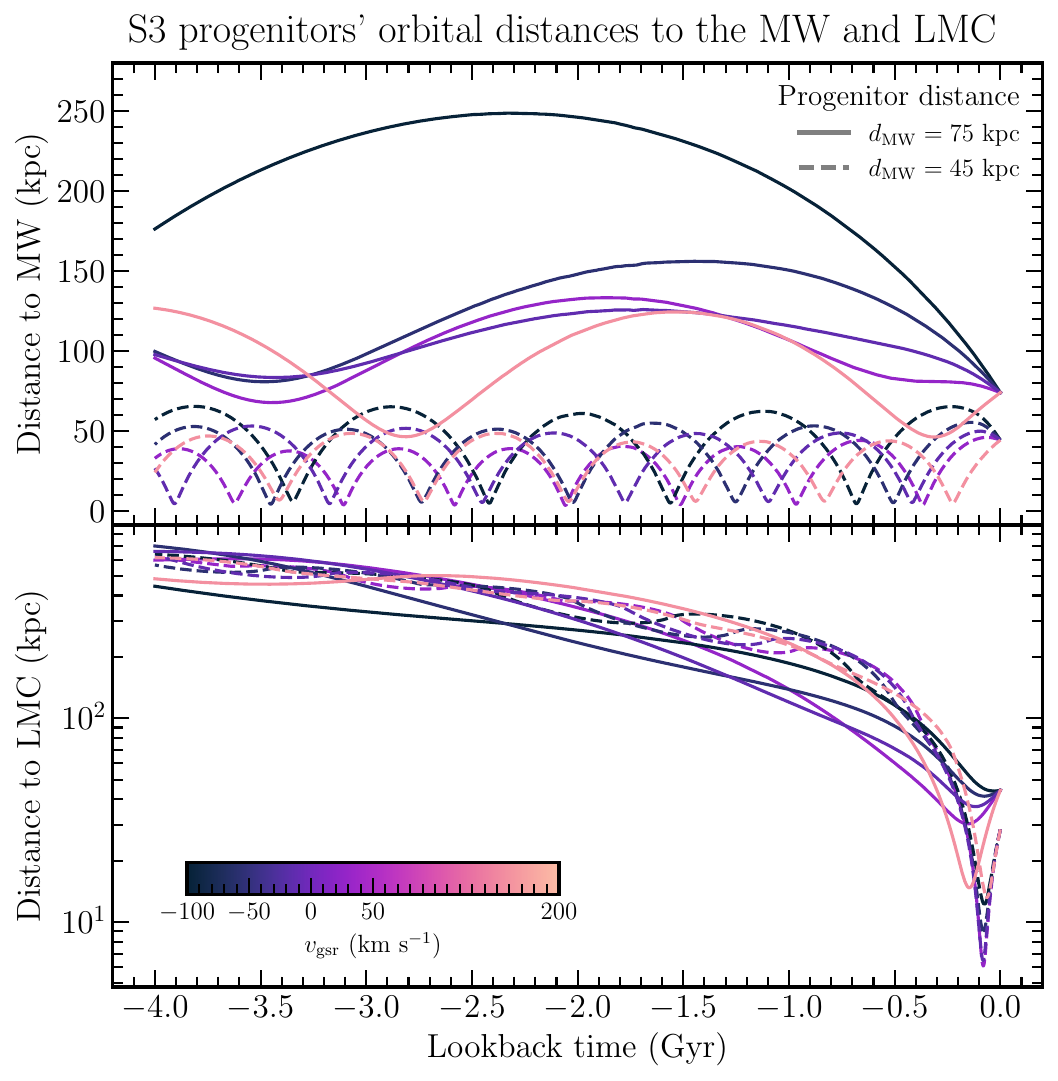}
    \caption{Distances of possible S3 stream progenitors to the MW (top panel) and LMC (bottom panel; log scale) over the past 4 Gyr. 
    The progenitors correspond to the streams shown in Fig.~\ref{fig:S3_stream_observables}.
    The distances of progenitors initialised at 75\,kpc are shown as solid lines, and the distances of progenitors at 45\,kpc are shown as dashed lines. 
    The colours indicate the progenitor's Galactic standard of rest radial velocities and are given in the colorbar in the second row.
    All realisations orbit around the MW, independent of chosen distances and radial velocities. 
    Their closest approach to the LMC is $\sim100$\,Myr ago, where the progenitors of streams at closer distances get within 15\,kpc of the LMC, and the farther progenitors between $15-50$\,kpc.
    }
    \label{fig:S3_progenitor_distances}
\end{figure*}

None of the models exactly match the data; however, they are still helpful to understand more about the S3 stream and inform follow-up observations that will enable stream fitting.
While the streams at larger distances match the data, particularly with progenitor radial velocities of 50 and 0 km\,s$^{-1}$ (light and dark purple points in the third row in Fig.~\ref{fig:S3_stream_observables}), streams at closer distances have a turning point near the progenitor and fail to produce streams that cover the whole data range. 
This is because the progenitors at smaller distances are near apocenter, with pericenters close to the MW center, shown in Fig.~\ref{fig:S3_progenitor_distances}.
The top panel shows the distance to the MW, and the dashed lines are the progenitors at smaller distances.

The observed radial velocities (white points and squares) do not follow any of the models, which show a strong gradient along the stream. 
None of the models at either distance can explain the variety in the observations.
To confirm S3 as a stream, we need spectroscopic follow-up observations to obtain radial velocities of the new S3 candidates provided by the neural network.

\subsubsection{Is S3 a MW or LMC stream?}
Figure~\ref{fig:S3_progenitor_distances} shows the distances between the progenitor and the MW (top panel) and the LMC (bottom panel).
The orbits of the more distant progenitors at 75\,kpc are in solid lines, while the orbits of the closer progenitors at 45\,kpc are in dashed lines.
All progenitors are bound to the MW, with longer orbital periods for the farther streams and shorter orbital periods for the closer streams. 
They have their closest approach to the LMC around 100\, Myr ago, where the closest distances for the farther streams are a few tens kpc.
Figure~\ref{fig:S3_stream_observables} shows that it is unlikely for S3 to be a stream at 45\,kpc. 

We also check whether the SMC could have an effect on any of the stream models.
We first integrate the SMC backwards as a tracer particle in the MW--LMC simulation, where its orbit is bound to the LMC.
Then, we calculate the distance between the SMC particle and the progenitors similar to the panels in Fig.~\ref{fig:S3_progenitor_distances}. 
For all streams, the distance to the SMC is further than to the LMC, indicating that the SMC does not significantly affect the S3 stream.

We conclude that the S3 stream is a distant ($\sim75$\,kpc) MW stream.
The progenitor of the model that best matches the data (light purple line with $d=75$\,kpc, $v_r=50$\,km\,s$^{-1}$) has a closest approach distance of 30\,kpc approximately 150\,Myr ago.
This stream could be affected by the MW--LMC interaction, similar to the Orphan-Chenab (OC) stream (see \citetalias{Lilleengen2023}).
In Sect.~\ref{subsec:S3_context}, we discuss how this makes S3 an exciting prospect for measuring the MW halo and the MW--LMC interaction at large distances pending spectroscopic follow-up observations (Sect.~\ref{subsec:follow_up_obs}).

\section{Discussion}
\label{sec:discussion}

A considerable amount of effort has been devoted to understanding the impact of the LMC on the dynamics and morphology of MW stellar streams \citep[e.g.,][]{erkal19,koposov19,shipp21,vasiliev21,koposov23,Lilleengen2023,brooks24}. These studies have provided important insights into how the infall of the LMC perturbs the Galactic halo and influences the trajectories of MW substructures. In contrast, much less attention has been paid to stellar streams that are thought to be associated with the LMC, but whose membership remains largely tentative -- such as the streams S1--S4 found by \citetalias{belokurov16} and later characterised by \citetalias{navarrete19}. Despite their potential to offer direct constraints on the LMC’s mass distribution, orbital history, and interaction with the MW, systematic efforts to identify and characterise potential LMC streams remain relatively scarce.

To address this gap, in this work, we first use \gaia DR3 proper motions to kinematically characterise and confirm the existence of the S3 stellar stream, which had previously been identified only through photometric data. Building on this, we apply a neural network classifier to search for new S3 candidate members, identifying 1,542 stars. This represents a substantial increase over the $\sim$10 stars previously known and provides a valuable foundation for future studies of the stream’s origin, properties, and possible association with the LMC. In Sect.~\ref{subsec:S3_context}, we place the S3 system in context by addressing its MW or LMC origin and outlining potential use-cases for its study, while in Sect.~\ref{subsec:follow_up_obs}, we discuss possible follow-up observations aimed at confirming candidate members and further constraining the nature of the stream.

\subsection{S3 in context: MW vs LMC stream and potential use-cases} \label{subsec:S3_context}

We have explored a range of possible dynamical models for S3 in Sect.~\ref{sec:simulations}.
These have revealed two results: (1) while the distance is ambiguous, S3 is likely at a large distance, and (2) all explored progenitor orbits are bound to the MW.
The distance ambiguity stems from small number statistics and the uncertain classification of the 10 \citetalias{navarrete19} stars into BHB and BS stars. 
The three clearly classified BS stars are at distances $<50$\,kpc.
If they were wrongly classified, and instead were BHB stars, we can assume a factor of two increase in their distance, putting them at distances between 80 and 100\,kpc, more in line with the other stars.

\begin{figure}
    \centering
    \includegraphics[width=\columnwidth]{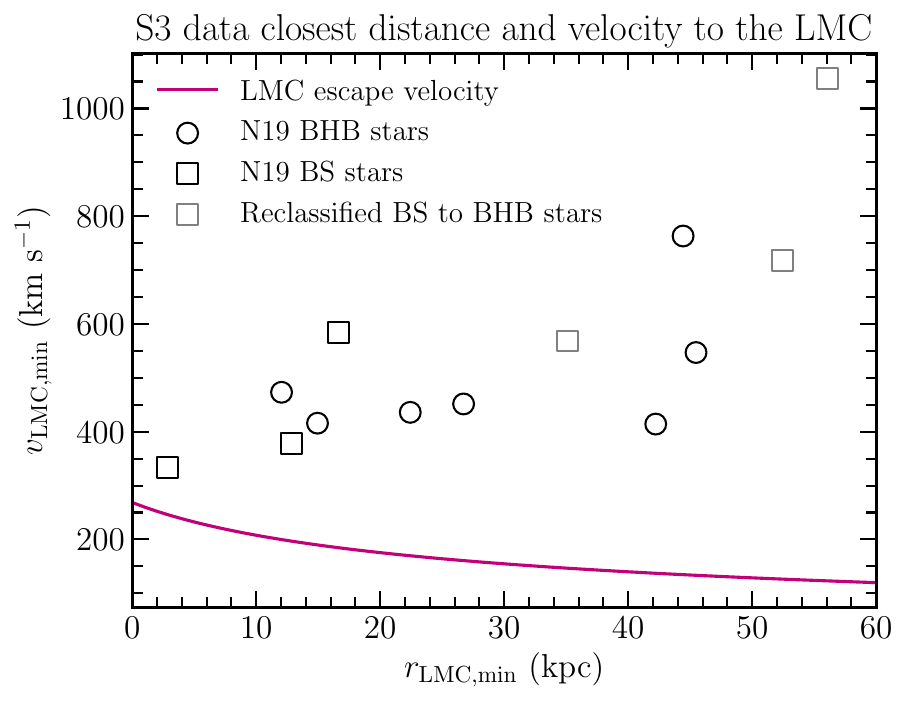}
    \caption{The closest distance to the LMC and velocity at closest approach for the S3 BHB (BS) stars from \citetalias{navarrete19}, represented by white point (square) symbols.
    The magenta line is the escape velocity curve of the LMC for a Hernquist sphere with $M_\mathrm{LMC}=1.25\times10^{11}\mathrm{M}_\odot$ and $r_{s,\mathrm{LMC}} = 14.9$\,kpc. 
    Any stars above the magenta line are unlikely to be bound to the LMC. 
    One BS has a close approach ($\sim3$\,kpc) and a relatively low velocity with respect to the LMC ($\sim330$\,km\,s$^{-1}$). 
    The other two BS stars are closer in distance and velocity to the LMC than most other BHB stars. 
    However, if the BS stars were misclassified BHB stars (grey squares), they would be among the most distant stars with the highest velocity offsets.
    This suggests, independently of the stream models, that S3 is not of LMC origin. 
    }
    \label{fig:S3_LMC_rmin_vmin}
\end{figure}

While the progenitor orbits show a clear association with the MW, a possible association of the S3 stars with the LMC can be investigated by calculating their closest approach distance and velocity.
If that velocity is larger than the LMC's escape velocity at that distance, the star is unlikely to be associated with the LMC.
We do this by backwards integrating the \citetalias{navarrete19} stars in the MW--LMC simulation described in Sect.~\ref{subsec:modelling} and recording the closest approach.
Figure~\ref{fig:S3_LMC_rmin_vmin} shows the distance and velocity relative to the LMC for the BHB (BS) stars marked as circles (squares).
The magenta line shows the escape velocity curve for the LMC used in the simulation, a Hernquist sphere with $M_\mathrm{LMC}=1.25\times10^{11}\mathrm{M}_\odot$ and $r_{s,\mathrm{LMC}} = 14.9$\,kpc. 
None of the stars are under the escape velocity line, which would indicate a clear association with the LMC. 
Some BS stars are at close distances and only slightly larger relative velocities; however, if they were reclassified as BHB stars, they would be at the distant end of the distribution (grey squares).
This shows that the S3 stars are unlikely to be associated with the LMC.

While identifying S3 as an LMC stream would have opened up a new way of investigating the LMC's present and past, S3 as an MW stream is interesting for both observers and theorists.
S3 is remarkable for its combination of large Galactocentric distance and extended morphology. Two streams often used to model the halos of the MW and the LMC, and to understand their interaction, are the OC and the Sagittarius stream. While most of the OC stream is at a distance of $\sim20-30$\,kpc, its edges reach up to 60\,kpc and are among its most informative parts \citep[e.g.,][]{erkal19, shipp21, Lilleengen2023, koposov23}. The Sagittarius stream reaches distances of close to 100\,kpc, but most of the stream is within 60\,kpc. It is notoriously difficult to model and needs the presence of the LMC \citep{vasiliev21}. We estimate the median distance of S3 to be $\sim 75$ kpc, and while Eridanus–M17 is the only known stream catalogued in \texttt{galstreams} \citep{mateu23} that surpasses S3 in distance ($\sim$95 kpc), the latter is extremely compact in projection compared to S3. Moreover, previous work by \citetalias{belokurov16} reported a stream width of 1.2$^\circ$; with our expanded candidate sample, we find S3 to be nearly $\sim3-4^\circ$ thick, indicating a substantial increase in its inferred physical width. Given the estimated median distance of S3 ($\sim$75 kpc), this angular width would translate to a physical thickness of roughly $\sim$4-5 kpc, making S3 one of the thickest stellar streams discovered to date. Whether this thickness is driven by an interaction with the LMC or follows from properties of the progenitor is a question for future research.

These properties make S3 the most distant ($\sim$75 kpc) extended ($\sim30^\circ$ long, $\sim3-4^\circ$ thick) stellar stream currently known in the MW, providing a unique opportunity to probe the outer Galactic halo and the recent dynamical influence of the LMC.
Fitting the MW halo with the S3 stream will provide measurements at unprecedented distances that we have not gained from streams before, as streams are most informative in the distances they span \citep{Bonaca2018}. 
It will likely be informative on the LMC halo as well. 
Moreover, the MW--LMC interaction affects stellar streams. 
With its distance and extent, the S3 stream could become a very useful tracer of the halo deformations, which could ultimately help make predictions and constraints on the nature of dark matter (see discussion in \citetalias{Lilleengen2023}).
However, to carry out these types of investigations, we need spectroscopic follow-up observations to build a detailed 6D picture of the S3 stream.

\subsection{Follow-up observations}\label{subsec:follow_up_obs}

To advance our understanding of S3, spectroscopic follow-up of the newly identified candidates is essential. Line-of-sight velocity measurements will provide the missing sixth dimension of phase-space information, allowing us to reconstruct the stream’s orbit with far greater accuracy. Expanding coverage beyond the currently sparse and scattered measurements will help define the velocity profile, reduce contamination, constrain orbital parameters such as pericenter, apocenter, and angular momentum, and probe the stream’s internal kinematics as well as possible perturbations induced by the LMC. Beyond kinematics, spectroscopy will also deliver independent distance estimates for red clump stars, which make up roughly 30\% of the new candidates identified in the catalogue of S3 candidates released alongside this work. These measurements, reliable even at 60–80 kpc, would sharpen the 3D map of the stream, refine its physical width and line-of-sight structure, and enable the detection of metallicity and age gradients. As a preliminary check on distances, we also crossmatched the sample with the \gaia DR3 RR Lyrae catalogue, but only a handful of stars overlapped, yielding too few matches to provide strong constraints (see Appendix \ref{sec:appendix_rrl}). Finally, spectroscopic abundances would provide a crucial chemical fingerprint of S3, helping to distinguish between a dwarf galaxy or globular cluster progenitor, assess the presence of multiple stellar populations, and compare the stream’s properties with those of other halo substructures.

Together, these observations would anchor high-fidelity dynamical and chemical models of S3, enabling a detailed reconstruction of its orbital history, progenitor properties, and disruption timescale. They would also provide new constraints on the mass distribution of the outer MW and on the dynamical influence of the LMC.

\section{Conclusions}
\label{sec:conclusions}

Despite extensive study of the LMC’s influence on MW stellar streams, those directly associated with the LMC and SMC have received little attention. Only a few candidate stellar streams (S1--S4) beyond the Magellanic Stream have been reported. Identified before \gaia, these streams remain largely unexplored with modern astrometric data and current state-of-the-art modelling.

In this work, we build upon these previous analyses, focusing on the S3 stream, a long ($\sim$30\degrnospace) and narrow ($\sim$1.2\degrnospace) stream at distances ranging from 60 to 80 kpc that points nearly exactly in the direction of the LMC. We pursue three primary objectives: 1) to extend the kinematic analysis of the S3 stellar candidates identified by \citetalias{navarrete19} by incorporating astrometric data from \gaia DR3, with the goal of reassessing and validating the stream’s existence; 2) to expand the sample of potential S3 members using machine learning techniques; and 3) to generate stream models to determine S3's association to either the MW or the LMC and to gain a better understanding of the data and future observation needs. Our main findings and conclusions are the following:

\begin{itemize}
    \item We report 1,542 new high-confidence S3 stream candidates, greatly enlarging the previous sample of $\sim$10 BHB and BS stars from \citetalias{navarrete19}. 
    \item Among these, we find 440 potential red clump stars, which provide valuable targets for spectroscopic follow-up thanks to their well-defined luminosities and precise distance measurements. 
    \item Compared to earlier studies, S3’s apparent width has increased from $\sim$1.2$^\circ$ to $\sim$3–4$^\circ$.
    \item Through modelling and by comparing the closest-approach distance and velocity of S3 stars with the LMC’s escape velocity, S3 is identified as a distant stream ($\sim$75 kpc) linked to the MW, having undergone a recent ($\sim$100 Myr) close encounter with the LMC.
    \item Stream models are not yet particularly constraining of the precise orbit or the progenitor properties; future observations will help to break degeneracies.
\end{itemize}

To sum up, we find that S3 is the most distant ($\sim$75 kpc) and extended ($\sim$30$^\circ$ long, $\sim$3–4$^\circ$ thick) MW stream known, offering a unique window into the outer Galactic halo and the recent dynamical influence of the LMC. Its angular width corresponds to a physical thickness of $\sim$4–5 kpc, making it one of the thickest stellar streams discovered to date. Fully exploiting the potential of S3 will require improved astrometric data; in particular, \gaia DR4, expected by the end of 2026, will provide proper motions at least twice as accurate as the current catalogue. Combined with spectroscopic follow-up from future surveys such as \textsl{4MOST} \citep{dejong19} or \textsl{SDSS}-V \citep{kollmeier17,kollmeier25}, these data will enable a more precise characterization of the stream’s kinematics and chemistry, confirmation of member stars, and a deeper understanding of its origin and dynamical history.

\begin{acknowledgements}

We thank Sergey Koposov for instructive conversations on the data.
We are grateful to Marcel Bernet for his always valuable suggestions that improved the clarity and aesthetics of the figures.
OJA acknowledges funding from ``Swedish National Space Agency 2023-00154 David Hobbs The GaiaNIR Mission'' and ``Swedish National Space Agency 2023-00137 David Hobbs The Extended Gaia Mission''. 
MSP acknowledges support from a UKRI Stephen Hawking Fellowship.
This work has made use of data from the European Space Agency (ESA) mission {\it Gaia} (\url{https://www.cosmos.esa.int/gaia}), processed by the {\it Gaia} Data Processing and Analysis Consortium (DPAC, \url{https://www.cosmos.esa.int/web/gaia/dpac/consortium}). Funding for the DPAC has been provided by national institutions, in particular the institutions participating in the {\it Gaia} Multilateral Agreement.

\\
\textit{Software:} \textsc{astropy} \citep{astropy:2013, astropy:2018, astropy:2022}, \textsc{cmasher} \citep{cmasher}, \textsc{exp} \citep{Petersen2022, Petesen.joss.2025}, \textsc{ipython} \citep{ipython}, \textsc{jupyter} \citep{jupyter}, \textsc{matplotlib} \citep{matplotlib}, \textsc{numpy} \citep{numpy}, \textsc{pandas} \citep{pandas...paper, pandas...software}, \textsc{scipy} \citep{scipy}, \textsc{shap} \citep{shap}, \textsc{sklearn} \citep{sklearn}

\end{acknowledgements}

\bibliographystyle{aa}
\bibliography{mylmcbib} 

\begin{thebibliography}{114}
\expandafter\ifx\csname natexlab\endcsname\relax\def\natexlab#1{#1}\fi

\bibitem[{{Arora} {et~al.}(2024){Arora}, {Garavito-Camargo}, {Sanderson}, {Cunningham}, {Wetzel}, {Panithanpaisal}, \& {Barry}}]{Arora2024}
{Arora}, A., {Garavito-Camargo}, N., {Sanderson}, R.~E., {et~al.} 2024, \apj, 974, 286

\bibitem[{{Astropy Collaboration} {et~al.}(2022){Astropy Collaboration}, {Price-Whelan}, {Lim}, {Earl}, {Starkman}, {Bradley}, {Shupe}, {Patil}, {Corrales}, {Brasseur}, {N{"o}the}, {Donath}, {Tollerud}, {Morris}, {Ginsburg}, {Vaher}, {Weaver}, {Tocknell}, {Jamieson}, {van Kerkwijk}, {Robitaille}, {Merry}, {Bachetti}, {G{"u}nther}, {Aldcroft}, {Alvarado-Montes}, {Archibald}, {B{'o}di}, {Bapat}, {Barentsen}, {Baz{'a}n}, {Biswas}, {Boquien}, {Burke}, {Cara}, {Cara}, {Conroy}, {Conseil}, {Craig}, {Cross}, {Cruz}, {D'Eugenio}, {Dencheva}, {Devillepoix}, {Dietrich}, {Eigenbrot}, {Erben}, {Ferreira}, {Foreman-Mackey}, {Fox}, {Freij}, {Garg}, {Geda}, {Glattly}, {Gondhalekar}, {Gordon}, {Grant}, {Greenfield}, {Groener}, {Guest}, {Gurovich}, {Handberg}, {Hart}, {Hatfield-Dodds}, {Homeier}, {Hosseinzadeh}, {Jenness}, {Jones}, {Joseph}, {Kalmbach}, {Karamehmetoglu}, {Ka{l}uszy{'n}ski}, {Kelley}, {Kern}, {Kerzendorf}, {Koch}, {Kulumani}, {Lee}, {Ly}, {Ma}, {MacBride}, {Maljaars}, {Muna}, {Murphy}, {Norman}, {O'Steen},
  {Oman}, {Pacifici}, {Pascual}, {Pascual-Granado}, {Patil}, {Perren}, {Pickering}, {Rastogi}, {Roulston}, {Ryan}, {Rykoff}, {Sabater}, {Sakurikar}, {Salgado}, {Sanghi}, {Saunders}, {Savchenko}, {Schwardt}, {Seifert-Eckert}, {Shih}, {Jain}, {Shukla}, {Sick}, {Simpson}, {Singanamalla}, {Singer}, {Singhal}, {Sinha}, {Sip{H{o}}cz}, {Spitler}, {Stansby}, {Streicher}, {{{S}}umak}, {Swinbank}, {Taranu}, {Tewary}, {Tremblay}, {Val-Borro}, {Van Kooten}, {Vasovi{'c}}, {Verma}, {de Miranda Cardoso}, {Williams}, {Wilson}, {Winkel}, {Wood-Vasey}, {Xue}, {Yoachim}, {Zhang}, {Zonca}, \& {Astropy Project Contributors}}]{astropy:2022}
{Astropy Collaboration}, {Price-Whelan}, A.~M., {Lim}, P.~L., {et~al.} 2022, \apj, 935, 167

\bibitem[{{Astropy Collaboration} {et~al.}(2018){Astropy Collaboration}, {Price-Whelan}, {Sip{\H{o}}cz}, {G{\"u}nther}, {Lim}, {Crawford}, {Conseil}, {Shupe}, {Craig}, {Dencheva}, {Ginsburg}, {Vand erPlas}, {Bradley}, {P{\'e}rez-Su{\'a}rez}, {de Val-Borro}, {Aldcroft}, {Cruz}, {Robitaille}, {Tollerud}, {Ardelean}, {Babej}, {Bach}, {Bachetti}, {Bakanov}, {Bamford}, {Barentsen}, {Barmby}, {Baumbach}, {Berry}, {Biscani}, {Boquien}, {Bostroem}, {Bouma}, {Brammer}, {Bray}, {Breytenbach}, {Buddelmeijer}, {Burke}, {Calderone}, {Cano Rodr{\'\i}guez}, {Cara}, {Cardoso}, {Cheedella}, {Copin}, {Corrales}, {Crichton}, {D'Avella}, {Deil}, {Depagne}, {Dietrich}, {Donath}, {Droettboom}, {Earl}, {Erben}, {Fabbro}, {Ferreira}, {Finethy}, {Fox}, {Garrison}, {Gibbons}, {Goldstein}, {Gommers}, {Greco}, {Greenfield}, {Groener}, {Grollier}, {Hagen}, {Hirst}, {Homeier}, {Horton}, {Hosseinzadeh}, {Hu}, {Hunkeler}, {Ivezi{\'c}}, {Jain}, {Jenness}, {Kanarek}, {Kendrew}, {Kern}, {Kerzendorf}, {Khvalko}, {King}, {Kirkby}, {Kulkarni},
  {Kumar}, {Lee}, {Lenz}, {Littlefair}, {Ma}, {Macleod}, {Mastropietro}, {McCully}, {Montagnac}, {Morris}, {Mueller}, {Mumford}, {Muna}, {Murphy}, {Nelson}, {Nguyen}, {Ninan}, {N{\"o}the}, {Ogaz}, {Oh}, {Parejko}, {Parley}, {Pascual}, {Patil}, {Patil}, {Plunkett}, {Prochaska}, {Rastogi}, {Reddy Janga}, {Sabater}, {Sakurikar}, {Seifert}, {Sherbert}, {Sherwood-Taylor}, {Shih}, {Sick}, {Silbiger}, {Singanamalla}, {Singer}, {Sladen}, {Sooley}, {Sornarajah}, {Streicher}, {Teuben}, {Thomas}, {Tremblay}, {Turner}, {Terr{\'o}n}, {van Kerkwijk}, {de la Vega}, {Watkins}, {Weaver}, {Whitmore}, {Woillez}, {Zabalza}, \& {Astropy Contributors}}]{astropy:2018}
{Astropy Collaboration}, {Price-Whelan}, A.~M., {Sip{\H{o}}cz}, B.~M., {et~al.} 2018, \aj, 156, 123

\bibitem[{{Astropy Collaboration} {et~al.}(2013){Astropy Collaboration}, {Robitaille}, {Tollerud}, {Greenfield}, {Droettboom}, {Bray}, {Aldcroft}, {Davis}, {Ginsburg}, {Price-Whelan}, {Kerzendorf}, {Conley}, {Crighton}, {Barbary}, {Muna}, {Ferguson}, {Grollier}, {Parikh}, {Nair}, {Unther}, {Deil}, {Woillez}, {Conseil}, {Kramer}, {Turner}, {Singer}, {Fox}, {Weaver}, {Zabalza}, {Edwards}, {Azalee Bostroem}, {Burke}, {Casey}, {Crawford}, {Dencheva}, {Ely}, {Jenness}, {Labrie}, {Lim}, {Pierfederici}, {Pontzen}, {Ptak}, {Refsdal}, {Servillat}, \& {Streicher}}]{astropy:2013}
{Astropy Collaboration}, {Robitaille}, T.~P., {Tollerud}, E.~J., {et~al.} 2013, \aap, 558, A33

\bibitem[{{Bajaja} {et~al.}(1985){Bajaja}, {Cappa de Nicolau}, {Cersosimo}, {Martin}, {Loiseau}, {Morras}, {Olano}, \& {Poeppel}}]{bajaja85}
{Bajaja}, E., {Cappa de Nicolau}, C.~E., {Cersosimo}, J.~C., {et~al.} 1985, \apjs, 58, 143

\bibitem[{{Belokurov} \& {Koposov}(2016)}]{belokurov16}
{Belokurov}, V. \& {Koposov}, S.~E. 2016, \mnras, 456, 602

\bibitem[{{Belokurov} {et~al.}(2006){Belokurov}, {Zucker}, {Evans}, {Gilmore}, {Vidrih}, {Bramich}, {Newberg}, {Wyse}, {Irwin}, {Fellhauer}, {Hewett}, {Walton}, {Wilkinson}, {Cole}, {Yanny}, {Rockosi}, {Beers}, {Bell}, {Brinkmann}, {Ivezi{\'c}}, \& {Lupton}}]{belokurov06}
{Belokurov}, V., {Zucker}, D.~B., {Evans}, N.~W., {et~al.} 2006, \apjl, 642, L137

\bibitem[{{Besla} {et~al.}(2012){Besla}, {Kallivayalil}, {Hernquist}, {van der Marel}, {Cox}, \& {Kere{\v{s}}}}]{besla12}
{Besla}, G., {Kallivayalil}, N., {Hernquist}, L., {et~al.} 2012, \mnras, 421, 2109

\bibitem[{{B{\'\i}lek} {et~al.}(2020){B{\'\i}lek}, {Duc}, {Cuillandre}, {Gwyn}, {Cappellari}, {Bekaert}, {Bonfini}, {Bitsakis}, {Paudel}, {Krajnovi{\'c}}, {Durrell}, \& {Marleau}}]{bilek20}
{B{\'\i}lek}, M., {Duc}, P.-A., {Cuillandre}, J.-C., {et~al.} 2020, \mnras, 498, 2138

\bibitem[{{Binney}(2008)}]{binney08}
{Binney}, J. 2008, \mnras, 386, L47

\bibitem[{{Bonaca} \& {Hogg}(2018)}]{Bonaca2018}
{Bonaca}, A. \& {Hogg}, D.~W. 2018, \apj, 867, 101

\bibitem[{{Bonaca} \& {Price-Whelan}(2025)}]{bonaca_pricewhelan25}
{Bonaca}, A. \& {Price-Whelan}, A.~M. 2025, \nar, 100, 101713

\bibitem[{{Borsato} {et~al.}(2020){Borsato}, {Martell}, \& {Simpson}}]{borsato20}
{Borsato}, N.~W., {Martell}, S.~L., \& {Simpson}, J.~D. 2020, \mnras, 492, 1370

\bibitem[{{Bovy} {et~al.}(2016){Bovy}, {Bahmanyar}, {Fritz}, \& {Kallivayalil}}]{bovy16}
{Bovy}, J., {Bahmanyar}, A., {Fritz}, T.~K., \& {Kallivayalil}, N. 2016, \apj, 833, 31

\bibitem[{{Brooks} {et~al.}(2025{\natexlab{a}}){Brooks}, {Garavito-Camargo}, {Johnston}, {Price-Whelan}, {Sanders}, \& {Lilleengen}}]{brooks25a}
{Brooks}, R. A.~N., {Garavito-Camargo}, N., {Johnston}, K.~V., {et~al.} 2025{\natexlab{a}}, \apj, 978, 79

\bibitem[{{Brooks} {et~al.}(2025{\natexlab{b}}){Brooks}, {Sanders}, {Dillamore}, {Garavito-Camargo}, \& {Price-Whelan}}]{brooks25b}
{Brooks}, R. A.~N., {Sanders}, J.~L., {Dillamore}, A.~M., {Garavito-Camargo}, N., \& {Price-Whelan}, A.~M. 2025{\natexlab{b}}, arXiv e-prints, arXiv:2507.10667

\bibitem[{{Brooks} {et~al.}(2024){Brooks}, {Sanders}, {Lilleengen}, {Petersen}, \& {Pontzen}}]{brooks24}
{Brooks}, R. A.~N., {Sanders}, J.~L., {Lilleengen}, S., {Petersen}, M.~S., \& {Pontzen}, A. 2024, \mnras, 532, 2657

\bibitem[{{Chandra} {et~al.}(2023){Chandra}, {Naidu}, {Conroy}, {Bonaca}, {Zaritsky}, {Cargile}, {Caldwell}, {Johnson}, {Han}, \& {Ting}}]{chandra23}
{Chandra}, V., {Naidu}, R.~P., {Conroy}, C., {et~al.} 2023, \apj, 956, 110

\bibitem[{{Clementini} {et~al.}(2023){Clementini}, {Ripepi}, {Garofalo}, {Molinaro}, {Muraveva}, {Leccia}, {Rimoldini}, {Holl}, {Jevardat de Fombelle}, {Sartoretti}, {Marchal}, {Audard}, {Nienartowicz}, {Andrae}, {Marconi}, {Szabados}, {Evans}, {Lecoeur-Taibi}, {Mowlavi}, {Musella}, \& {Eyer}}]{clementini23}
{Clementini}, G., {Ripepi}, V., {Garofalo}, A., {et~al.} 2023, \aap, 674, A18

\bibitem[{{de Jong} {et~al.}(2019){de Jong}, {Agertz}, {Berbel}, {Aird}, {Alexander}, {Amarsi}, {Anders}, {Andrae}, {Ansarinejad}, {Ansorge}, {Antilogus}, {Anwand-Heerwart}, {Arentsen}, {Arnadottir}, {Asplund}, {Auger}, {Azais}, {Baade}, {Baker}, {Baker}, {Balbinot}, {Baldry}, {Banerji}, {Barden}, {Barklem}, {Barth{\'e}l{\'e}my-Mazot}, {Battistini}, {Bauer}, {Bell}, {Bellido-Tirado}, {Bellstedt}, {Belokurov}, {Bensby}, {Bergemann}, {Bestenlehner}, {Bielby}, {Bilicki}, {Blake}, {Bland-Hawthorn}, {Boeche}, {Boland}, {Boller}, {Bongard}, {Bongiorno}, {Bonifacio}, {Boudon}, {Brooks}, {Brown}, {Brown}, {Br{\"u}ggen}, {Brynnel}, {Brzeski}, {Buchert}, {Buschkamp}, {Caffau}, {Caillier}, {Carrick}, {Casagrande}, {Case}, {Casey}, {Cesarini}, {Cescutti}, {Chapuis}, {Chiappini}, {Childress}, {Christlieb}, {Church}, {Cioni}, {Cluver}, {Colless}, {Collett}, {Comparat}, {Cooper}, {Couch}, {Courbin}, {Croom}, {Croton}, {Daguis{\'e}}, {Dalton}, {Davies}, {Davis}, {de Laverny}, {Deason}, {Dionies}, {Disseau}, {Doel},
  {D{\"o}scher}, {Driver}, {Dwelly}, {Eckert}, {Edge}, {Edvardsson}, {Youssoufi}, {Elhaddad}, {Enke}, {Erfanianfar}, {Farrell}, {Fechner}, {Feiz}, {Feltzing}, {Ferreras}, {Feuerstein}, {Feuillet}, {Finoguenov}, {Ford}, {Fotopoulou}, {Fouesneau}, {Frenk}, {Frey}, {Gaessler}, {Geier}, {Gentile Fusillo}, {Gerhard}, {Giannantonio}, {Giannone}, {Gibson}, {Gillingham}, {Gonz{\'a}lez-Fern{\'a}ndez}, {Gonzalez-Solares}, {Gottloeber}, {Gould}, {Grebel}, {Gueguen}, {Guiglion}, {Haehnelt}, {Hahn}, {Hansen}, {Hartman}, {Hauptner}, {Hawkins}, {Haynes}, {Haynes}, {Heiter}, {Helmi}, {Aguayo}, {Hewett}, {Hinton}, {Hobbs}, {Hoenig}, {Hofman}, {Hook}, {Hopgood}, {Hopkins}, {Hourihane}, {Howes}, {Howlett}, {Huet}, {Irwin}, {Iwert}, {Jablonka}, {Jahn}, {Jahnke}, {Jarno}, {Jin}, {Jofre}, {Johl}, {Jones}, {J{\"o}nsson}, {Jordan}, {Karovicova}, {Khalatyan}, {Kelz}, {Kennicutt}, {King}, {Kitaura}, {Klar}, {Klauser}, {Kneib}, {Koch}, {Koposov}, {Kordopatis}, {Korn}, {Kosmalski}, {Kotak}, {Kovalev}, {Kreckel}, {Kripak}, {Krumpe},
  {Kuijken}, {Kunder}, {Kushniruk}, {Lam}, {Lamer}, {Laurent}, {Lawrence}, {Lehmitz}, {Lemasle}, {Lewis}, {Li}, {Lidman}, {Lind}, {Liske}, {Lizon}, {Loveday}, {Ludwig}, {McDermid}, {Maguire}, {Mainieri}, {Mali}, {Mandel}, {Mandel}, {Mannering}, {Martell}, {Martinez Delgado}, {Matijevic}, {McGregor}, {McMahon}, {McMillan}, {Mena}, {Merloni}, {Meyer}, {Michel}, {Micheva}, {Migniau}, {Minchev}, {Monari}, {Muller}, {Murphy}, {Muthukrishna}, {Nandra}, {Navarro}, {Ness}, {Nichani}, {Nichol}, {Nicklas}, {Niederhofer}, {Norberg}, {Obreschkow}, {Oliver}, {Owers}, {Pai}, {Pankratow}, {Parkinson}, {Paschke}, {Paterson}, {Pecontal}, {Parry}, {Phillips}, {Pillepich}, {Pinard}, {Pirard}, {Piskunov}, {Plank}, {Pl{\"u}schke}, {Pons}, {Popesso}, {Power}, {Pragt}, {Pramskiy}, {Pryer}, {Quattri}, {Queiroz}, {Quirrenbach}, {Rahurkar}, {Raichoor}, {Ramstedt}, {Rau}, {Recio-Blanco}, {Reiss}, {Renaud}, {Revaz}, {Rhode}, {Richard}, {Richter}, {Rix}, {Robotham}, {Roelfsema}, {Romaniello}, {Rosario}, {Rothmaier}, {Roukema}, {Ruchti},
  {Rupprecht}, {Rybizki}, {Ryde}, {Saar}, {Sadler}, {Sahl{\'e}n}, {Salvato}, {Sassolas}, {Saunders}, {Saviauk}, {Sbordone}, {Schmidt}, {Schnurr}, {Scholz}, {Schwope}, {Seifert}, {Shanks}, {Sheinis}, {Sivov}, {Sk{\'u}lad{\'o}ttir}, {Smartt}, {Smedley}, {Smith}, {Smith}, {Sorce}, {Spitler}, {Starkenburg}, {Steinmetz}, {Stilz}, {Storm}, {Sullivan}, {Sutherland}, {Swann}, {Tamone}, {Taylor}, {Teillon}, {Tempel}, {ter Horst}, {Thi}, {Tolstoy}, {Trager}, {Traven}, {Tremblay}, {Tresse}, {Valentini}, {van de Weygaert}, {van den Ancker}, {Veljanoski}, {Venkatesan}, {Wagner}, {Wagner}, {Walcher}, {Waller}, {Walton}, {Wang}, {Winkler}, {Wisotzki}, {Worley}, {Worseck}, {Xiang}, {Xu}, {Yong}, {Zhao}, {Zheng}, {Zscheyge}, \& {Zucker}}]{dejong19}
{de Jong}, R.~S., {Agertz}, O., {Berbel}, A.~A., {et~al.} 2019, The Messenger, 175, 3

\bibitem[{{Dhanush} {et~al.}(2024){Dhanush}, {Subramaniam}, \& {Subramanian}}]{dhanush2024}
{Dhanush}, S.~R., {Subramaniam}, A., \& {Subramanian}, S. 2024, \apj, 968, 103

\bibitem[{{Diehl} {et~al.}(2014){Diehl}, {Abbott}, {Annis}, {Armstrong}, {Baruah}, {Bermeo}, {Bernstein}, {Beynon}, {Bruderer}, {Buckley-Geer}, {Campbell}, {Capozzi}, {Carter}, {Casas}, {Clerkin}, {Covarrubias}, {Cuhna}, {D'Andrea}, {da Costa}, {Das}, {DePoy}, {Dietrich}, {Drlica-Wagner}, {Elliott}, {Eifler}, {Estrada}, {Etherington}, {Flaugher}, {Frieman}, {Fausti Neto}, {Gelman}, {Gerdes}, {Gruen}, {Gruendl}, {Hao}, {Head}, {Helsby}, {Hoffman}, {Honscheid}, {James}, {Johnson}, {Kacprzac}, {Katsaros}, {Kennedy}, {Kent}, {Kessler}, {Kim}, {Krause}, {Kron}, {Kuhlmann}, {Kunder}, {Li}, {Lin}, {Maccrann}, {March}, {Marshall}, {Neilsen}, {Nugent}, {Martini}, {Melchior}, {Menanteau}, {Nichol}, {Nord}, {Ogando}, {Old}, {Papadopoulos}, {Patton}, {Petravick}, {Plazas}, {Poulton}, {Pujol}, {Reil}, {Rigby}, {Romer}, {Roodman}, {Rooney}, {Sanchez Alvaro}, {Serrano}, {Sheldon}, {Smith}, {Smith}, {Soares-Santos}, {Soumagnac}, {Spinka}, {Suchyta}, {Tucker}, {Walker}, {Wester}, {Wiesner}, {Wilcox}, {Williams}, {Yanny}, \&
  {Zhang}}]{diehl15}
{Diehl}, H.~T., {Abbott}, T.~M.~C., {Annis}, J., {et~al.} 2014, in Society of Photo-Optical Instrumentation Engineers (SPIE) Conference Series, Vol. 9149, Observatory Operations: Strategies, Processes, and Systems V, ed. A.~B. {Peck}, C.~R. {Benn}, \& R.~L. {Seaman}, 91490V

\bibitem[{{D'Onghia} \& {Fox}(2016)}]{donghia-fox16}
{D'Onghia}, E. \& {Fox}, A.~J. 2016, \araa, 54, 363

\bibitem[{{Dubinski} {et~al.}(1999){Dubinski}, {Mihos}, \& {Hernquist}}]{dubinski99}
{Dubinski}, J., {Mihos}, J.~C., \& {Hernquist}, L. 1999, \apj, 526, 607

\bibitem[{{Erkal} {et~al.}(2016){Erkal}, {Belokurov}, {Bovy}, \& {Sanders}}]{erkal16}
{Erkal}, D., {Belokurov}, V., {Bovy}, J., \& {Sanders}, J.~L. 2016, \mnras, 463, 102

\bibitem[{{Erkal} {et~al.}(2019){Erkal}, {Belokurov}, {Laporte}, {Koposov}, {Li}, {Grillmair}, {Kallivayalil}, {Price-Whelan}, {Evans}, {Hawkins}, {Hendel}, {Mateu}, {Navarro}, {del Pino}, {Slater}, {Sohn}, \& {Orphan Aspen Treasury Collaboration}}]{erkal19}
{Erkal}, D., {Belokurov}, V., {Laporte}, C.~F.~P., {et~al.} 2019, \mnras, 487, 2685

\bibitem[{{Ferguson} {et~al.}(2022){Ferguson}, {Shipp}, {Drlica-Wagner}, {Li}, {Cerny}, {Tavangar}, {Pace}, {Marshall}, {Riley}, {Adam{\'o}w}, {Carlin}, {Choi}, {Erkal}, {James}, {Koposov}, {Kuropatkin}, {Mart{\'\i}nez-V{\'a}zquez}, {Mau}, {Mutlu-Pakdil}, {Olsen}, {Sakowska}, {Stringfellow}, {Yanny}, \& {Yanny}}]{ferguson22}
{Ferguson}, P.~S., {Shipp}, N., {Drlica-Wagner}, A., {et~al.} 2022, \aj, 163, 18

\bibitem[{{Gaia Collaboration} {et~al.}(2018){Gaia Collaboration}, {Brown}, {Vallenari}, {Prusti}, {de Bruijne}, {Babusiaux}, {Bailer-Jones}, {Biermann}, {Evans}, {Eyer}, {Jansen}, {Jordi}, {Klioner}, {Lammers}, {Lindegren}, {Luri}, {Mignard}, {Panem}, {Pourbaix}, {Randich}, {Sartoretti}, {Siddiqui}, {Soubiran}, {van Leeuwen}, {Walton}, {Arenou}, {Bastian}, {Cropper}, {Drimmel}, {Katz}, {Lattanzi}, {Bakker}, {Cacciari}, {Casta{\~n}eda}, {Chaoul}, {Cheek}, {De Angeli}, {Fabricius}, {Guerra}, {Holl}, {Masana}, {Messineo}, {Mowlavi}, {Nienartowicz}, {Panuzzo}, {Portell}, {Riello}, {Seabroke}, {Tanga}, {Th{\'e}venin}, {Gracia-Abril}, {Comoretto}, {Garcia-Reinaldos}, {Teyssier}, {Altmann}, {Andrae}, {Audard}, {Bellas-Velidis}, {Benson}, {Berthier}, {Blomme}, {Burgess}, {Busso}, {Carry}, {Cellino}, {Clementini}, {Clotet}, {Creevey}, {Davidson}, {De Ridder}, {Delchambre}, {Dell'Oro}, {Ducourant}, {Fern{\'a}ndez-Hern{\'a}ndez}, {Fouesneau}, {Fr{\'e}mat}, {Galluccio}, {Garc{\'\i}a-Torres},
  {Gonz{\'a}lez-N{\'u}{\~n}ez}, {Gonz{\'a}lez-Vidal}, {Gosset}, {Guy}, {Halbwachs}, {Hambly}, {Harrison}, {Hern{\'a}ndez}, {Hestroffer}, {Hodgkin}, {Hutton}, {Jasniewicz}, {Jean-Antoine-Piccolo}, {Jordan}, {Korn}, {Krone-Martins}, {Lanzafame}, {Lebzelter}, {L{\"o}ffler}, {Manteiga}, {Marrese}, {Mart{\'\i}n-Fleitas}, {Moitinho}, {Mora}, {Muinonen}, {Osinde}, {Pancino}, {Pauwels}, {Petit}, {Recio-Blanco}, {Richards}, {Rimoldini}, {Robin}, {Sarro}, {Siopis}, {Smith}, {Sozzetti}, {S{\"u}veges}, {Torra}, {van Reeven}, {Abbas}, {Abreu Aramburu}, {Accart}, {Aerts}, {Altavilla}, {{\'A}lvarez}, {Alvarez}, {Alves}, {Anderson}, {Andrei}, {Anglada Varela}, {Antiche}, {Antoja}, {Arcay}, {Astraatmadja}, {Bach}, {Baker}, {Balaguer-N{\'u}{\~n}ez}, {Balm}, {Barache}, {Barata}, {Barbato}, {Barblan}, {Barklem}, {Barrado}, {Barros}, {Barstow}, {Bartholom{\'e} Mu{\~n}oz}, {Bassilana}, {Becciani}, {Bellazzini}, {Berihuete}, {Bertone}, {Bianchi}, {Bienaym{\'e}}, {Blanco-Cuaresma}, {Boch}, {Boeche}, {Bombrun}, {Borrachero},
  {Bossini}, {Bouquillon}, {Bourda}, {Bragaglia}, {Bramante}, {Breddels}, {Bressan}, {Brouillet}, {Br{\"u}semeister}, {Brugaletta}, {Bucciarelli}, {Burlacu}, {Busonero}, {Butkevich}, {Buzzi}, {Caffau}, {Cancelliere}, {Cannizzaro}, {Cantat-Gaudin}, {Carballo}, {Carlucci}, {Carrasco}, {Casamiquela}, {Castellani}, {Castro-Ginard}, {Charlot}, {Chemin}, {Chiavassa}, {Cocozza}, {Costigan}, {Cowell}, {Crifo}, {Crosta}, {Crowley}, {Cuypers}, {Dafonte}, {Damerdji}, {Dapergolas}, {David}, {David}, {de Laverny}, {De Luise}, {De March}, {de Martino}, {de Souza}, {de Torres}, {Debosscher}, {del Pozo}, {Delbo}, {Delgado}, {Delgado}, {Di Matteo}, {Diakite}, {Diener}, {Distefano}, {Dolding}, {Drazinos}, {Dur{\'a}n}, {Edvardsson}, {Enke}, {Eriksson}, {Esquej}, {Eynard Bontemps}, {Fabre}, {Fabrizio}, {Faigler}, {Falc{\~a}o}, {Farr{\`a}s Casas}, {Federici}, {Fedorets}, {Fernique}, {Figueras}, {Filippi}, {Findeisen}, {Fonti}, {Fraile}, {Fraser}, {Fr{\'e}zouls}, {Gai}, {Galleti}, {Garabato}, {Garc{\'\i}a-Sedano}, {Garofalo},
  {Garralda}, {Gavel}, {Gavras}, {Gerssen}, {Geyer}, {Giacobbe}, {Gilmore}, {Girona}, {Giuffrida}, {Glass}, {Gomes}, {Granvik}, {Gueguen}, {Guerrier}, {Guiraud}, {Guti{\'e}rrez-S{\'a}nchez}, {Haigron}, {Hatzidimitriou}, {Hauser}, {Haywood}, {Heiter}, {Helmi}, {Heu}, {Hilger}, {Hobbs}, {Hofmann}, {Holland}, {Huckle}, {Hypki}, {Icardi}, {Jan{\ss}en}, {Jevardat de Fombelle}, {Jonker}, {Juh{\'a}sz}, {Julbe}, {Karampelas}, {Kewley}, {Klar}, {Kochoska}, {Kohley}, {Kolenberg}, {Kontizas}, {Kontizas}, {Koposov}, {Kordopatis}, {Kostrzewa-Rutkowska}, {Koubsky}, {Lambert}, {Lanza}, {Lasne}, {Lavigne}, {Le Fustec}, {Le Poncin-Lafitte}, {Lebreton}, {Leccia}, {Leclerc}, {Lecoeur-Taibi}, {Lenhardt}, {Leroux}, {Liao}, {Licata}, {Lindstr{\o}m}, {Lister}, {Livanou}, {Lobel}, {L{\'o}pez}, {Managau}, {Mann}, {Mantelet}, {Marchal}, {Marchant}, {Marconi}, {Marinoni}, {Marschalk{\'o}}, {Marshall}, {Martino}, {Marton}, {Mary}, {Massari}, {Matijevi{\v{c}}}, {Mazeh}, {McMillan}, {Messina}, {Michalik}, {Millar}, {Molina}, {Molinaro},
  {Moln{\'a}r}, {Montegriffo}, {Mor}, {Morbidelli}, {Morel}, {Morris}, {Mulone}, {Muraveva}, {Musella}, {Nelemans}, {Nicastro}, {Noval}, {O'Mullane}, {Ord{\'e}novic}, {Ord{\'o}{\~n}ez-Blanco}, {Osborne}, {Pagani}, {Pagano}, {Pailler}, {Palacin}, {Palaversa}, {Panahi}, {Pawlak}, {Piersimoni}, {Pineau}, {Plachy}, {Plum}, {Poggio}, {Poujoulet}, {Pr{\v{s}}a}, {Pulone}, {Racero}, {Ragaini}, {Rambaux}, {Ramos-Lerate}, {Regibo}, {Reyl{\'e}}, {Riclet}, {Ripepi}, {Riva}, {Rivard}, {Rixon}, {Roegiers}, {Roelens}, {Romero-G{\'o}mez}, {Rowell}, {Royer}, {Ruiz-Dern}, {Sadowski}, {Sagrist{\`a} Sell{\'e}s}, {Sahlmann}, {Salgado}, {Salguero}, {Sanna}, {Santana-Ros}, {Sarasso}, {Savietto}, {Schultheis}, {Sciacca}, {Segol}, {Segovia}, {S{\'e}gransan}, {Shih}, {Siltala}, {Silva}, {Smart}, {Smith}, {Solano}, {Solitro}, {Sordo}, {Soria Nieto}, {Souchay}, {Spagna}, {Spoto}, {Stampa}, {Steele}, {Steidelm{\"u}ller}, {Stephenson}, {Stoev}, {Suess}, {Surdej}, {Szabados}, {Szegedi-Elek}, {Tapiador}, {Taris}, {Tauran}, {Taylor},
  {Teixeira}, {Terrett}, {Teyssandier}, {Thuillot}, {Titarenko}, {Torra Clotet}, {Turon}, {Ulla}, {Utrilla}, {Uzzi}, {Vaillant}, {Valentini}, {Valette}, {van Elteren}, {Van Hemelryck}, {van Leeuwen}, {Vaschetto}, {Vecchiato}, {Veljanoski}, {Viala}, {Vicente}, {Vogt}, {von Essen}, {Voss}, {Votruba}, {Voutsinas}, {Walmsley}, {Weiler}, {Wertz}, {Wevers}, {Wyrzykowski}, {Yoldas}, {{\v{Z}}erjal}, {Ziaeepour}, {Zorec}, {Zschocke}, {Zucker}, {Zurbach}, \& {Zwitter}}]{gaiadr2summary}
{Gaia Collaboration}, {Brown}, A.~G.~A., {Vallenari}, A., {et~al.} 2018, \aap, 616, A1

\bibitem[{{Gaia Collaboration} {et~al.}(2021{\natexlab{a}}){Gaia Collaboration}, {Brown}, {Vallenari}, {Prusti}, {de Bruijne}, {Babusiaux}, {Biermann}, {Creevey}, {Evans}, {Eyer}, {Hutton}, {Jansen}, {Jordi}, {Klioner}, {Lammers}, {Lindegren}, {Luri}, {Mignard}, {Panem}, {Pourbaix}, {Randich}, {Sartoretti}, {Soubiran}, {Walton}, {Arenou}, {Bailer-Jones}, {Bastian}, {Cropper}, {Drimmel}, {Katz}, {Lattanzi}, {van Leeuwen}, {Bakker}, {Cacciari}, {Casta{\~n}eda}, {De Angeli}, {Ducourant}, {Fabricius}, {Fouesneau}, {Fr{\'e}mat}, {Guerra}, {Guerrier}, {Guiraud}, {Jean-Antoine Piccolo}, {Masana}, {Messineo}, {Mowlavi}, {Nicolas}, {Nienartowicz}, {Pailler}, {Panuzzo}, {Riclet}, {Roux}, {Seabroke}, {Sordo}, {Tanga}, {Th{\'e}venin}, {Gracia-Abril}, {Portell}, {Teyssier}, {Altmann}, {Andrae}, {Bellas-Velidis}, {Benson}, {Berthier}, {Blomme}, {Brugaletta}, {Burgess}, {Busso}, {Carry}, {Cellino}, {Cheek}, {Clementini}, {Damerdji}, {Davidson}, {Delchambre}, {Dell'Oro}, {Fern{\'a}ndez-Hern{\'a}ndez}, {Galluccio},
  {Garc{\'\i}a-Lario}, {Garcia-Reinaldos}, {Gonz{\'a}lez-N{\'u}{\~n}ez}, {Gosset}, {Haigron}, {Halbwachs}, {Hambly}, {Harrison}, {Hatzidimitriou}, {Heiter}, {Hern{\'a}ndez}, {Hestroffer}, {Hodgkin}, {Holl}, {Jan{\ss}en}, {Jevardat de Fombelle}, {Jordan}, {Krone-Martins}, {Lanzafame}, {L{\"o}ffler}, {Lorca}, {Manteiga}, {Marchal}, {Marrese}, {Moitinho}, {Mora}, {Muinonen}, {Osborne}, {Pancino}, {Pauwels}, {Petit}, {Recio-Blanco}, {Richards}, {Riello}, {Rimoldini}, {Robin}, {Roegiers}, {Rybizki}, {Sarro}, {Siopis}, {Smith}, {Sozzetti}, {Ulla}, {Utrilla}, {van Leeuwen}, {van Reeven}, {Abbas}, {Abreu Aramburu}, {Accart}, {Aerts}, {Aguado}, {Ajaj}, {Altavilla}, {{\'A}lvarez}, {{\'A}lvarez Cid-Fuentes}, {Alves}, {Anderson}, {Anglada Varela}, {Antoja}, {Audard}, {Baines}, {Baker}, {Balaguer-N{\'u}{\~n}ez}, {Balbinot}, {Balog}, {Barache}, {Barbato}, {Barros}, {Barstow}, {Bartolom{\'e}}, {Bassilana}, {Bauchet}, {Baudesson-Stella}, {Becciani}, {Bellazzini}, {Bernet}, {Bertone}, {Bianchi}, {Blanco-Cuaresma}, {Boch},
  {Bombrun}, {Bossini}, {Bouquillon}, {Bragaglia}, {Bramante}, {Breedt}, {Bressan}, {Brouillet}, {Bucciarelli}, {Burlacu}, {Busonero}, {Butkevich}, {Buzzi}, {Caffau}, {Cancelliere}, {C{\'a}novas}, {Cantat-Gaudin}, {Carballo}, {Carlucci}, {Carnerero}, {Carrasco}, {Casamiquela}, {Castellani}, {Castro-Ginard}, {Castro Sampol}, {Chaoul}, {Charlot}, {Chemin}, {Chiavassa}, {Cioni}, {Comoretto}, {Cooper}, {Cornez}, {Cowell}, {Crifo}, {Crosta}, {Crowley}, {Dafonte}, {Dapergolas}, {David}, {David}, {de Laverny}, {De Luise}, {De March}, {De Ridder}, {de Souza}, {de Teodoro}, {de Torres}, {del Peloso}, {del Pozo}, {Delbo}, {Delgado}, {Delgado}, {Delisle}, {Di Matteo}, {Diakite}, {Diener}, {Distefano}, {Dolding}, {Eappachen}, {Edvardsson}, {Enke}, {Esquej}, {Fabre}, {Fabrizio}, {Faigler}, {Fedorets}, {Fernique}, {Fienga}, {Figueras}, {Fouron}, {Fragkoudi}, {Fraile}, {Franke}, {Gai}, {Garabato}, {Garcia-Gutierrez}, {Garc{\'\i}a-Torres}, {Garofalo}, {Gavras}, {Gerlach}, {Geyer}, {Giacobbe}, {Gilmore}, {Girona},
  {Giuffrida}, {Gomel}, {Gomez}, {Gonzalez-Santamaria}, {Gonz{\'a}lez-Vidal}, {Granvik}, {Guti{\'e}rrez-S{\'a}nchez}, {Guy}, {Hauser}, {Haywood}, {Helmi}, {Hidalgo}, {Hilger}, {H{\l}adczuk}, {Hobbs}, {Holland}, {Huckle}, {Jasniewicz}, {Jonker}, {Juaristi Campillo}, {Julbe}, {Karbevska}, {Kervella}, {Khanna}, {Kochoska}, {Kontizas}, {Kordopatis}, {Korn}, {Kostrzewa-Rutkowska}, {Kruszy{\'n}ska}, {Lambert}, {Lanza}, {Lasne}, {Le Campion}, {Le Fustec}, {Lebreton}, {Lebzelter}, {Leccia}, {Leclerc}, {Lecoeur-Taibi}, {Liao}, {Licata}, {Lindstr{\o}m}, {Lister}, {Livanou}, {Lobel}, {Madrero Pardo}, {Managau}, {Mann}, {Marchant}, {Marconi}, {Marcos Santos}, {Marinoni}, {Marocco}, {Marshall}, {Martin Polo}, {Mart{\'\i}n-Fleitas}, {Masip}, {Massari}, {Mastrobuono-Battisti}, {Mazeh}, {McMillan}, {Messina}, {Michalik}, {Millar}, {Mints}, {Molina}, {Molinaro}, {Moln{\'a}r}, {Montegriffo}, {Mor}, {Morbidelli}, {Morel}, {Morris}, {Mulone}, {Munoz}, {Muraveva}, {Murphy}, {Musella}, {Noval}, {Ord{\'e}novic}, {Orr{\`u}},
  {Osinde}, {Pagani}, {Pagano}, {Palaversa}, {Palicio}, {Panahi}, {Pawlak}, {Pe{\~n}alosa Esteller}, {Penttil{\"a}}, {Piersimoni}, {Pineau}, {Plachy}, {Plum}, {Poggio}, {Poretti}, {Poujoulet}, {Pr{\v{s}}a}, {Pulone}, {Racero}, {Ragaini}, {Rainer}, {Raiteri}, {Rambaux}, {Ramos}, {Ramos-Lerate}, {Re Fiorentin}, {Regibo}, {Reyl{\'e}}, {Ripepi}, {Riva}, {Rixon}, {Robichon}, {Robin}, {Roelens}, {Rohrbasser}, {Romero-G{\'o}mez}, {Rowell}, {Royer}, {Rybicki}, {Sadowski}, {Sagrist{\`a} Sell{\'e}s}, {Sahlmann}, {Salgado}, {Salguero}, {Samaras}, {Sanchez Gimenez}, {Sanna}, {Santove{\~n}a}, {Sarasso}, {Schultheis}, {Sciacca}, {Segol}, {Segovia}, {S{\'e}gransan}, {Semeux}, {Shahaf}, {Siddiqui}, {Siebert}, {Siltala}, {Slezak}, {Smart}, {Solano}, {Solitro}, {Souami}, {Souchay}, {Spagna}, {Spoto}, {Steele}, {Steidelm{\"u}ller}, {Stephenson}, {S{\"u}veges}, {Szabados}, {Szegedi-Elek}, {Taris}, {Tauran}, {Taylor}, {Teixeira}, {Thuillot}, {Tonello}, {Torra}, {Torra}, {Turon}, {Unger}, {Vaillant}, {van Dillen}, {Vanel},
  {Vecchiato}, {Viala}, {Vicente}, {Voutsinas}, {Weiler}, {Wevers}, {Wyrzykowski}, {Yoldas}, {Yvard}, {Zhao}, {Zorec}, {Zucker}, {Zurbach}, \& {Zwitter}}]{gaiaedr3summary}
{Gaia Collaboration}, {Brown}, A.~G.~A., {Vallenari}, A., {et~al.} 2021{\natexlab{a}}, \aap, 649, A1

\bibitem[{{Gaia Collaboration} {et~al.}(2021{\natexlab{b}}){Gaia Collaboration}, {Luri}, {Chemin}, {Clementini}, {Delgado}, {McMillan}, {Romero-G{\'o}mez}, {Balbinot}, {Castro-Ginard}, {Mor}, {Ripepi}, {Sarro}, {Cioni}, {Fabricius}, {Garofalo}, {Helmi}, {Muraveva}, {Brown}, {Vallenari}, {Prusti}, {de Bruijne}, {Babusiaux}, {Biermann}, {Creevey}, {Evans}, {Eyer}, {Hutton}, {Jansen}, {Jordi}, {Klioner}, {Lammers}, {Lindegren}, {Mignard}, {Panem}, {Pourbaix}, {Randich}, {Sartoretti}, {Soubiran}, {Walton}, {Arenou}, {Bailer-Jones}, {Bastian}, {Cropper}, {Drimmel}, {Katz}, {Lattanzi}, {van Leeuwen}, {Bakker}, {Casta{\~n}eda}, {De Angeli}, {Ducourant}, {Fouesneau}, {Fr{\'e}mat}, {Guerra}, {Guerrier}, {Guiraud}, {Jean-Antoine Piccolo}, {Masana}, {Messineo}, {Mowlavi}, {Nicolas}, {Nienartowicz}, {Pailler}, {Panuzzo}, {Riclet}, {Roux}, {Seabroke}, {Sordo}, {Tanga}, {Th{\'e}venin}, {Gracia-Abril}, {Portell}, {Teyssier}, {Altmann}, {Andrae}, {Bellas-Velidis}, {Benson}, {Berthier}, {Blomme}, {Brugaletta}, {Burgess},
  {Busso}, {Carry}, {Cellino}, {Cheek}, {Damerdji}, {Davidson}, {Delchambre}, {Dell'Oro}, {Fern{\'a}ndez-Hern{\'a}ndez}, {Galluccio}, {Garc{\'\i}a-Lario}, {Garcia-Reinaldos}, {Gonz{\'a}lez-N{\'u}{\~n}ez}, {Gosset}, {Haigron}, {Halbwachs}, {Hambly}, {Harrison}, {Hatzidimitriou}, {Heiter}, {Hern{\'a}ndez}, {Hestroffer}, {Hodgkin}, {Holl}, {Jan{\ss}en}, {Jevardat de Fombelle}, {Jordan}, {Krone-Martins}, {Lanzafame}, {L{\"o}ffler}, {Lorca}, {Manteiga}, {Marchal}, {Marrese}, {Moitinho}, {Mora}, {Muinonen}, {Osborne}, {Pancino}, {Pauwels}, {Recio-Blanco}, {Richards}, {Riello}, {Rimoldini}, {Robin}, {Roegiers}, {Rybizki}, {Siopis}, {Smith}, {Sozzetti}, {Ulla}, {Utrilla}, {van Leeuwen}, {van Reeven}, {Abbas}, {Abreu Aramburu}, {Accart}, {Aerts}, {Aguado}, {Ajaj}, {Altavilla}, {{\'A}lvarez}, {{\'A}lvarez Cid-Fuentes}, {Alves}, {Anderson}, {Anglada Varela}, {Antoja}, {Audard}, {Baines}, {Baker}, {Balaguer-N{\'u}{\~n}ez}, {Balog}, {Barache}, {Barbato}, {Barros}, {Barstow}, {Bartolom{\'e}}, {Bassilana}, {Bauchet},
  {Baudesson-Stella}, {Becciani}, {Bellazzini}, {Bernet}, {Bertone}, {Bianchi}, {Blanco-Cuaresma}, {Boch}, {Bombrun}, {Bossini}, {Bouquillon}, {Bragaglia}, {Bramante}, {Breedt}, {Bressan}, {Brouillet}, {Bucciarelli}, {Burlacu}, {Busonero}, {Butkevich}, {Buzzi}, {Caffau}, {Cancelliere}, {C{\'a}novas}, {Cantat-Gaudin}, {Carballo}, {Carlucci}, {Carnerero}, {Carrasco}, {Casamiquela}, {Castellani}, {Castro Sampol}, {Chaoul}, {Charlot}, {Chiavassa}, {Comoretto}, {Cooper}, {Cornez}, {Cowell}, {Crifo}, {Crosta}, {Crowley}, {Dafonte}, {Dapergolas}, {David}, {David}, {de Laverny}, {De Luise}, {De March}, {De Ridder}, {de Souza}, {de Teodoro}, {de Torres}, {del Peloso}, {del Pozo}, {Delgado}, {Delisle}, {Di Matteo}, {Diakite}, {Diener}, {Distefano}, {Dolding}, {Eappachen}, {Enke}, {Esquej}, {Fabre}, {Fabrizio}, {Faigler}, {Fedorets}, {Fernique}, {Fienga}, {Figueras}, {Fouron}, {Fragkoudi}, {Fraile}, {Franke}, {Gai}, {Garabato}, {Garcia-Gutierrez}, {Garc{\'\i}a-Torres}, {Gavras}, {Gerlach}, {Geyer}, {Giacobbe},
  {Gilmore}, {Girona}, {Giuffrida}, {Gomez}, {Gonzalez-Santamaria}, {Gonz{\'a}lez-Vidal}, {Granvik}, {Guti{\'e}rrez-S{\'a}nchez}, {Guy}, {Hauser}, {Haywood}, {Hidalgo}, {Hilger}, {H{\l}adczuk}, {Hobbs}, {Holland}, {Huckle}, {Jasniewicz}, {Jonker}, {Juaristi Campillo}, {Julbe}, {Karbevska}, {Kervella}, {Khanna}, {Kochoska}, {Kontizas}, {Kordopatis}, {Korn}, {Kostrzewa-Rutkowska}, {Kruszy{\'n}ska}, {Lambert}, {Lanza}, {Lasne}, {Le Campion}, {Le Fustec}, {Lebreton}, {Lebzelter}, {Leccia}, {Leclerc}, {Lecoeur-Taibi}, {Liao}, {Licata}, {Lindstr{\o}m}, {Lister}, {Livanou}, {Lobel}, {Madrero Pardo}, {Managau}, {Mann}, {Marchant}, {Marconi}, {Marcos Santos}, {Marinoni}, {Marocco}, {Marshall}, {Martin Polo}, {Mart{\'\i}n-Fleitas}, {Masip}, {Massari}, {Mastrobuono-Battisti}, {Mazeh}, {Messina}, {Michalik}, {Millar}, {Mints}, {Molina}, {Molinaro}, {Moln{\'a}r}, {Montegriffo}, {Morbidelli}, {Morel}, {Morris}, {Mulone}, {Munoz}, {Murphy}, {Musella}, {Noval}, {Ord{\'e}novic}, {Orr{\`u}}, {Osinde}, {Pagani}, {Pagano},
  {Palaversa}, {Palicio}, {Panahi}, {Pawlak}, {Pe{\~n}alosa Esteller}, {Penttil{\"a}}, {Piersimoni}, {Pineau}, {Plachy}, {Plum}, {Poggio}, {Poretti}, {Poujoulet}, {Pr{\v{s}}a}, {Pulone}, {Racero}, {Ragaini}, {Rainer}, {Raiteri}, {Rambaux}, {Ramos}, {Ramos-Lerate}, {Re Fiorentin}, {Regibo}, {Reyl{\'e}}, {Riva}, {Rixon}, {Robichon}, {Robin}, {Roelens}, {Rohrbasser}, {Rowell}, {Royer}, {Rybicki}, {Sadowski}, {Sagrist{\`a} Sell{\'e}s}, {Sahlmann}, {Salgado}, {Salguero}, {Samaras}, {Gimenez}, {Sanna}, {Santove{\~n}a}, {Sarasso}, {Schultheis}, {Sciacca}, {Segol}, {Segovia}, {S{\'e}gransan}, {Semeux}, {Siddiqui}, {Siebert}, {Siltala}, {Slezak}, {Smart}, {Solano}, {Solitro}, {Souami}, {Souchay}, {Spagna}, {Spoto}, {Steele}, {Steidelm{\"u}ller}, {Stephenson}, {S{\"u}veges}, {Szabados}, {Szegedi-Elek}, {Taris}, {Tauran}, {Taylor}, {Teixeira}, {Thuillot}, {Tonello}, {Torra}, {Torra}, {Turon}, {Unger}, {Vaillant}, {van Dillen}, {Vanel}, {Vecchiato}, {Viala}, {Vicente}, {Voutsinas}, {Weiler}, {Wevers}, {Wyrzykowski},
  {Yoldas}, {Yvard}, {Zhao}, {Zorec}, {Zucker}, {Zurbach}, \& {Zwitter}}]{luri20}
{Gaia Collaboration}, {Luri}, X., {Chemin}, L., {et~al.} 2021{\natexlab{b}}, \aap, 649, A7

\bibitem[{{Gaia Collaboration} {et~al.}(2016){Gaia Collaboration}, {Prusti}, {de Bruijne}, {Brown}, {Vallenari}, {Babusiaux}, {Bailer-Jones}, {Bastian}, {Biermann}, {Evans}, {Eyer}, {Jansen}, {Jordi}, {Klioner}, {Lammers}, {Lindegren}, {Luri}, {Mignard}, {Milligan}, {Panem}, {Poinsignon}, {Pourbaix}, {Randich}, {Sarri}, {Sartoretti}, {Siddiqui}, {Soubiran}, {Valette}, {van Leeuwen}, {Walton}, {Aerts}, {Arenou}, {Cropper}, {Drimmel}, {H{\o}g}, {Katz}, {Lattanzi}, {O'Mullane}, {Grebel}, {Holland}, {Huc}, {Passot}, {Bramante}, {Cacciari}, {Casta{\~n}eda}, {Chaoul}, {Cheek}, {De Angeli}, {Fabricius}, {Guerra}, {Hern{\'a}ndez}, {Jean-Antoine-Piccolo}, {Masana}, {Messineo}, {Mowlavi}, {Nienartowicz}, {Ord{\'o}{\~n}ez-Blanco}, {Panuzzo}, {Portell}, {Richards}, {Riello}, {Seabroke}, {Tanga}, {Th{\'e}venin}, {Torra}, {Els}, {Gracia-Abril}, {Comoretto}, {Garcia-Reinaldos}, {Lock}, {Mercier}, {Altmann}, {Andrae}, {Astraatmadja}, {Bellas-Velidis}, {Benson}, {Berthier}, {Blomme}, {Busso}, {Carry}, {Cellino}, {Clementini},
  {Cowell}, {Creevey}, {Cuypers}, {Davidson}, {De Ridder}, {de Torres}, {Delchambre}, {Dell'Oro}, {Ducourant}, {Fr{\'e}mat}, {Garc{\'\i}a-Torres}, {Gosset}, {Halbwachs}, {Hambly}, {Harrison}, {Hauser}, {Hestroffer}, {Hodgkin}, {Huckle}, {Hutton}, {Jasniewicz}, {Jordan}, {Kontizas}, {Korn}, {Lanzafame}, {Manteiga}, {Moitinho}, {Muinonen}, {Osinde}, {Pancino}, {Pauwels}, {Petit}, {Recio-Blanco}, {Robin}, {Sarro}, {Siopis}, {Smith}, {Smith}, {Sozzetti}, {Thuillot}, {van Reeven}, {Viala}, {Abbas}, {Abreu Aramburu}, {Accart}, {Aguado}, {Allan}, {Allasia}, {Altavilla}, {{\'A}lvarez}, {Alves}, {Anderson}, {Andrei}, {Anglada Varela}, {Antiche}, {Antoja}, {Ant{\'o}n}, {Arcay}, {Atzei}, {Ayache}, {Bach}, {Baker}, {Balaguer-N{\'u}{\~n}ez}, {Barache}, {Barata}, {Barbier}, {Barblan}, {Baroni}, {Barrado y Navascu{\'e}s}, {Barros}, {Barstow}, {Becciani}, {Bellazzini}, {Bellei}, {Bello Garc{\'\i}a}, {Belokurov}, {Bendjoya}, {Berihuete}, {Bianchi}, {Bienaym{\'e}}, {Billebaud}, {Blagorodnova}, {Blanco-Cuaresma}, {Boch},
  {Bombrun}, {Borrachero}, {Bouquillon}, {Bourda}, {Bouy}, {Bragaglia}, {Breddels}, {Brouillet}, {Br{\"u}semeister}, {Bucciarelli}, {Budnik}, {Burgess}, {Burgon}, {Burlacu}, {Busonero}, {Buzzi}, {Caffau}, {Cambras}, {Campbell}, {Cancelliere}, {Cantat-Gaudin}, {Carlucci}, {Carrasco}, {Castellani}, {Charlot}, {Charnas}, {Charvet}, {Chassat}, {Chiavassa}, {Clotet}, {Cocozza}, {Collins}, {Collins}, {Costigan}, {Crifo}, {Cross}, {Crosta}, {Crowley}, {Dafonte}, {Damerdji}, {Dapergolas}, {David}, {David}, {De Cat}, {de Felice}, {de Laverny}, {De Luise}, {De March}, {de Martino}, {de Souza}, {Debosscher}, {del Pozo}, {Delbo}, {Delgado}, {Delgado}, {di Marco}, {Di Matteo}, {Diakite}, {Distefano}, {Dolding}, {Dos Anjos}, {Drazinos}, {Dur{\'a}n}, {Dzigan}, {Ecale}, {Edvardsson}, {Enke}, {Erdmann}, {Escolar}, {Espina}, {Evans}, {Eynard Bontemps}, {Fabre}, {Fabrizio}, {Faigler}, {Falc{\~a}o}, {Farr{\`a}s Casas}, {Faye}, {Federici}, {Fedorets}, {Fern{\'a}ndez-Hern{\'a}ndez}, {Fernique}, {Fienga}, {Figueras}, {Filippi},
  {Findeisen}, {Fonti}, {Fouesneau}, {Fraile}, {Fraser}, {Fuchs}, {Furnell}, {Gai}, {Galleti}, {Galluccio}, {Garabato}, {Garc{\'\i}a-Sedano}, {Gar{\'e}}, {Garofalo}, {Garralda}, {Gavras}, {Gerssen}, {Geyer}, {Gilmore}, {Girona}, {Giuffrida}, {Gomes}, {Gonz{\'a}lez-Marcos}, {Gonz{\'a}lez-N{\'u}{\~n}ez}, {Gonz{\'a}lez-Vidal}, {Granvik}, {Guerrier}, {Guillout}, {Guiraud}, {G{\'u}rpide}, {Guti{\'e}rrez-S{\'a}nchez}, {Guy}, {Haigron}, {Hatzidimitriou}, {Haywood}, {Heiter}, {Helmi}, {Hobbs}, {Hofmann}, {Holl}, {Holland}, {Hunt}, {Hypki}, {Icardi}, {Irwin}, {Jevardat de Fombelle}, {Jofr{\'e}}, {Jonker}, {Jorissen}, {Julbe}, {Karampelas}, {Kochoska}, {Kohley}, {Kolenberg}, {Kontizas}, {Koposov}, {Kordopatis}, {Koubsky}, {Kowalczyk}, {Krone-Martins}, {Kudryashova}, {Kull}, {Bachchan}, {Lacoste-Seris}, {Lanza}, {Lavigne}, {Le Poncin-Lafitte}, {Lebreton}, {Lebzelter}, {Leccia}, {Leclerc}, {Lecoeur-Taibi}, {Lemaitre}, {Lenhardt}, {Leroux}, {Liao}, {Licata}, {Lindstr{\o}m}, {Lister}, {Livanou}, {Lobel}, {L{\"o}ffler},
  {L{\'o}pez}, {Lopez-Lozano}, {Lorenz}, {Loureiro}, {MacDonald}, {Magalh{\~a}es Fernandes}, {Managau}, {Mann}, {Mantelet}, {Marchal}, {Marchant}, {Marconi}, {Marie}, {Marinoni}, {Marrese}, {Marschalk{\'o}}, {Marshall}, {Mart{\'\i}n-Fleitas}, {Martino}, {Mary}, {Matijevi{\v{c}}}, {Mazeh}, {McMillan}, {Messina}, {Mestre}, {Michalik}, {Millar}, {Miranda}, {Molina}, {Molinaro}, {Molinaro}, {Moln{\'a}r}, {Moniez}, {Montegriffo}, {Monteiro}, {Mor}, {Mora}, {Morbidelli}, {Morel}, {Morgenthaler}, {Morley}, {Morris}, {Mulone}, {Muraveva}, {Musella}, {Narbonne}, {Nelemans}, {Nicastro}, {Noval}, {Ord{\'e}novic}, {Ordieres-Mer{\'e}}, {Osborne}, {Pagani}, {Pagano}, {Pailler}, {Palacin}, {Palaversa}, {Parsons}, {Paulsen}, {Pecoraro}, {Pedrosa}, {Pentik{\"a}inen}, {Pereira}, {Pichon}, {Piersimoni}, {Pineau}, {Plachy}, {Plum}, {Poujoulet}, {Pr{\v{s}}a}, {Pulone}, {Ragaini}, {Rago}, {Rambaux}, {Ramos-Lerate}, {Ranalli}, {Rauw}, {Read}, {Regibo}, {Renk}, {Reyl{\'e}}, {Ribeiro}, {Rimoldini}, {Ripepi}, {Riva}, {Rixon},
  {Roelens}, {Romero-G{\'o}mez}, {Rowell}, {Royer}, {Rudolph}, {Ruiz-Dern}, {Sadowski}, {Sagrist{\`a} Sell{\'e}s}, {Sahlmann}, {Salgado}, {Salguero}, {Sarasso}, {Savietto}, {Schnorhk}, {Schultheis}, {Sciacca}, {Segol}, {Segovia}, {Segransan}, {Serpell}, {Shih}, {Smareglia}, {Smart}, {Smith}, {Solano}, {Solitro}, {Sordo}, {Soria Nieto}, {Souchay}, {Spagna}, {Spoto}, {Stampa}, {Steele}, {Steidelm{\"u}ller}, {Stephenson}, {Stoev}, {Suess}, {S{\"u}veges}, {Surdej}, {Szabados}, {Szegedi-Elek}, {Tapiador}, {Taris}, {Tauran}, {Taylor}, {Teixeira}, {Terrett}, {Tingley}, {Trager}, {Turon}, {Ulla}, {Utrilla}, {Valentini}, {van Elteren}, {Van Hemelryck}, {van Leeuwen}, {Varadi}, {Vecchiato}, {Veljanoski}, {Via}, {Vicente}, {Vogt}, {Voss}, {Votruba}, {Voutsinas}, {Walmsley}, {Weiler}, {Weingrill}, {Werner}, {Wevers}, {Whitehead}, {Wyrzykowski}, {Yoldas}, {{\v{Z}}erjal}, {Zucker}, {Zurbach}, {Zwitter}, {Alecu}, {Allen}, {Allende Prieto}, {Amorim}, {Anglada-Escud{\'e}}, {Arsenijevic}, {Azaz}, {Balm}, {Beck}, {Bernstein},
  {Bigot}, {Bijaoui}, {Blasco}, {Bonfigli}, {Bono}, {Boudreault}, {Bressan}, {Brown}, {Brunet}, {Bunclark}, {Buonanno}, {Butkevich}, {Carret}, {Carrion}, {Chemin}, {Ch{\'e}reau}, {Corcione}, {Darmigny}, {de Boer}, {de Teodoro}, {de Zeeuw}, {Delle Luche}, {Domingues}, {Dubath}, {Fodor}, {Fr{\'e}zouls}, {Fries}, {Fustes}, {Fyfe}, {Gallardo}, {Gallegos}, {Gardiol}, {Gebran}, {Gomboc}, {G{\'o}mez}, {Grux}, {Gueguen}, {Heyrovsky}, {Hoar}, {Iannicola}, {Isasi Parache}, {Janotto}, {Joliet}, {Jonckheere}, {Keil}, {Kim}, {Klagyivik}, {Klar}, {Knude}, {Kochukhov}, {Kolka}, {Kos}, {Kutka}, {Lainey}, {LeBouquin}, {Liu}, {Loreggia}, {Makarov}, {Marseille}, {Martayan}, {Martinez-Rubi}, {Massart}, {Meynadier}, {Mignot}, {Munari}, {Nguyen}, {Nordlander}, {Ocvirk}, {O'Flaherty}, {Olias Sanz}, {Ortiz}, {Osorio}, {Oszkiewicz}, {Ouzounis}, {Palmer}, {Park}, {Pasquato}, {Peltzer}, {Peralta}, {P{\'e}turaud}, {Pieniluoma}, {Pigozzi}, {Poels}, {Prat}, {Prod'homme}, {Raison}, {Rebordao}, {Risquez}, {Rocca-Volmerange}, {Rosen},
  {Ruiz-Fuertes}, {Russo}, {Sembay}, {Serraller Vizcaino}, {Short}, {Siebert}, {Silva}, {Sinachopoulos}, {Slezak}, {Soffel}, {Sosnowska}, {Strai{\v{z}}ys}, {ter Linden}, {Terrell}, {Theil}, {Tiede}, {Troisi}, {Tsalmantza}, {Tur}, {Vaccari}, {Vachier}, {Valles}, {Van Hamme}, {Veltz}, {Virtanen}, {Wallut}, {Wichmann}, {Wilkinson}, {Ziaeepour}, \& {Zschocke}}]{gaiadr2mission}
{Gaia Collaboration}, {Prusti}, T., {de Bruijne}, J.~H.~J., {et~al.} 2016, \aap, 595, A1

\bibitem[{{Gaia Collaboration} {et~al.}(2023){Gaia Collaboration}, {Vallenari}, {Brown}, {Prusti}, {de Bruijne}, {Arenou}, {Babusiaux}, {Biermann}, {Creevey}, {Ducourant}, {Evans}, {Eyer}, {Guerra}, {Hutton}, {Jordi}, {Klioner}, {Lammers}, {Lindegren}, {Luri}, {Mignard}, {Panem}, {Pourbaix}, {Randich}, {Sartoretti}, {Soubiran}, {Tanga}, {Walton}, {Bailer-Jones}, {Bastian}, {Drimmel}, {Jansen}, {Katz}, {Lattanzi}, {van Leeuwen}, {Bakker}, {Cacciari}, {Casta{\~n}eda}, {De Angeli}, {Fabricius}, {Fouesneau}, {Fr{\'e}mat}, {Galluccio}, {Guerrier}, {Heiter}, {Masana}, {Messineo}, {Mowlavi}, {Nicolas}, {Nienartowicz}, {Pailler}, {Panuzzo}, {Riclet}, {Roux}, {Seabroke}, {Sordo}, {Th{\'e}venin}, {Gracia-Abril}, {Portell}, {Teyssier}, {Altmann}, {Andrae}, {Audard}, {Bellas-Velidis}, {Benson}, {Berthier}, {Blomme}, {Burgess}, {Busonero}, {Busso}, {C{\'a}novas}, {Carry}, {Cellino}, {Cheek}, {Clementini}, {Damerdji}, {Davidson}, {de Teodoro}, {Nu{\~n}ez Campos}, {Delchambre}, {Dell'Oro}, {Esquej},
  {Fern{\'a}ndez-Hern{\'a}ndez}, {Fraile}, {Garabato}, {Garc{\'\i}a-Lario}, {Gosset}, {Haigron}, {Halbwachs}, {Hambly}, {Harrison}, {Hern{\'a}ndez}, {Hestroffer}, {Hodgkin}, {Holl}, {Jan{\ss}en}, {Jevardat de Fombelle}, {Jordan}, {Krone-Martins}, {Lanzafame}, {L{\"o}ffler}, {Marchal}, {Marrese}, {Moitinho}, {Muinonen}, {Osborne}, {Pancino}, {Pauwels}, {Recio-Blanco}, {Reyl{\'e}}, {Riello}, {Rimoldini}, {Roegiers}, {Rybizki}, {Sarro}, {Siopis}, {Smith}, {Sozzetti}, {Utrilla}, {van Leeuwen}, {Abbas}, {{\'A}brah{\'a}m}, {Abreu Aramburu}, {Aerts}, {Aguado}, {Ajaj}, {Aldea-Montero}, {Altavilla}, {{\'A}lvarez}, {Alves}, {Anders}, {Anderson}, {Anglada Varela}, {Antoja}, {Baines}, {Baker}, {Balaguer-N{\'u}{\~n}ez}, {Balbinot}, {Balog}, {Barache}, {Barbato}, {Barros}, {Barstow}, {Bartolom{\'e}}, {Bassilana}, {Bauchet}, {Becciani}, {Bellazzini}, {Berihuete}, {Bernet}, {Bertone}, {Bianchi}, {Binnenfeld}, {Blanco-Cuaresma}, {Blazere}, {Boch}, {Bombrun}, {Bossini}, {Bouquillon}, {Bragaglia}, {Bramante}, {Breedt},
  {Bressan}, {Brouillet}, {Brugaletta}, {Bucciarelli}, {Burlacu}, {Butkevich}, {Buzzi}, {Caffau}, {Cancelliere}, {Cantat-Gaudin}, {Carballo}, {Carlucci}, {Carnerero}, {Carrasco}, {Casamiquela}, {Castellani}, {Castro-Ginard}, {Chaoul}, {Charlot}, {Chemin}, {Chiaramida}, {Chiavassa}, {Chornay}, {Comoretto}, {Contursi}, {Cooper}, {Cornez}, {Cowell}, {Crifo}, {Cropper}, {Crosta}, {Crowley}, {Dafonte}, {Dapergolas}, {David}, {David}, {de Laverny}, {De Luise}, \& {De March}}]{gaiadr3summary}
{Gaia Collaboration}, {Vallenari}, A., {Brown}, A.~G.~A., {et~al.} 2023, \aap, 674, A1

\bibitem[{{Garavito-Camargo} {et~al.}(2019){Garavito-Camargo}, {Besla}, {Laporte}, {Johnston}, {G{\'o}mez}, \& {Watkins}}]{garavito-camargo19}
{Garavito-Camargo}, N., {Besla}, G., {Laporte}, C. F.~P., {et~al.} 2019, \apj, 884, 51

\bibitem[{{Garavito-Camargo} {et~al.}(2021){Garavito-Camargo}, {Besla}, {Laporte}, {Price-Whelan}, {Cunningham}, {Johnston}, {Weinberg}, \& {G{\'o}mez}}]{garavito-camargo21}
{Garavito-Camargo}, N., {Besla}, G., {Laporte}, C. F.~P., {et~al.} 2021, \apj, 919, 109

\bibitem[{{Gatto} {et~al.}(2020){Gatto}, {Napolitano}, {Spiniello}, {Longo}, \& {Paolillo}}]{gatto20}
{Gatto}, M., {Napolitano}, N.~R., {Spiniello}, C., {Longo}, G., \& {Paolillo}, M. 2020, \aap, 644, A134

\bibitem[{{Gibbons} {et~al.}(2014){Gibbons}, {Belokurov}, \& {Evans}}]{Gibbons2014}
{Gibbons}, S.~L.~J., {Belokurov}, V., \& {Evans}, N.~W. 2014, \mnras, 445, 3788

\bibitem[{{Graczyk} {et~al.}(2014){Graczyk}, {Pietrzy{\'n}ski}, {Thompson}, {Gieren}, {Pilecki}, {Konorski}, {Udalski}, {Soszy{\'n}ski}, {Villanova}, {G{\'o}rski}, {Suchomska}, {Karczmarek}, {Kudritzki}, {Bresolin}, \& {Gallenne}}]{graczyk14}
{Graczyk}, D., {Pietrzy{\'n}ski}, G., {Thompson}, I.~B., {et~al.} 2014, \apj, 780, 59

\bibitem[{{Grillmair} \& {Dionatos}(2006)}]{grillmair_dionatos06}
{Grillmair}, C.~J. \& {Dionatos}, O. 2006, \apjl, 643, L17

\bibitem[{Harris {et~al.}(2020)Harris, Millman, van~der Walt, Gommers, Virtanen, Cournapeau, Wieser, Taylor, Berg, Smith, Kern, Picus, Hoyer, van Kerkwijk, Brett, Haldane, del R{\'{i}}o, Wiebe, Peterson, G{\'{e}}rard-Marchant, Sheppard, Reddy, Weckesser, Abbasi, Gohlke, \& Oliphant}]{numpy}
Harris, C.~R., Millman, K.~J., van~der Walt, S.~J., {et~al.} 2020, Nature, 585, 357

\bibitem[{Hunter(2007)}]{matplotlib}
Hunter, J.~D. 2007, Computing in Science \& Engineering, 9, 90

\bibitem[{{Ibata} {et~al.}(2024){Ibata}, {Malhan}, {Tenachi}, {Ardern-Arentsen}, {Bellazzini}, {Bianchini}, {Bonifacio}, {Caffau}, {Diakogiannis}, {Errani}, {Famaey}, {Ferrone}, {Martin}, {di Matteo}, {Monari}, {Renaud}, {Starkenburg}, {Thomas}, {Viswanathan}, \& {Yuan}}]{ibata24}
{Ibata}, R., {Malhan}, K., {Tenachi}, W., {et~al.} 2024, \apj, 967, 89

\bibitem[{{Iorio} \& {Belokurov}(2021)}]{Iorio.RRlyrae.2021}
{Iorio}, G. \& {Belokurov}, V. 2021, \mnras, 502, 5686

\bibitem[{{Ivezi{\'c}} {et~al.}(2019){Ivezi{\'c}}, {Kahn}, {Tyson}, {Abel}, {Acosta}, {Allsman}, {Alonso}, {AlSayyad}, {Anderson}, {Andrew}, {Angel}, {Angeli}, {Ansari}, {Antilogus}, {Araujo}, {Armstrong}, {Arndt}, {Astier}, {Aubourg}, {Auza}, {Axelrod}, {Bard}, {Barr}, {Barrau}, {Bartlett}, {Bauer}, {Bauman}, {Baumont}, {Bechtol}, {Bechtol}, {Becker}, {Becla}, {Beldica}, {Bellavia}, {Bianco}, {Biswas}, {Blanc}, {Blazek}, {Blandford}, {Bloom}, {Bogart}, {Bond}, {Booth}, {Borgland}, {Borne}, {Bosch}, {Boutigny}, {Brackett}, {Bradshaw}, {Brandt}, {Brown}, {Bullock}, {Burchat}, {Burke}, {Cagnoli}, {Calabrese}, {Callahan}, {Callen}, {Carlin}, {Carlson}, {Chandrasekharan}, {Charles-Emerson}, {Chesley}, {Cheu}, {Chiang}, {Chiang}, {Chirino}, {Chow}, {Ciardi}, {Claver}, {Cohen-Tanugi}, {Cockrum}, {Coles}, {Connolly}, {Cook}, {Cooray}, {Covey}, {Cribbs}, {Cui}, {Cutri}, {Daly}, {Daniel}, {Daruich}, {Daubard}, {Daues}, {Dawson}, {Delgado}, {Dellapenna}, {de Peyster}, {de Val-Borro}, {Digel}, {Doherty}, {Dubois},
  {Dubois-Felsmann}, {Durech}, {Economou}, {Eifler}, {Eracleous}, {Emmons}, {Fausti Neto}, {Ferguson}, {Figueroa}, {Fisher-Levine}, {Focke}, {Foss}, {Frank}, {Freemon}, {Gangler}, {Gawiser}, {Geary}, {Gee}, {Geha}, {Gessner}, {Gibson}, {Gilmore}, {Glanzman}, {Glick}, {Goldina}, {Goldstein}, {Goodenow}, {Graham}, {Gressler}, {Gris}, {Guy}, {Guyonnet}, {Haller}, {Harris}, {Hascall}, {Haupt}, {Hernandez}, {Herrmann}, {Hileman}, {Hoblitt}, {Hodgson}, {Hogan}, {Howard}, {Huang}, {Huffer}, {Ingraham}, {Innes}, {Jacoby}, {Jain}, {Jammes}, {Jee}, {Jenness}, {Jernigan}, {Jevremovi{\'c}}, {Johns}, {Johnson}, {Johnson}, {Jones}, {Juramy-Gilles}, {Juri{\'c}}, {Kalirai}, {Kallivayalil}, {Kalmbach}, {Kantor}, {Karst}, {Kasliwal}, {Kelly}, {Kessler}, {Kinnison}, {Kirkby}, {Knox}, {Kotov}, {Krabbendam}, {Krughoff}, {Kub{\'a}nek}, {Kuczewski}, {Kulkarni}, {Ku}, {Kurita}, {Lage}, {Lambert}, {Lange}, {Langton}, {Le Guillou}, {Levine}, {Liang}, {Lim}, {Lintott}, {Long}, {Lopez}, {Lotz}, {Lupton}, {Lust}, {MacArthur}, {Mahabal},
  {Mandelbaum}, {Markiewicz}, {Marsh}, {Marshall}, {Marshall}, {May}, {McKercher}, {McQueen}, {Meyers}, {Migliore}, {Miller}, {Mills}, {Miraval}, {Moeyens}, {Moolekamp}, {Monet}, {Moniez}, {Monkewitz}, {Montgomery}, {Morrison}, {Mueller}, {Muller}, {Mu{\~n}oz Arancibia}, {Neill}, {Newbry}, {Nief}, {Nomerotski}, {Nordby}, {O'Connor}, {Oliver}, {Olivier}, {Olsen}, {O'Mullane}, {Ortiz}, {Osier}, {Owen}, {Pain}, {Palecek}, {Parejko}, {Parsons}, {Pease}, {Peterson}, {Peterson}, {Petravick}, {Libby Petrick}, {Petry}, {Pierfederici}, {Pietrowicz}, {Pike}, {Pinto}, {Plante}, {Plate}, {Plutchak}, {Price}, {Prouza}, {Radeka}, {Rajagopal}, {Rasmussen}, {Regnault}, {Reil}, {Reiss}, {Reuter}, {Ridgway}, {Riot}, {Ritz}, {Robinson}, {Roby}, {Roodman}, {Rosing}, {Roucelle}, {Rumore}, {Russo}, {Saha}, {Sassolas}, {Schalk}, {Schellart}, {Schindler}, {Schmidt}, {Schneider}, {Schneider}, {Schoening}, {Schumacher}, {Schwamb}, {Sebag}, {Selvy}, {Sembroski}, {Seppala}, {Serio}, {Serrano}, {Shaw}, {Shipsey}, {Sick}, {Silvestri},
  {Slater}, {Smith}, {Smith}, {Sobhani}, {Soldahl}, {Storrie-Lombardi}, {Stover}, {Strauss}, {Street}, {Stubbs}, {Sullivan}, {Sweeney}, {Swinbank}, {Szalay}, {Takacs}, {Tether}, {Thaler}, {Thayer}, {Thomas}, {Thornton}, {Thukral}, {Tice}, {Trilling}, {Turri}, {Van Berg}, {Vanden Berk}, {Vetter}, {Virieux}, {Vucina}, {Wahl}, {Walkowicz}, {Walsh}, {Walter}, {Wang}, {Wang}, {Warner}, {Wiecha}, {Willman}, {Winters}, {Wittman}, {Wolff}, {Wood-Vasey}, {Wu}, {Xin}, {Yoachim}, \& {Zhan}}]{ivezic19}
{Ivezi{\'c}}, {\v{Z}}., {Kahn}, S.~M., {Tyson}, J.~A., {et~al.} 2019, \apj, 873, 111

\bibitem[{{Jim{\'e}nez-Arranz} {et~al.}(2024{\natexlab{a}}){Jim{\'e}nez-Arranz}, {Chemin}, {Romero-G{\'o}mez}, {Luri}, {Adamczyk}, {Castro-Ginard}, {Roca-F{\`a}brega}, {McMillan}, \& {Cioni}}]{jimenez-arranz24a}
{Jim{\'e}nez-Arranz}, {\'O}., {Chemin}, L., {Romero-G{\'o}mez}, M., {et~al.} 2024{\natexlab{a}}, \aap, 683, A102

\bibitem[{{Jim{\'e}nez-Arranz} {et~al.}(2025){Jim{\'e}nez-Arranz}, {Horta}, {van der Marel}, {Nidever}, {Laporte}, {Patel}, \& {Rix}}]{jimenez-arranz25a}
{Jim{\'e}nez-Arranz}, {\'O}., {Horta}, D., {van der Marel}, R.~P., {et~al.} 2025, arXiv e-prints, arXiv:2501.04616

\bibitem[{{Jim{\'e}nez-Arranz} \& {Roca-F{\`a}brega}(2025)}]{jimenez-arranz25b}
{Jim{\'e}nez-Arranz}, {\'O}. \& {Roca-F{\`a}brega}, S. 2025, \aap, 698, L7

\bibitem[{{Jim{\'e}nez-Arranz} {et~al.}(2024{\natexlab{b}}){Jim{\'e}nez-Arranz}, {Roca-F{\`a}brega}, {Romero-G{\'o}mez}, {Luri}, {Bernet}, {McMillan}, \& {Chemin}}]{jimenez-arranz24b}
{Jim{\'e}nez-Arranz}, {\'O}., {Roca-F{\`a}brega}, S., {Romero-G{\'o}mez}, M., {et~al.} 2024{\natexlab{b}}, \aap, 688, A51

\bibitem[{{Jim{\'e}nez-Arranz} {et~al.}(2023{\natexlab{a}}){Jim{\'e}nez-Arranz}, {Romero-G{\'o}mez}, {Luri}, \& {Masana}}]{jimenez-arranz23b}
{Jim{\'e}nez-Arranz}, {\'O}., {Romero-G{\'o}mez}, M., {Luri}, X., \& {Masana}, E. 2023{\natexlab{a}}, \aap, 672, A65

\bibitem[{{Jim{\'e}nez-Arranz} {et~al.}(2023{\natexlab{b}}){Jim{\'e}nez-Arranz}, {Romero-G{\'o}mez}, {Luri}, {McMillan}, {Antoja}, {Chemin}, {Roca-F{\`a}brega}, {Masana}, \& {Muros}}]{jimenez-arranz23a}
{Jim{\'e}nez-Arranz}, {\'O}., {Romero-G{\'o}mez}, M., {Luri}, X., {et~al.} 2023{\natexlab{b}}, \aap, 669, A91

\bibitem[{{Johnston} {et~al.}(2005){Johnston}, {Law}, \& {Majewski}}]{johnston05}
{Johnston}, K.~V., {Law}, D.~R., \& {Majewski}, S.~R. 2005, \apj, 619, 800

\bibitem[{{Johnston} {et~al.}(1999){Johnston}, {Majewski}, {Siegel}, {Reid}, \& {Kunkel}}]{johnston99}
{Johnston}, K.~V., {Majewski}, S.~R., {Siegel}, M.~H., {Reid}, I.~N., \& {Kunkel}, W.~E. 1999, \aj, 118, 1719

\bibitem[{{Kacharov} {et~al.}(2024){Kacharov}, {Tahmasebzadeh}, {Cioni}, {van de Ven}, {Zhu}, \& {Khoperskov}}]{kacharov24}
{Kacharov}, N., {Tahmasebzadeh}, B., {Cioni}, M.-R.~L., {et~al.} 2024, \aap, 692, A40

\bibitem[{{Kado-Fong} {et~al.}(2018){Kado-Fong}, {Greene}, {Hendel}, {Price-Whelan}, {Greco}, {Goulding}, {Huang}, {Johnston}, {Komiyama}, {Lee}, {Lust}, {Strauss}, \& {Tanaka}}]{kado-fong18}
{Kado-Fong}, E., {Greene}, J.~E., {Hendel}, D., {et~al.} 2018, \apj, 866, 103

\bibitem[{Kingma \& Ba(2017)}]{kingma17}
Kingma, D.~P. \& Ba, J. 2017, Adam: A Method for Stochastic Optimization

\bibitem[{Kluyver {et~al.}(2016)Kluyver, Ragan-Kelley, P{\'e}rez, Granger, Bussonnier, Frederic, Kelley, Hamrick, Grout, Corlay, Ivanov, Avila, Abdalla, Willing, \& development team}]{jupyter}
Kluyver, T., Ragan-Kelley, B., P{\'e}rez, F., {et~al.} 2016, in Positioning and Power in Academic Publishing: Players, Agents and Agendas, ed. F.~Loizides \& B.~Scmidt (Netherlands: IOS Press), 87--90

\bibitem[{{Kollmeier} {et~al.}(2025){Kollmeier}, {Rix}, {Aerts}, {Aird}, {Alfaro}, {Almeida}, {Anderson}, {Jim{\'e}nez Arranz}, {Arseneau}, {Assef}, {Aviram}, {Aydar}, {Badenes}, {Bandyopadhyay}, {Barger}, {Barkhouser}, {Bauer}, {Bender}, {Besser}, {Bhattarai}, {Bilgi}, {Bird}, {Bizyaev}, {Blanc}, {Blanton}, {Bochanski}, {Bovy}, {Brandon}, {Brandt}, {Brownstein}, {Buchner}, {Burchett}, {Carlberg}, {Casey}, {Castaneda-Carlos}, {Chakraborty}, {Chanam{\'e}}, {Chandra}, {Cherinka}, {Chilingarian}, {Comparat}, {Cosens}, {Covey}, {Crane}, {Crumpler}, {Cunha}, {Cunningham}, {Dai}, {Darling}, {Davidson}, {Davis}, {De Lee}, {Deacon}, {M{\'e}ndez Delgado}, {Demasi}, {Demianenko}, {Derwent}, {D'Onghia}, {Di Mille}, {Dias}, {Donor}, {Drory}, {Dwelly}, {Egorov}, {Egorova}, {El-Badry}, {Engelman}, {Eracleous}, {Fan}, {Farr}, {Fries}, {Frinchaboy}, {Froning}, {G{\"a}nsicke}, {Garc{\'\i}a}, {Gelfand}, {Gentile Fusillo}, {Glover}, {Grabowski}, {Grebel}, {Green}, {Grier}, {Gupta}, {Gray}, {H{\"a}berle}, {Hall}, {Hammond},
  {Hawkins}, {Harding}, {Heged{\H{u}}s}, {Herbst}, {Hermes}, {Rodr{\'\i}guez Hidalgo}, {Hilder}, {Hogg}, {Holtzman}, {Horta}, {Huang}, {Hwang}, {Ibarra-Medel}, {Imig}, {Inight}, {Jana}, {Ji}, {Jofre}, {Johns}, {Johnson}, {Johnson}, {Johnston}, {Jones}, {Katkov}, {Koekemoer}, {Kounkel}, {Kreckel}, {Krishnarao}, {Krumpe}, {Kumari}, {Kupfer}, {Lacerna}, {Laporte}, {Lepine}, {Li}, {Liu}, {Loebman}, {Long}, {Roman-Lopes}, {Lu}, {Majewski}, {Maoz}, {McKinnon}, {Medan}, {Merloni}, {Minniti}, {Morrison}, {Myers}, {M{\'e}sz{\'a}ros}, {Nandra}, {Nayak}, {Ness}, {Nidever}, {O'Brien}, {Oeur}, {Oravetz}, {Oravetz}, {Otto}, {Adamane Pallathadka}, {Palunas}, {Pan}, {Pappalardo}, {Pandey}, {Negrete Pe{\~n}aloza}, {Pinsonneault}, {Pogge}, {Taghizadeh Popp}, {Price-Whelan}, {Pulatova}, {Qiu}, {Ramirez}, {Rankine}, {Ricci}, {Runnoe}, {Sanchez}, {Salvato}, {Sattler}, {Saydjari}, {Sayres}, {Schlaufman}, {Schneider}, {Schreiber}, {Schwope}, {Serna}, {Shen}, {Sif{\'o}n}, {Singh}, {Sinha}, {Smee}, {Song}, {Souto}, {Stassun},
  {Steinmetz}, {Stone-Martinez}, {Stringfellow}, {Stutz}, {Jos{\'e}}, {S{\'a}}, {nchez-Gallego}, {Tan}, {Tayar}, {Thai}, {Thakar}, {Ting}, {Tkachenko}, {Tovmasian}, {Trakhtenbrot}, {Fern{\'a}ndez-Trincado}, {Troup}, {Trump}, {Tuttle}, {van der Marel}, \& {Villanova}}]{kollmeier25}
{Kollmeier}, J.~A., {Rix}, H.-W., {Aerts}, C., {et~al.} 2025, arXiv e-prints, arXiv:2507.06989

\bibitem[{{Kollmeier} {et~al.}(2017){Kollmeier}, {Zasowski}, {Rix}, {Johns}, {Anderson}, {Drory}, {Johnson}, {Pogge}, {Bird}, {Blanc}, {Brownstein}, {Crane}, {De Lee}, {Klaene}, {Kreckel}, {MacDonald}, {Merloni}, {Ness}, {O'Brien}, {Sanchez-Gallego}, {Sayres}, {Shen}, {Thakar}, {Tkachenko}, {Aerts}, {Blanton}, {Eisenstein}, {Holtzman}, {Maoz}, {Nandra}, {Rockosi}, {Weinberg}, {Bovy}, {Casey}, {Chaname}, {Clerc}, {Conroy}, {Eracleous}, {G{\"a}nsicke}, {Hekker}, {Horne}, {Kauffmann}, {McQuinn}, {Pellegrini}, {Schinnerer}, {Schlafly}, {Schwope}, {Seibert}, {Teske}, \& {van Saders}}]{kollmeier17}
{Kollmeier}, J.~A., {Zasowski}, G., {Rix}, H.-W., {et~al.} 2017, arXiv e-prints, arXiv:1711.03234

\bibitem[{{Koposov} {et~al.}(2019){Koposov}, {Belokurov}, {Li}, {Mateu}, {Erkal}, {Grillmair}, {Hendel}, {Price-Whelan}, {Laporte}, {Hawkins}, {Sohn}, {del Pino}, {Evans}, {Slater}, {Kallivayalil}, {Navarro}, \& {Orphan Aspen Treasury Collaboration}}]{koposov19}
{Koposov}, S.~E., {Belokurov}, V., {Li}, T.~S., {et~al.} 2019, \mnras, 485, 4726

\bibitem[{{Koposov} {et~al.}(2015){Koposov}, {Belokurov}, {Torrealba}, \& {Evans}}]{koposov15}
{Koposov}, S.~E., {Belokurov}, V., {Torrealba}, G., \& {Evans}, N.~W. 2015, \apj, 805, 130

\bibitem[{{Koposov} {et~al.}(2023){Koposov}, {Erkal}, {Li}, {Da Costa}, {Cullinane}, {Ji}, {Kuehn}, {Lewis}, {Pace}, {Shipp}, {Zucker}, {Bland-Hawthorn}, {Lilleengen}, {Martell}, \& {S5 Collaboration}}]{koposov23}
{Koposov}, S.~E., {Erkal}, D., {Li}, T.~S., {et~al.} 2023, \mnras, 521, 4936

\bibitem[{{Koposov} {et~al.}(2010){Koposov}, {Rix}, \& {Hogg}}]{koposov10}
{Koposov}, S.~E., {Rix}, H.-W., \& {Hogg}, D.~W. 2010, \apj, 712, 260

\bibitem[{{Law} \& {Majewski}(2010)}]{law-majewski10}
{Law}, D.~R. \& {Majewski}, S.~R. 2010, \apj, 714, 229

\bibitem[{{Lilleengen} {et~al.}(2023){Lilleengen}, {Petersen}, {Erkal}, {Pe{\~n}arrubia}, {Koposov}, {Li}, {Cullinane}, {Ji}, {Kuehn}, {Lewis}, {Mackey}, {Pace}, {Shipp}, {Zucker}, {Bland-Hawthorn}, {Hilmi}, \& {S5 Collaboration}}]{Lilleengen2023}
{Lilleengen}, S., {Petersen}, M.~S., {Erkal}, D., {et~al.} 2023, \mnras, 518, 774

\bibitem[{{Lucchini} {et~al.}(2021){Lucchini}, {D'Onghia}, \& {Fox}}]{lucchini21}
{Lucchini}, S., {D'Onghia}, E., \& {Fox}, A.~J. 2021, \apjl, 921, L36

\bibitem[{{Lucchini} {et~al.}(2020){Lucchini}, {D'Onghia}, {Fox}, {Bustard}, {Bland-Hawthorn}, \& {Zweibel}}]{lucchini20}
{Lucchini}, S., {D'Onghia}, E., {Fox}, A.~J., {et~al.} 2020, \nat, 585, 203

\bibitem[{Lundberg \& Lee(2017)}]{shap}
Lundberg, S.~M. \& Lee, S.-I. 2017, in Advances in Neural Information Processing Systems, Vol.~30 (Curran Associates, Inc.)

\bibitem[{{Malhan} \& {Ibata}(2018)}]{malhan-ibata18}
{Malhan}, K. \& {Ibata}, R.~A. 2018, \mnras, 477, 4063

\bibitem[{{Mart{\'\i}nez-Delgado} {et~al.}(2023){Mart{\'\i}nez-Delgado}, {Cooper}, {Rom{\'a}n}, {Pillepich}, {Erkal}, {Pearson}, {Moustakas}, {Laporte}, {Laine}, {Akhlaghi}, {Lang}, {Makarov}, {Borlaff}, {Donatiello}, {Pearson}, {Mir{\'o}-Carretero}, {Cuillandre}, {Dom{\'\i}nguez}, {Roca-F{\`a}brega}, {Frenk}, {Schmidt}, {G{\'o}mez-Flechoso}, {Guzman}, {Libeskind}, {Dey}, {Weaver}, {Schlegel}, {Myers}, \& {Valdes}}]{martínez-delgado23}
{Mart{\'\i}nez-Delgado}, D., {Cooper}, A.~P., {Rom{\'a}n}, J., {et~al.} 2023, \aap, 671, A141

\bibitem[{{Mart{\'\i}nez-Delgado} {et~al.}(2010){Mart{\'\i}nez-Delgado}, {Gabany}, {Crawford}, {Zibetti}, {Majewski}, {Rix}, {Fliri}, {Carballo-Bello}, {Bardalez-Gagliuffi}, {Pe{\~n}arrubia}, {Chonis}, {Madore}, {Trujillo}, {Schirmer}, \& {McDavid}}]{martínez-delgado10}
{Mart{\'\i}nez-Delgado}, D., {Gabany}, R.~J., {Crawford}, K., {et~al.} 2010, \aj, 140, 962

\bibitem[{{Martinez-Delgado} {et~al.}(2025){Martinez-Delgado}, {Stein}, {Sakowska}, {Weigelt}, {Roman}, {Donatiello}, {Roca-Fabrega}, {Schirmer}, {Grebel}, {Saifollahi}, {Kanipe}, {Gomez-Flechoso}, {Akhlaghi}, {Javanmardi}, {Wu}, {Eskandarlou}, {Bomans}, {Henkel}, {Block}, {Hanson}, {Schedler}, {Teuwen}, {GaBany}, {Iba{\~n}ez Perez}, {Crawford}, {Promper}, {Jimenez}, {Farras-Aloy}, \& {Miro-Carretero}}]{martínez-delgado25}
{Martinez-Delgado}, D., {Stein}, M., {Sakowska}, J.~D., {et~al.} 2025, arXiv e-prints, arXiv:2504.02071

\bibitem[{{Mateu}(2023)}]{mateu23}
{Mateu}, C. 2023, \mnras, 520, 5225

\bibitem[{McKinney(2010)}]{pandas...paper}
McKinney, W. 2010, in Proceedings of the 9th Python in Science Conference, ed. S.~van~der Walt \& J.~Millman, 56 -- 61

\bibitem[{{Mir{\'o}-Carretero} {et~al.}(2024){Mir{\'o}-Carretero}, {Mart{\'\i}nez-Delgado}, {G{\'o}mez-Flechoso}, {Cooper}, {Akhlaghi}, {Donatiello}, {Kuijken}, {Lang}, {Makarov}, {Laine}, \& {Roca-F{\`a}brega}}]{miro-carretero24}
{Mir{\'o}-Carretero}, J., {Mart{\'\i}nez-Delgado}, D., {G{\'o}mez-Flechoso}, M.~A., {et~al.} 2024, \aap, 691, A196

\bibitem[{{Navarrete} {et~al.}(2023){Navarrete}, {Aguado}, {Belokurov}, {Erkal}, {Deason}, {Cullinane}, \& {Carballo-Bello}}]{navarrete23}
{Navarrete}, C., {Aguado}, D.~S., {Belokurov}, V., {et~al.} 2023, \mnras, 523, 4720

\bibitem[{{Navarrete} {et~al.}(2019){Navarrete}, {Belokurov}, {Catelan}, {Jethwa}, {Koposov}, {Carballo-Bello}, {Jofr{\'e}}, {Erkal}, {Duffau}, \& {Corral-Santana}}]{navarrete19}
{Navarrete}, C., {Belokurov}, V., {Catelan}, M., {et~al.} 2019, \mnras, 483, 4160

\bibitem[{{Necib} {et~al.}(2020){Necib}, {Ostdiek}, {Lisanti}, {Cohen}, {Freytsis}, \& {Garrison-Kimmel}}]{necib20}
{Necib}, L., {Ostdiek}, B., {Lisanti}, M., {et~al.} 2020, \apj, 903, 25

\bibitem[{{Newberg} {et~al.}(2002){Newberg}, {Yanny}, {Rockosi}, {Grebel}, {Rix}, {Brinkmann}, {Csabai}, {Hennessy}, {Hindsley}, {Ibata}, {Ivezi{\'c}}, {Lamb}, {Nash}, {Odenkirchen}, {Rave}, {Schneider}, {Smith}, {Stolte}, \& {York}}]{newberg02}
{Newberg}, H.~J., {Yanny}, B., {Rockosi}, C., {et~al.} 2002, \apj, 569, 245

\bibitem[{{Nidever} {et~al.}(2008){Nidever}, {Majewski}, \& {Butler Burton}}]{nidever08}
{Nidever}, D.~L., {Majewski}, S.~R., \& {Butler Burton}, W. 2008, \apj, 679, 432

\bibitem[{{Patrick} {et~al.}(2022){Patrick}, {Koposov}, \& {Walker}}]{patrick22}
{Patrick}, J.~M., {Koposov}, S.~E., \& {Walker}, M.~G. 2022, \mnras, 514, 1757

\bibitem[{{Pearson} {et~al.}(2022){Pearson}, {Price-Whelan}, {Hogg}, {Seth}, {Sand}, {Hunt}, \& {Crnojevi{\'c}}}]{pearson22}
{Pearson}, S., {Price-Whelan}, A.~M., {Hogg}, D.~W., {et~al.} 2022, \apj, 941, 19

\bibitem[{Pedregosa {et~al.}(2011)Pedregosa, Varoquaux, Gramfort, Michel, Thirion, Grisel, Blondel, Prettenhofer, Weiss, Dubourg, Vanderplas, Passos, Cournapeau, Brucher, Perrot, \& Duchesnay}]{sklearn}
Pedregosa, F., Varoquaux, G., Gramfort, A., {et~al.} 2011, Journal of Machine Learning Research, 12, 2825

\bibitem[{P\'erez \& Granger(2007)}]{ipython}
P\'erez, F. \& Granger, B.~E. 2007, Computing in Science and Engineering, 9, 21

\bibitem[{{Petersen} \& {Weinberg}(2025)}]{Petesen.joss.2025}
{Petersen}, M. \& {Weinberg}, M. 2025, The Journal of Open Source Software, 10, 7302

\bibitem[{{Petersen} \& {Pe{\~n}arrubia}(2020)}]{petersen.lmc.2020}
{Petersen}, M.~S. \& {Pe{\~n}arrubia}, J. 2020, \mnras, 494, L11

\bibitem[{{Petersen} \& {Pe{\~n}arrubia}(2021)}]{petersen.lmc.2021}
{Petersen}, M.~S. \& {Pe{\~n}arrubia}, J. 2021, Nature Astronomy, 5, 251

\bibitem[{{Petersen} {et~al.}(2022{\natexlab{a}}){Petersen}, {Pe{\~n}arrubia}, \& {Jones}}]{Petersen.lmc.2022}
{Petersen}, M.~S., {Pe{\~n}arrubia}, J., \& {Jones}, E. 2022{\natexlab{a}}, \mnras, 514, 1266

\bibitem[{{Petersen} {et~al.}(2022{\natexlab{b}}){Petersen}, {Weinberg}, \& {Katz}}]{Petersen2022}
{Petersen}, M.~S., {Weinberg}, M.~D., \& {Katz}, N. 2022{\natexlab{b}}, \mnras, 510, 6201

\bibitem[{{Pettee} {et~al.}(2024){Pettee}, {Thanvantri}, {Nachman}, {Shih}, {Buckley}, \& {Collins}}]{pettee24}
{Pettee}, M., {Thanvantri}, S., {Nachman}, B., {et~al.} 2024, \mnras, 527, 8459

\bibitem[{{Pietrzy{\'n}ski} {et~al.}(2019){Pietrzy{\'n}ski}, {Graczyk}, {Gallenne}, {Gieren}, {Thompson}, {Pilecki}, {Karczmarek}, {G{\'o}rski}, {Suchomska}, {Taormina}, {Zgirski}, {Wielg{\'o}rski}, {Ko{\l}aczkowski}, {Konorski}, {Villanova}, {Nardetto}, {Kervella}, {Bresolin}, {Kudritzki}, {Storm}, {Smolec}, \& {Narloch}}]{pietrzynski19}
{Pietrzy{\'n}ski}, G., {Graczyk}, D., {Gallenne}, A., {et~al.} 2019, \nat, 567, 200

\bibitem[{{Price-Whelan} \& {Bonaca}(2018)}]{price_whelan_bonaca18}
{Price-Whelan}, A.~M. \& {Bonaca}, A. 2018, \apjl, 863, L20

\bibitem[{{Price-Whelan} {et~al.}(2014){Price-Whelan}, {Hogg}, {Johnston}, \& {Hendel}}]{price_whelan14}
{Price-Whelan}, A.~M., {Hogg}, D.~W., {Johnston}, K.~V., \& {Hendel}, D. 2014, \apj, 794, 4

\bibitem[{{Putman}(2000)}]{putman00}
{Putman}, M.~E. 2000, \pasa, 17, 1

\bibitem[{{Putman} {et~al.}(2003){Putman}, {Staveley-Smith}, {Freeman}, {Gibson}, \& {Barnes}}]{putman03b}
{Putman}, M.~E., {Staveley-Smith}, L., {Freeman}, K.~C., {Gibson}, B.~K., \& {Barnes}, D.~G. 2003, \apj, 586, 170

\bibitem[{{Rathore} {et~al.}(2025){Rathore}, {Choi}, {Olsen}, \& {Besla}}]{rathore25}
{Rathore}, H., {Choi}, Y., {Olsen}, K. A.~G., \& {Besla}, G. 2025, \apj, 978, 55

\bibitem[{Reback {et~al.}(2020)Reback, McKinney, jbrockmendel, den Bossche, Augspurger, Cloud, gfyoung, Sinhrks, Klein, Roeschke, Hawkins, Tratner, She, Ayd, Petersen, Garcia, Schendel, Hayden, MomIsBestFriend, Jancauskas, Battiston, Seabold, chris b1, h~vetinari, Hoyer, Overmeire, alimcmaster1, Dong, Whelan, \& Mehyar}]{pandas...software}
Reback, J., McKinney, W., jbrockmendel, {et~al.} 2020, pandas-dev/pandas: Pandas 1.0.3

\bibitem[{{Rockosi} {et~al.}(2002){Rockosi}, {Odenkirchen}, {Grebel}, {Dehnen}, {Cudworth}, {Gunn}, {York}, {Brinkmann}, {Hennessy}, \& {Ivezi{\'c}}}]{rockosi02}
{Rockosi}, C.~M., {Odenkirchen}, M., {Grebel}, E.~K., {et~al.} 2002, \aj, 124, 349

\bibitem[{{Sch{\"o}lch} {et~al.}(2025){Sch{\"o}lch}, {Jim{\'e}nez-Arranz}, {Romero-G{\'o}mez}, {Luri}, {Hobbs}, {Salmer{\'o}n-Larraz}, \& {L{\'o}pez Vilamaj{\'o}}}]{scholch25}
{Sch{\"o}lch}, M., {Jim{\'e}nez-Arranz}, {\'O}., {Romero-G{\'o}mez}, M., {et~al.} 2025, arXiv e-prints, arXiv:2508.01434

\bibitem[{{Shih} {et~al.}(2022){Shih}, {Buckley}, {Necib}, \& {Tamanas}}]{shih22}
{Shih}, D., {Buckley}, M.~R., {Necib}, L., \& {Tamanas}, J. 2022, \mnras, 509, 5992

\bibitem[{{Shipp} {et~al.}(2021){Shipp}, {Erkal}, {Drlica-Wagner}, {Li}, {Pace}, {Koposov}, {Cullinane}, {Da Costa}, {Ji}, {Kuehn}, {Lewis}, {Mackey}, {Simpson}, {Wan}, {Zucker}, {Bland-Hawthorn}, {Ferguson}, {Lilleengen}, \& {Lilleengen}}]{shipp21}
{Shipp}, N., {Erkal}, D., {Drlica-Wagner}, A., {et~al.} 2021, \apj, 923, 149

\bibitem[{{Spergel} {et~al.}(2015){Spergel}, {Gehrels}, {Baltay}, {Bennett}, {Breckinridge}, {Donahue}, {Dressler}, {Gaudi}, {Greene}, {Guyon}, {Hirata}, {Kalirai}, {Kasdin}, {Macintosh}, {Moos}, {Perlmutter}, {Postman}, {Rauscher}, {Rhodes}, {Wang}, {Weinberg}, {Benford}, {Hudson}, {Jeong}, {Mellier}, {Traub}, {Yamada}, {Capak}, {Colbert}, {Masters}, {Penny}, {Savransky}, {Stern}, {Zimmerman}, {Barry}, {Bartusek}, {Carpenter}, {Cheng}, {Content}, {Dekens}, {Demers}, {Grady}, {Jackson}, {Kuan}, {Kruk}, {Melton}, {Nemati}, {Parvin}, {Poberezhskiy}, {Peddie}, {Ruffa}, {Wallace}, {Whipple}, {Wollack}, \& {Zhao}}]{spergel15}
{Spergel}, D., {Gehrels}, N., {Baltay}, C., {et~al.} 2015, arXiv e-prints, arXiv:1503.03757

\bibitem[{{Springel} \& {White}(1999)}]{springel_white99}
{Springel}, V. \& {White}, S. D.~M. 1999, \mnras, 307, 162

\bibitem[{{Starkman} {et~al.}(2023){Starkman}, {Bovy}, {Webb}, {Calvetti}, \& {Somersalo}}]{starkman23}
{Starkman}, N., {Bovy}, J., {Webb}, J.~J., {Calvetti}, D., \& {Somersalo}, E. 2023, \mnras, 522, 5022

\bibitem[{{Tavangar} {et~al.}(2022){Tavangar}, {Ferguson}, {Shipp}, {Drlica-Wagner}, {Koposov}, {Erkal}, {Balbinot}, {Garc{\'\i}a-Bellido}, {Kuehn}, {Lewis}, {Li}, {Mau}, {Pace}, {Riley}, {Abbott}, {Aguena}, {Allam}, {Andrade-Oliveira}, {Annis}, {Bertin}, {Brooks}, {Burke}, {Carnero Rosell}, {Carrasco Kind}, {Carretero}, {Costanzi}, {da Costa}, {Pereira}, {De Vicente}, {Diehl}, {Everett}, {Ferrero}, {Flaugher}, {Frieman}, {Gaztanaga}, {Gerdes}, {Gruen}, {Gruendl}, {Gschwend}, {Gutierrez}, {Hinton}, {Hollowood}, {Honscheid}, {James}, {Kuropatkin}, {Maia}, {Marshall}, {Menanteau}, {Miquel}, {Morgan}, {Ogando}, {Palmese}, {Paz-Chinch{\'o}n}, {Pieres}, {Plazas Malag{\'o}n}, {Rodriguez-Monroy}, {Sanchez}, {Scarpine}, {Serrano}, {Sevilla-Noarbe}, {Smith}, {Suchyta}, {Swanson}, {Tarle}, {To}, {Varga}, \& {Walker}}]{tavangar22}
{Tavangar}, K., {Ferguson}, P., {Shipp}, N., {et~al.} 2022, \apj, 925, 118

\bibitem[{{Tavangar} \& {Price-Whelan}(2025)}]{tavangar25}
{Tavangar}, K. \& {Price-Whelan}, A.~M. 2025, arXiv e-prints, arXiv:2502.13236

\bibitem[{{The Dark Energy Survey Collaboration}(2005)}]{dark_energy_survey05}
{The Dark Energy Survey Collaboration}. 2005, arXiv e-prints, astro

\bibitem[{{van der Marel}(2001)}]{vandermarel01}
{van der Marel}, R.~P. 2001, \aj, 122, 1827

\bibitem[{{van der Velden}(2020)}]{cmasher}
{van der Velden}, E. 2020, The Journal of Open Source Software, 5, 2004

\bibitem[{{Vasiliev}(2018)}]{Vasiliev2018}
{Vasiliev}, E. 2018, \mnras, 481, L100

\bibitem[{{Vasiliev}(2024)}]{vasiliev23b}
{Vasiliev}, E. 2024, \mnras, 527, 437

\bibitem[{{Vasiliev} {et~al.}(2021){Vasiliev}, {Belokurov}, \& {Erkal}}]{vasiliev21}
{Vasiliev}, E., {Belokurov}, V., \& {Erkal}, D. 2021, \mnras, 501, 2279

\bibitem[{Virtanen {et~al.}(2020)Virtanen, Gommers, Oliphant, Haberland, Reddy, Cournapeau, Burovski, Peterson, Weckesser, Bright, {van der Walt}, Brett, Wilson, Millman, Mayorov, Nelson, Jones, Kern, Larson, Carey, Polat, Feng, Moore, {VanderPlas}, Laxalde, Perktold, Cimrman, Henriksen, Quintero, Harris, Archibald, Ribeiro, Pedregosa, {van Mulbregt}, \& {SciPy 1.0 Contributors}}]{scipy}
Virtanen, P., Gommers, R., Oliphant, T.~E., {et~al.} 2020, Nature Methods, 17, 261

\bibitem[{{Weerasooriya} {et~al.}(2025){Weerasooriya}, {Starkenburg}, {Cunningham}, \& {Johnston}}]{Weerasooriya2025}
{Weerasooriya}, S., {Starkenburg}, T., {Cunningham}, E.~C., \& {Johnston}, K.~V. 2025, arXiv e-prints, arXiv:2505.14792

\bibitem[{{Yaaqib} {et~al.}(2025){Yaaqib}, {Petersen}, \& {Pe{\~n}arrubia}}]{Yaaqib.lmc.2025}
{Yaaqib}, R., {Petersen}, M., \& {Pe{\~n}arrubia}, J. 2025, arXiv e-prints, arXiv:2508.04781

\bibitem[{{Zaritsky} {et~al.}(2025){Zaritsky}, {Chandra}, {Conroy}, {Bonaca}, {Cargile}, \& {Naidu}}]{zaritsky25}
{Zaritsky}, D., {Chandra}, V., {Conroy}, C., {et~al.} 2025, The Open Journal of Astrophysics, 8, 16

\end{thebibliography}
\label{lastpage}

\begin{appendix}

\section{S3 clean sample on-sky distribution for $P_{\rm cut} =0.5$}
\label{sec:appendix_p05}

\begin{figure*}
    \centering
    \includegraphics[width=0.9\textwidth]{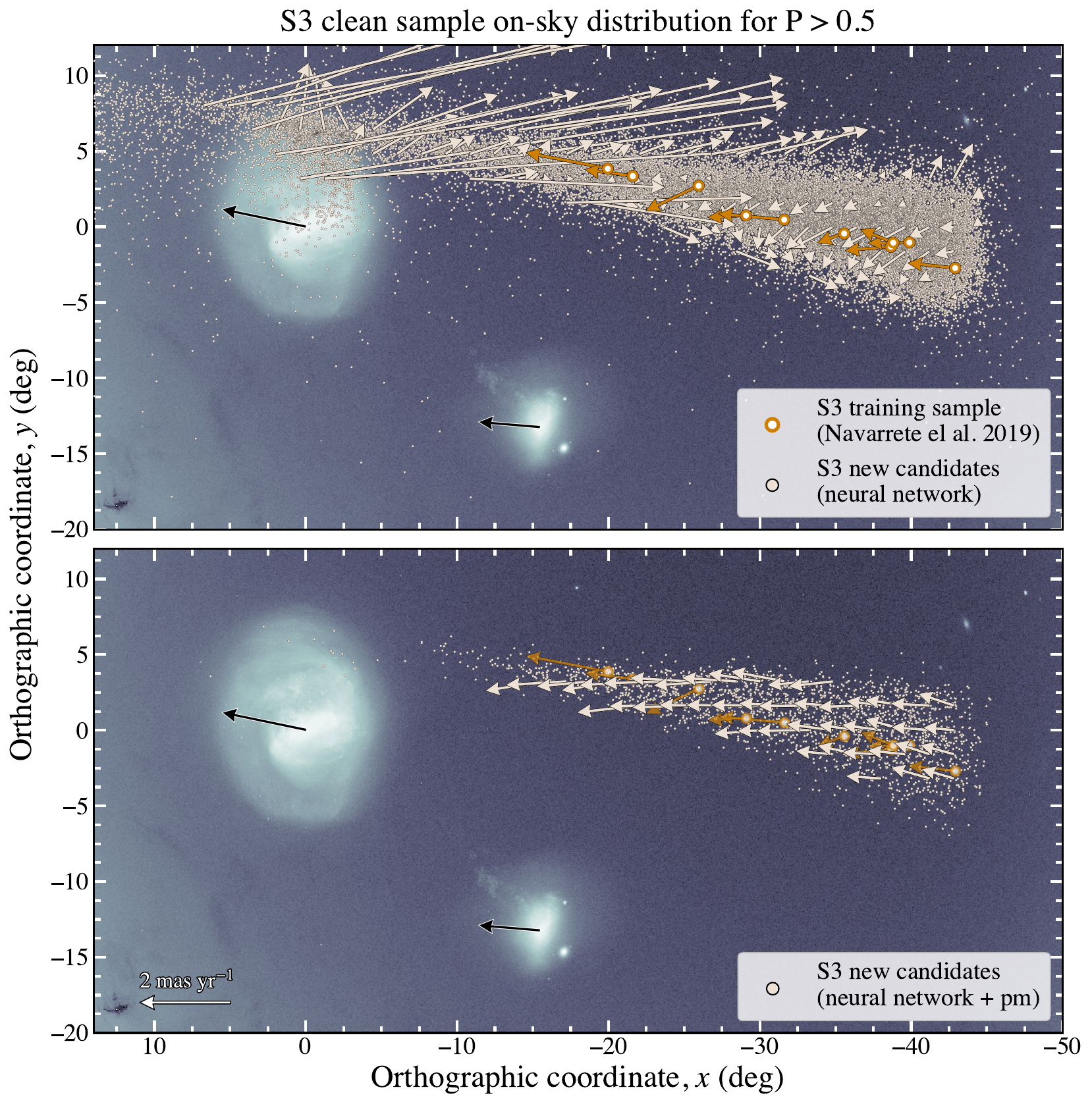}
    \caption{Same as Fig. \ref{fig:S3_skymap}, but for $P_{\rm cut} =0.5$.}
    \label{fig:S3_skymap_p05}
\end{figure*}

In this study, we adopt a probability threshold of  $P_{\text{cut}}=0.8$ for the neural network classifier, prioritizing a cleaner and less contaminated sample of S3 candidates, even at the expense of excluding some genuine members. Nevertheless, the main results presented in this work are robust across the probability threshold range of $P_{\text{cut}}=0.5-0.8$. To illustrate this, Fig. \ref{fig:S3_skymap_p05} presents the on-sky distribution of the S3 clean samples obtained with $P_{\text{cut}}=0.5$ , following the same format as Fig. \ref{fig:S3_skymap}. While the proper motion distribution of the neural network-selected sample (top panel) appears more irregular compared to the $P_{\text{cut}}=0.8$ case (see Fig. \ref{fig:S3_skymap}), the application of the polygon selection (bottom panel) yields results broadly consistent with those discussed in the main text. To facilitate further studies of the S3 stream by other researchers -- allowing them to adjust the balance between completeness and purity according to their specific goals -- the released S3 star catalogue includes the 2,177 stars with $P>0.5$ that also meet the polygon selection criteria in proper motion space.

\section{Validation and explanation of the neural network classifier}
\label{sec:appendix_nn}

\begin{figure}
    \centering
    \includegraphics[width=\columnwidth]{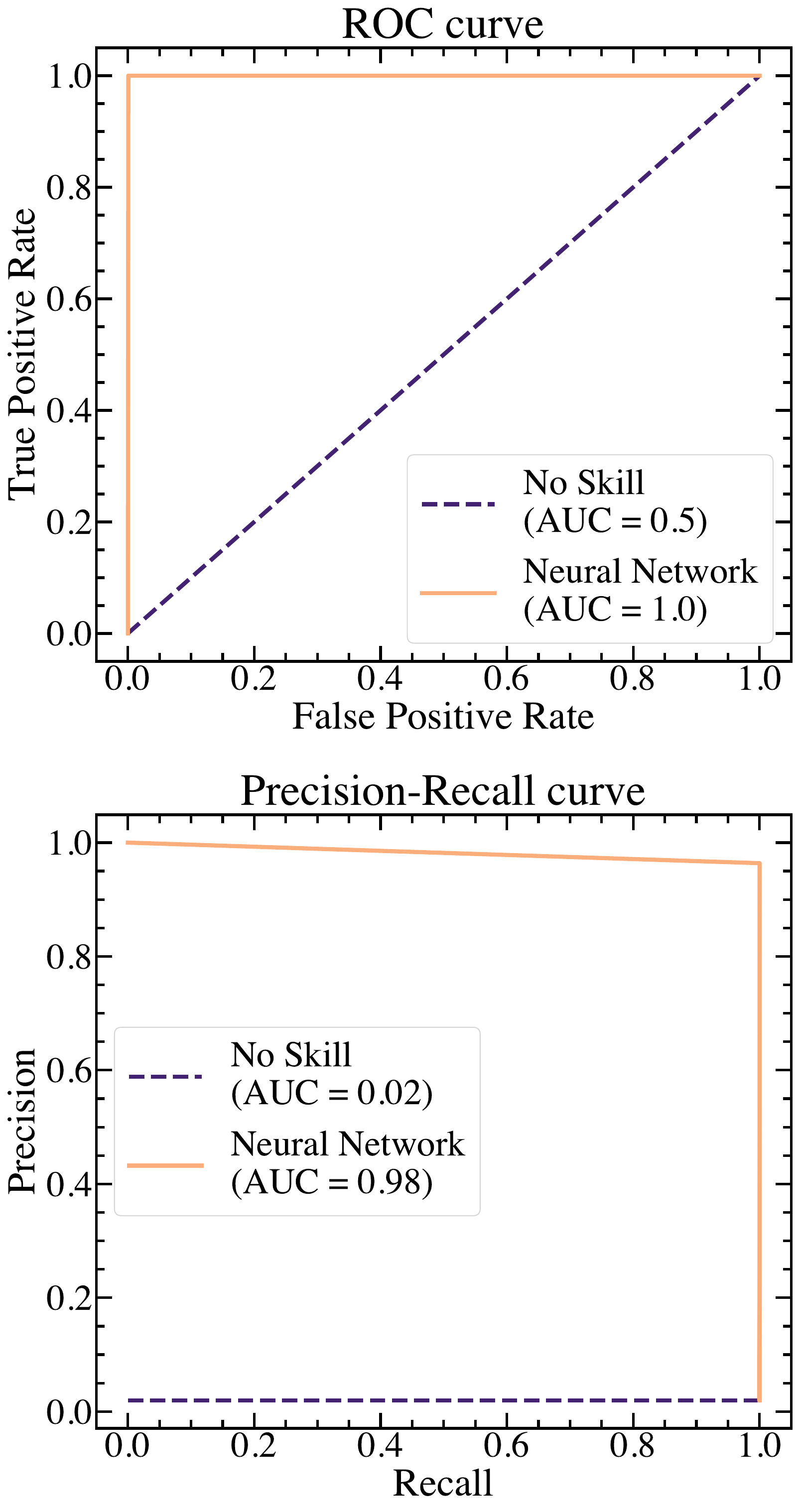}
    \caption{Evaluation metrics for the neural network classifier (see Sect. \ref{subsec:neural_network}) performance. Top: ROC curve. Bottom: precision-recall curve. In both cases, we compare our model (orange solid curve) with a classifier that has no class separation capacity (purple dashed curve).}
    \label{fig:roc}
\end{figure}

To train and assess the performance of the classifier, we divide the sample of 10,200 (including both S3 and field stars) stars into two subsets: 60\% for training the algorithm and 40\% for testing its performance. We evaluate the classifier by generating the receiver ROC curve, the precision-recall curve, and calculating their respective AUCs. The ROC curve is a key metric for evaluating classification models, illustrating the trade-off between the true positive rate and false positive rate across different probability thresholds $P_{\text{cut}}$. Its AUC value reflects the model's ability to distinguish between classes: the closer the AUC is to 1, the better the model performs. An AUC of 0.5 indicates no discriminative power. The precision-recall curve is particularly useful in scenarios with highly imbalanced classes, as in this case. Precision (the ratio of true positives to all stars classified as S3) indicates the relevance of the results, while recall (the ratio of true positives to all actual S3 stars) measures the completeness of the relevant results identified. Similar to the ROC curve, the precision-recall curve illustrates the trade-off between precision and recall across varying probability thresholds $P_{\text{cut}}$. Both the ROC curve, the precision-recall curve, and their corresponding AUC values indicate an almost perfect classifier (see Fig. \ref{fig:roc}). However, these results should be interpreted with caution, as they reflect performance on the subset of our simulated sample used for testing, rather than on the full \gaia DR3 dataset.

\begin{figure*}
    \centering
    \includegraphics[width=0.8\textwidth]{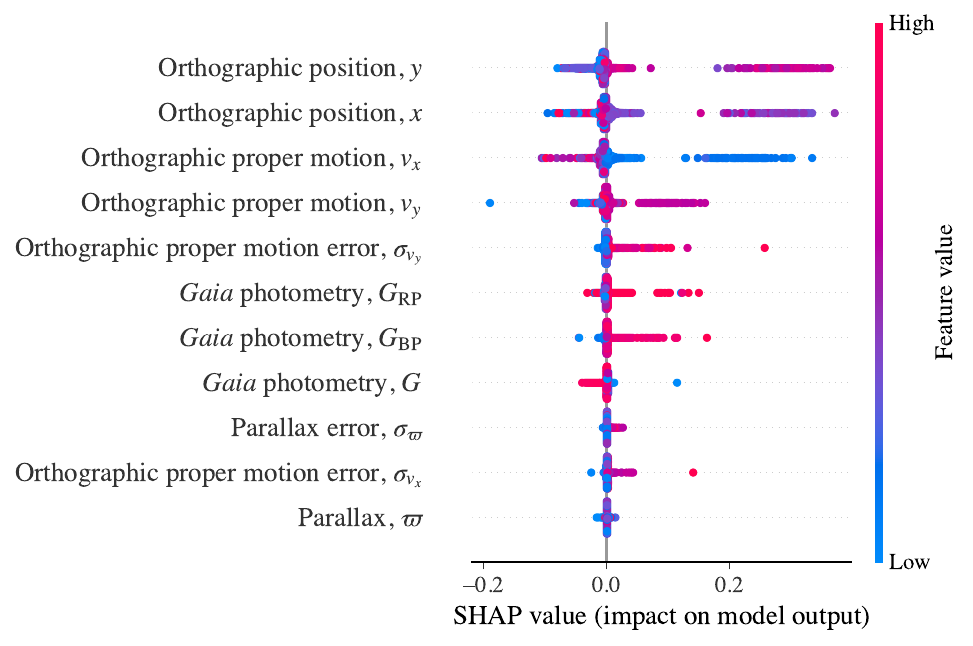}
    \caption{SHAP summary plot showing the impact of \gaia DR3 astrometric and photometric features on the S3 neural network model (see Sect. \ref{subsec:neural_network}) output. Each dot represents an individual star from the test sample, which includes 4,080 stars—accounting for 40\% of the total dataset of 10,200 stars used for training and testing. Dot color indicates the feature value (red for high, blue for low), while the position along the $x$-axis shows the SHAP value -- that is, the feature’s contribution to the classification decision. Features are ordered from top to bottom by their overall importance, with those higher in the list having greater influence on the model’s output.}
    \label{fig:shap}
\end{figure*}

The SHAP summary plot shown in Fig. \ref{fig:shap} illustrates the impact of each input feature on the S3 classifier’s output, providing a detailed view of feature importance and the direction of their influence. Each point corresponds to an individual star from the test sample -- consisting of 4,080 stars, which make up 40\% of the total 10,200 star training and testing dataset. The color indicates the feature value (blue for low values, red for high), while the position along the $x$-axis shows the SHAP value, representing that feature’s contribution to the model’s classification output. Features are ranked by their overall importance (mean absolute SHAP value), with spatial position ($x$ and $y$) and proper motion ($v_x$ and $v_y$) emerging as the most influential variables. The spread of SHAP values along the $x$-axis for each feature indicates how much variation in model output is attributable to that feature. For instance, low values (blue) of $v_x$ tend to push the prediction toward one class (positive SHAP values), while high values (red) push it in the opposite direction. This analysis highlights which features are driving the classifier's decisions and provides a level of interpretability often lacking in complex models. 

\section{Stellar populations within the S3 new candidate sample}
\label{sec:appendix_stellar_types}

In Sect. \ref{subsec:characterisation}, we analyse the different stellar populations within the S3 candidate sample using photometry. This is done by comparing the CMD of our sample with the LMC evolutionary phases defined in \cite{luri20}, shifted to a distance of 73.5 kpc -- the median distance of the \citetalias{navarrete19} training sample -- from the LMC mean distance of 49.5 kpc \citep{pietrzynski19}. Shifting the distance affects only the apparent magnitude in the CMD, as color, being related to temperature and composition, remains unchanged. The shift is applied using the distance modulus:
\begin{equation}
    m - M = 5 \log_{10}(d/10~\rm pc)
\end{equation}

The difference in apparent magnitude between the two distances is:
\begin{equation}
    \Delta m = 5 \log_{10} \left( \frac{73.5\,\mathrm{kpc}}{49.5\,\mathrm{kpc}} \right) = 0.86~\rm mag
\end{equation}

\begin{figure}
    \centering
    \includegraphics[width=\columnwidth]{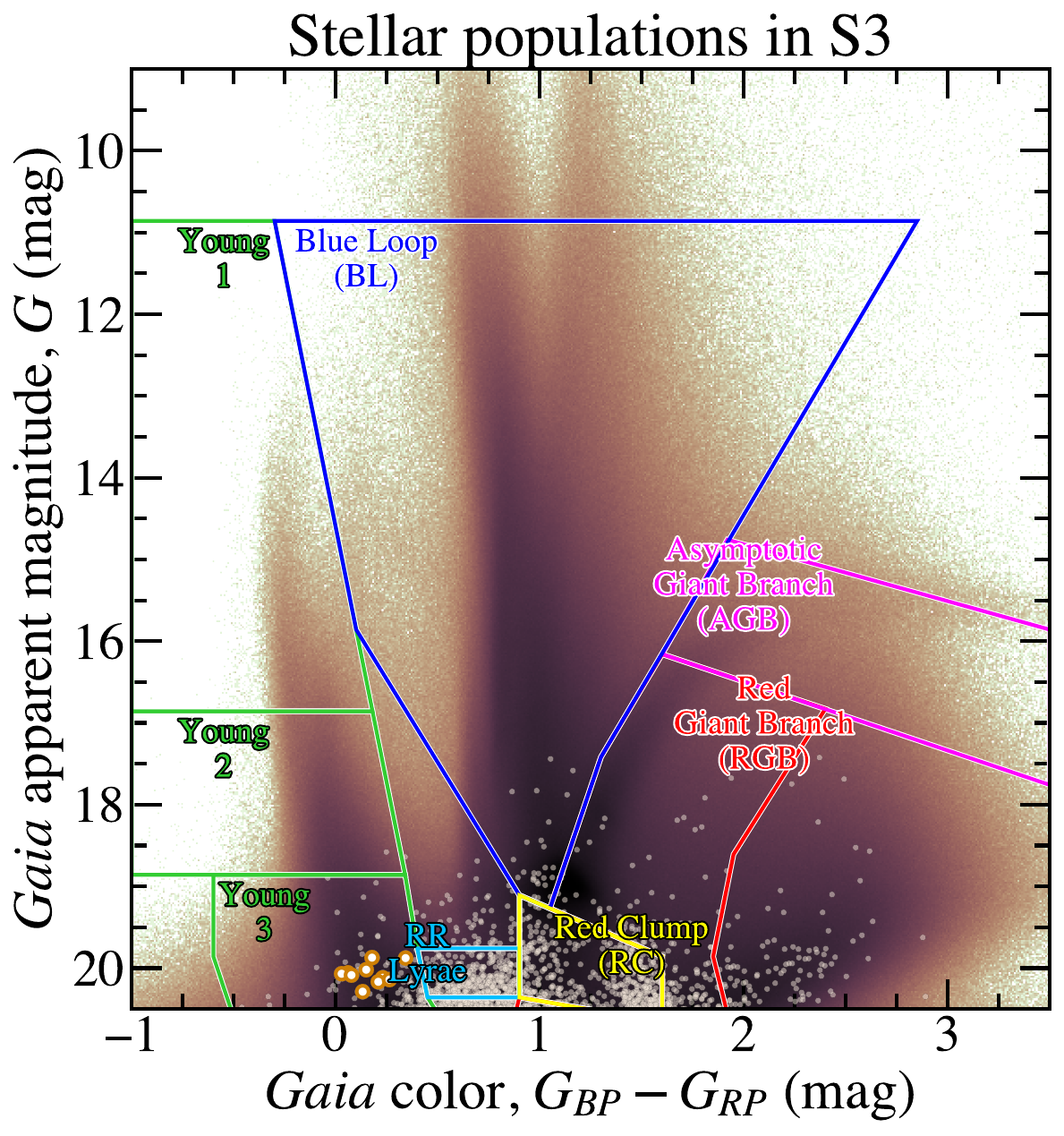}
    \caption{CMD showing the S3 training sample from \citetalias{navarrete19} (orange circles) and the 1,542 new S3 candidates (beige circles), with the LMC CMD polygons from \citet{luri20} shifted to 73.5 kpc -- the median distance of the \citetalias{navarrete19} training sample. The background image corresponds to the CMD of the \gaia DR3 sample utilized in this study (see Sect. \ref{subsec:gaiadr3}), consisting of 28 million stars that include both stars from the Clouds and foreground halo stars of the MW.}
    \label{fig:cmd_stellar_pop}
\end{figure}

Thus, the effect on the polygon selection is that the color remains unchanged, while all apparent magnitudes in the CMD shift fainter by $+0.86$ mag, as the increased distance makes the stars appear dimmer. Figure \ref{fig:cmd_stellar_pop} shows the CMD of the S3 training sample from \citetalias{navarrete19} (orange circles) and the 1,542 new S3 candidates (beige circles), with the LMC CMD polygons from \citet{luri20} shifted to 73.5 kpc -- the median distance of the \citetalias{navarrete19} training sample. We find that the sample of 1,542 new S3 stellar candidates is (as a first indication) predominantly composed of RC ($29\%$) and RR Lyrae ($25\%$) stars.

\section{RR Lyrae stars in the S3 new candidate sample}
\label{sec:appendix_rrl}

\begin{figure*}
    \centering
    \includegraphics[width=0.9\textwidth]{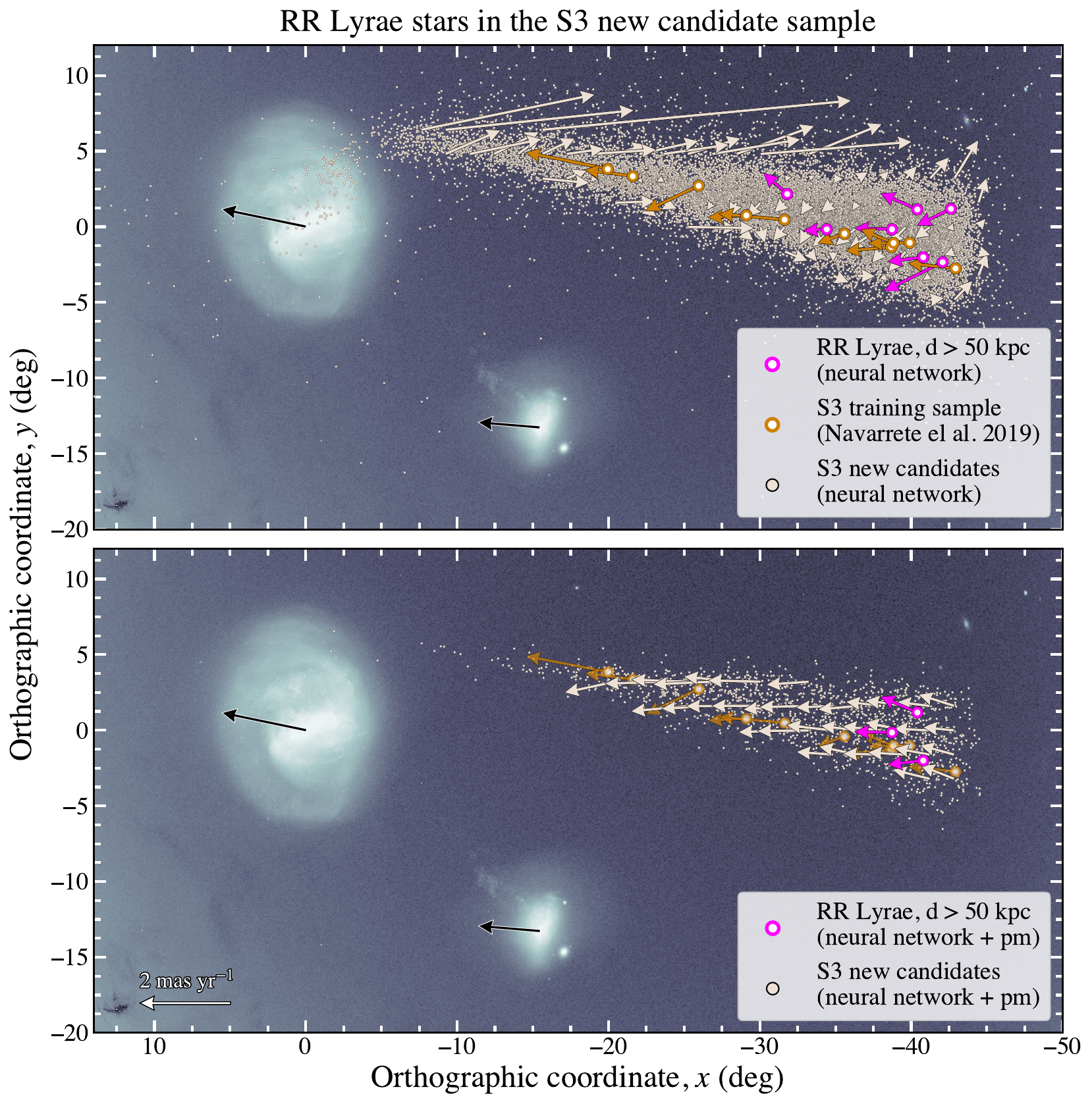}
    \caption{Same as Fig. \ref{fig:S3_skymap}, but highlighting in pink the 3 (7) RR Lyrae stars located beyond 50 kpc within the neural network S3 sample, after (before) applying the proper motion cut in the bottom (top) panel.}
    \label{fig:rrlyrae}
\end{figure*}

Given that the CMD polygon cut proposed by \citet{luri20} indicated the possible presence of RR Lyrae stars within our S3 clean sample, we attempted to crossmatch this sample with the \gaia DR3 RR Lyrae catalogue \citep[\texttt{gaiadr3.vari\_rrlyrae};][]{clementini23}, aiming to identify any overlap between the datasets. However, we found only 3 (7) RR Lyrae stars at distances greater than 50 kpc within the neural network S3 sample after (before) applying the proper motion cut. The individual distances of those 3 RR Lyrae stars are 60, 55, and 73 kpc, placing some of them on the nearer side of the S3 stream. Figure \ref{fig:rrlyrae} displays the on-sky distribution of the RR Lyrae samples (pink circles), with arrows indicating their respective proper motions, in a similar manner to Fig. \ref{fig:S3_skymap}. The orientation and length of the arrows represent the direction and magnitude of the stars’ motion across the sky. This visualization underscores both the spatial alignment and the coherent motion of the candidate members, illustrating that the stream is not only continuous in position but also coherent in proper motion space. 

\end{appendix}

\end{document}